\pdfoutput=1

\documentclass[11pt,twoside,a4paper,cmspaper,final,collab]{cms-tdr}
\def\svnVersion{347385}\def\cmsCernNoTag{CERN-PH-EP/2015-221}\def\cmsCernDate{\today}\def\cmsMessage{Published in the Journal of High Energy Physics as \href{http://dx.doi.org/10.1007/JHEP11(2015)018}{\doi{10.1007/JHEP11(2015)018.}}}

\begin{document}\cmsNoteHeader{HIG-14-023}

\hyphenation{had-ron-i-za-tion}
\hyphenation{cal-or-i-me-ter}
\hyphenation{de-vices}
\RCS$Revision: 347325 $
\RCS$HeadURL: svn+ssh://svn.cern.ch/reps/tdr2/papers/HIG-14-023/trunk/HIG-14-023.tex $
\RCS$Id: HIG-14-023.tex 347325 2016-06-15 14:45:19Z vischia $
\newlength\cmsFigWidth
\ifthenelse{\boolean{cms@external}}{\setlength\cmsFigWidth{0.85\columnwidth}}{\setlength\cmsFigWidth{0.4\textwidth}}
\ifthenelse{\boolean{cms@external}}{\providecommand{\cmsLeft}{top\xspace}}{\providecommand{\cmsLeft}{left\xspace}}
\ifthenelse{\boolean{cms@external}}{\providecommand{\cmsRight}{bottom\xspace}}{\providecommand{\cmsRight}{right\xspace}}
\cmsNoteHeader{HIG-14-023}
\title{Search for a charged Higgs boson in pp collisions at $\sqrt{s}=8\TeV$}

\date{\today}

\abstract{
A search for a charged Higgs boson is performed with a data sample corresponding to an integrated luminosity
of $19.7 \pm 0.5\fbinv$ collected with the CMS detector in proton-proton collisions at $\sqrt{s}=8\TeV$.
The charged Higgs boson is searched for in top quark decays for $m_{\mathrm H^\pm} < m_{\mathrm t} - m_{\mathrm b}$,
and in the direct production ${\rm pp \rightarrow t (b) H^\pm}$ for $m_{\mathrm H^\pm} > m_{\mathrm t} - m_{\mathrm b}$.
The ${\mathrm H^\pm\rightarrow \tau^\pm\nu_\tau}$ and ${\rm H^\pm\rightarrow t b}$ decay modes in the final states $\tau_{\rm h}$+jets,
$\mu\tau_{\rm h}$, $\ell$+jets, and $\ell\ell$' ($\ell =$e, $\mu$) are considered in the search.
No signal is observed and 95\% confidence level upper limits are set on the charged Higgs boson production.
A model-independent upper limit on the product branching fraction 
$\mathcal{B}({\rm t\rightarrow H^\pm b}) \, \mathcal{B}({\mathrm H^\pm \rightarrow \tau^\pm \nu_\tau})= 1.2$--$0.15\%$
is obtained in the mass range $m_{\mathrm H^\pm} = 80$--$160\GeV$, while the upper limit on the cross section times branching fraction
$\sigma({\rm pp \rightarrow  t (b) H^\pm}) \, \mathcal{B} ({\mathrm H^\pm \rightarrow \tau^\pm \nu_\tau})= 0.38$--$0.025\unit{pb}$ is set 
in the mass range $m_{\mathrm H^+} = 180$--$600\GeV$.
Here, $\sigma({\rm pp} \to {\rm t(b)H}^{\pm})$ stands for the cross section sum $\sigma({\rm pp} \to \overline{\rm t}({\rm b}){\rm H}^{+}) + \sigma({\rm pp} \to {\rm t}(\overline{\rm b}){\rm H}^{-})$.
Assuming $\mathcal{B}({\rm H}^\pm \rightarrow {\rm t b})=1$, an upper limit on
$\sigma ({\rm pp} \rightarrow {\rm t} ({\rm b}) {\rm H}^\pm)$ of $2.0$--$0.13\unit{pb}$ is set for $m_{\mathrm H^\pm} = 180$--$600\GeV$.
The combination of all considered decay modes and final states is used to set exclusion limits in the $m_{\mathrm H^\pm}$--$\tan\beta$ parameter space in different MSSM benchmark scenarios.
}

\hypersetup{%
pdfauthor={CMS Collaboration},%
pdftitle={Search for a charged Higgs boson in pp collisions at sqrt(s)=8 TeV},%
pdfsubject={CMS},%
pdfkeywords={CMS, physics, Higgs boson, 2HDM, MSSM, taus, leptons}}

\maketitle

\newcommand{\sqrtsSeven}{\ensuremath{\sqrt{s}=7\TeV}\xspace}
\newcommand{\sqrtsEight}{\ensuremath{\sqrt{s}=8\TeV}\xspace}
\newcommand{\tauh}{\ensuremath{\Pgt_{\mathrm{h}}}\xspace}
\newcommand{\ptTauh}{\ensuremath{\Pgt_{\mathrm{h}}}\xspace}
\newcommand{\mT}{\ensuremath{m_\mathrm{T}}\xspace}
\newcommand{\Rcollmin}{\ensuremath{R_{\mathrm{coll}}^\text{min}}\xspace}
\newcommand{\Rbbmin}{\ensuremath{R_{\mathrm{bb}}^\text{min}}\xspace}
\newcommand{\tanbeta}{\ensuremath{\tanb}\xspace}
\newcommand{\mHp}{\ensuremath{m_{\PH^+}}\xspace}
\newcommand{\mh}{\ensuremath{m_{\cmsSymbolFace{h}}}\xspace}
\newcommand{\mH}{\ensuremath{m_{\cmsSymbolFace{H}}}\xspace}
\newcommand{\mA}{\ensuremath{m_{\cmsSymbolFace{A}}}\xspace}
\newcommand{\mb}{\ensuremath{m_{\cPqb}}\xspace}
\newcommand{\mt}{\ensuremath{m_{\cPqt}}\xspace}
\newcommand{\tbHp}{\ensuremath{\cPqt\to\PH^{+}\cPqb}\xspace}
\newcommand{\Hptaunu}{\ensuremath{\PH^{+}\to\Pgt^{+}\Pgngt}\xspace}
\newcommand{\Hpcsbar}{\ensuremath{\PH^{+}\to\cPqc\cPaqs}\xspace}
\newcommand{\Hptb}{\ensuremath{\PH^{+}\to\cPqt\cPaqb}\xspace}
\newcommand{\DrellYan}{\ensuremath{\cPZ/\gamma^{*}}\xspace}
\newcommand{\DY}{\DrellYan}
\newcommand{\ttbb}{$\ttbar$+$\cPqb\cPaqb$\xspace}
\newcommand{\tauhjets}{\ensuremath{\tauh}+jets\xspace}
\newcommand{\mutauh}{\ensuremath{\Pgm\tauh}\xspace}
\newcommand{\ljets}{\ensuremath{\ell}+jets\xspace}

\newcommand{\ProdHpFourandFiveFS}{\ensuremath{\Pp\Pp \to \cPaqt(\cPqb)\PH^{+}}\xspace}
\newcommand{\ProdHpFourandFiveFScc}{\ensuremath{\Pp\Pp \to \cPqt(\cPaqb)\PH^{-}}\xspace}

\newcommand{\SigmaHpFourandFiveFS}{\ensuremath{\sigma({\Pp\Pp \to \cPaqt(\cPqb)\PH^{+}})}\xspace}
\newcommand{\SigmaHpFourandFiveFScc}{\ensuremath{\sigma({\Pp\Pp \to \cPqt(\cPaqb)\PH^{-}})}\xspace}
\newcommand{\BtbHp}{\ensuremath{\mathcal{B}(\tbHp)}\xspace}
\newcommand{\BHptaunu}{\ensuremath{\mathcal{B}(\Hptaunu)}\xspace}
\newcommand{\BHpcsbar}{\ensuremath{\mathcal{B}(\Hpcsbar)}\xspace}
\newcommand{\BHptb}{\ensuremath{\mathcal{B}(\Hptb)}\xspace}
\newcommand{\lightLimitTaunuHadr}{\ensuremath{\BtbHp \, \BHptaunu}\xspace}
\newcommand{\heavyLimitTaunuHadr}{\ensuremath{\SigmaHpFourandFiveFS \, \BHptaunu}\xspace}
\newcommand{\heavyLimitTb}{\ensuremath{\SigmaHpFourandFiveFS \, \BHptb}\xspace}
\newcommand{\refsec}[1]{Section~\ref{sec:#1}\xspace}
\newcommand{\refdisec}[2]{Sections~\ref{sec:#1} and~\ref{sec:#2}\xspace}
\newcommand{\refmultisec}[2]{Sections~\ref{sec:#1}--\ref{sec:#2}\xspace}
\newcommand{\refapp}[1]{Appendix~\ref{app:#1}\xspace}
\newcommand{\refdiapp}[2]{Appendices~\ref{app:#1} and~\ref{app:#2}\xspace}
\newcommand{\reffig}[1]{Fig.~\ref{fig:#1}\xspace}
\newcommand{\reffigplural}{Figs.}
\newcommand{\refdifig}[2]{\reffigplural~\ref{fig:#1} and~\ref{fig:#2}\xspace}
\newcommand{\reftrifig}[3]{\reffigplural~\ref{fig:#1}, \ref{fig:#2}, and~\ref{fig:#3}\xspace}
\newcommand{\reffigrange}[2]{\reffigplural~\ref{fig:#1}--\ref{fig:#2}\xspace}
\newcommand{\reffigbegin}[1]{Figure~\ref{fig:#1}\xspace}
\newcommand{\refdifigbegin}[2]{Figures~\ref{fig:#1} and~\ref{fig:#2}\xspace}
\newcommand{\reftrifigbegin}[3]{Figures~\ref{fig:#1}, \ref{fig:#2}, and~\ref{fig:#3}\xspace}
\newcommand{\reftab}[1]{Table~\ref{tab:#1}\xspace}
\newcommand{\refditab}[2]{Tables~\ref{tab:#1} and~\ref{tab:#2}\xspace}
\newcommand{\reftritab}[3]{Tables~\ref{tab:#1}, \ref{tab:#2}, and~\ref{tab:#3}\xspace}
\newcommand{\refquadtab}[4]{Tables~\ref{tab:#1}, \ref{tab:#2}, \ref{tab:#3}, and~\ref{tab:#4}\xspace}
\newcommand{\reftabrange}[2]{Tables~\ref{tab:#1}--\ref{tab:#2}\xspace}
\newcommand{\refcite}[1]{Ref.~\cite{#1}\xspace}
\newcommand{\refcites}[1]{Refs.~\cite{#1}\xspace}
\newcommand{\refeq}[1]{Eq.~\eqref{eq:#1}}
\newcommand{\refdieq}[2]{Eq.~\eqref{eq:#1} and~\eqref{eq:#2}}
\newcommand{\CLs}{\ensuremath{\text{CL}_{\text{s}}}\xspace}
\newcommand{\mhmax}{\ensuremath{m_\mathrm{h}^\mathrm{max}}\xspace}
\newcommand{\mhmodp}{\ensuremath{m_\mathrm{h}^\mathrm{mod+}}\xspace}
\newcommand{\mhmodm}{\ensuremath{m_\mathrm{h}^\mathrm{mod-}}\xspace}

\section{Introduction}
\label{sec:introduction}

In 2012, a neutral boson with a mass of approximately 125\GeV was discovered 
by the CMS and ATLAS experiments~\cite{Aad:2012tfa,Chatrchyan:2012ufa,Chatrchyan:2013higgsProperties1} at the CERN LHC.
The properties of the new boson are consistent with those predicted for the standard model (SM) Higgs
boson~\cite{Aad:2013higgsProperties1,Aad:2013xqa,Chatrchyan:2013higgsProperties2,Chatrchyan:2014higgsProperties1,Khachatryan:2014iha,Khachatryan:1979247}.
Models with an extended Higgs sector are constrained by the measured mass, CP quantum numbers, and production rates 
of the new boson. The discovery of another scalar boson, neutral or charged, would represent unambiguous evidence for the presence of physics beyond the SM. 

Charged Higgs bosons are predicted in models including at least two Higgs doublets. The simplest
of such models are the two-Higgs-doublet models (2HDM)~\cite{Lee:1973iz}.
Two Higgs doublets result in five physical Higgs bosons: light and heavy CP-even Higgs bosons $\Ph$ and $\PH$, a CP-odd Higgs boson $\mathrm{A}$, 
plus two charged Higgs bosons $\PH^\pm$.
Throughout this paper, charge conjugate states are implied,
the cross section $\sigma(\Pp\Pp \to \cPaqt(\cPqb)\PH^{+})$ denotes the sum $\sigma(\Pp\Pp \to \cPaqt(\cPqb)\PH^{+}) + \sigma(\Pp\Pp \to \cPqt(\cPaqb)\PH^{-})$,
and the branching fractions $\mathcal{B} (\PH^+ \to {\rm X})$ stand for $\mathcal{B} (\PH^\pm \to {\rm X})$.
The minimal supersymmetric SM
(MSSM)~\cite{Fayet:1974pd,Fayet:1976et,Fayet:1977yc,Dimopoulos:1981zb,Sakai:1981gr,Inoue:1982ej,Inoue:1982pi,Inoue:1983pp}
used as a benchmark in this paper is a special case of a Type-II 2HDM scenario. 
In such a scenario, the couplings of the charged Higgs boson to up-type quarks is proportional 
to $\cot\beta$ while the charged Higgs boson couplings to the down-type quarks and charged leptons are proportional
to $\tan\beta$, where $\tan\beta$ is defined as the ratio of the vacuum expectation values of the two Higgs boson doublet fields.

\begin{figure*}[htb]
\begin{center}
{\includegraphics[width=0.27\textwidth]{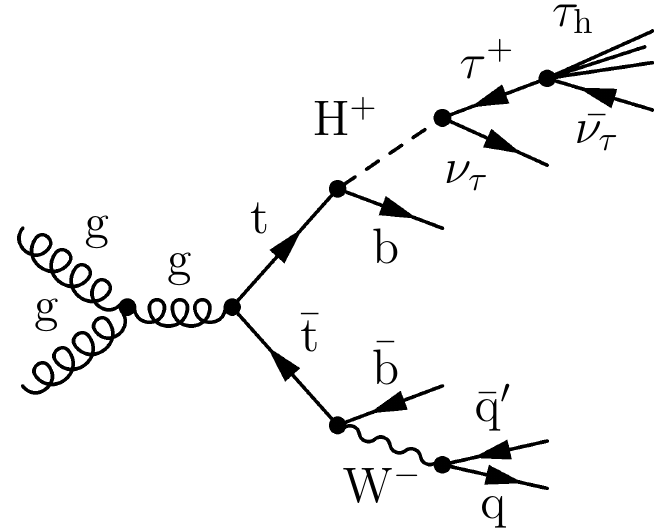}}
\hspace{0.03\textwidth}
{\includegraphics[width=0.27\textwidth]{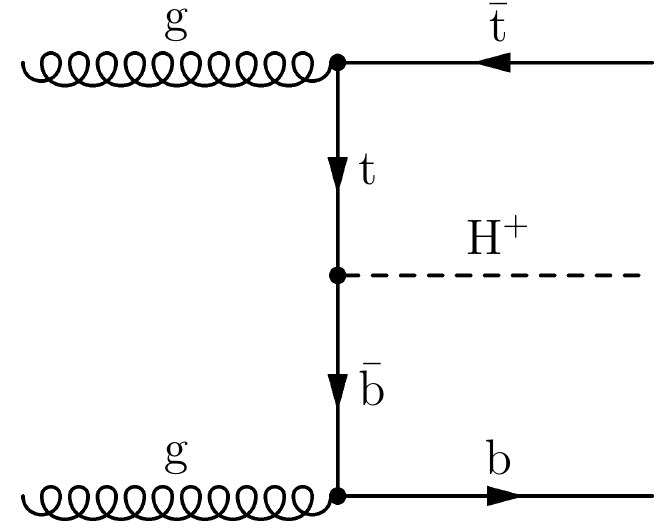}}
\hspace{0.03\textwidth}
{\includegraphics[width=0.27\textwidth]{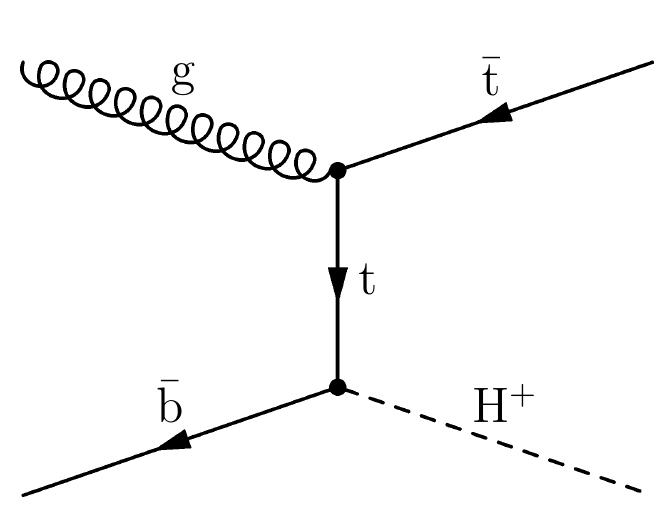}}
\caption{Left: A representative diagram for the production mode of the light charged Higgs boson through \ttbar production
with a subsequent decay to the \tauhjets final state.
Middle and right: Representative diagrams for the direct production of the charged Higgs boson in the four-flavour scheme and five-flavour scheme, respectively.}
\label{fig:feynmanDiagrams}
\end{center}
\end{figure*}

If the mass of the charged Higgs boson is smaller than the mass difference between the top and the bottom quarks, $\mHp < \mt-\mb$, 
the top quark can decay via $\tbHp$.
In this case, the charged Higgs boson is produced most frequently via \ttbar production.
In the MSSM scenarios considered, it preferentially decays to a $\Pgt$ lepton and the corresponding neutrino, \Hptaunu, for $\tan\beta > 5$~\cite{Heinemeyer:1998yj}.
A representative diagram for the production and decay mode for a low-mass charged Higgs boson is shown in \reffig{feynmanDiagrams} (left).
Compared to the SM prediction, 
the presence of the \Hptaunu decay modes would alter the $\Pgt$ yield in the decays of \ttbar pairs.

The Large Electron-Positron (LEP) collider experiments determined a model-independent lower limit of $78.6\GeV$
on the $\PH^+$ mass~\cite{Heister:2002ev,Abdallah:2003wd,Achard:2003gt,Abbiendi:2008aa} at a 95\% confidence level (CL).
The most sensitive 95\% CL upper limits on \BtbHp have been determined by the ATLAS and CMS experiments and are described in the following.
For the \Hptaunu decay mode with the hadronic decay of the $\tau$ lepton (\tauh) and hadronic $\PW$ boson decays (\tauhjets) final state,
95\% CL upper limits of 1.3--0.2\% have been set on \lightLimitTaunuHadr
 for $\mHp = 80$--160\GeV by the ATLAS experiment using data at $\sqrt{s}=8\TeV$~\cite{Aad:2014kga}.
For the $\ell\tauh$ ($\ell$=$\Pe$, $\Pgm$) final states 95\% CL upper limits of 3--9\% have been set by the ATLAS and CMS experiments on
\BtbHp in the \Hptaunu decay mode for $\mHp = 80$--160\GeV
assuming $\BHptaunu = 1$ and using data at $\sqrt{s}=7\TeV$~\cite{HIG-11-019,atlaslightchargedhiggs2013}.
The \Hpcsbar decay mode, whose branching fraction dominates for $\tanbeta < 5$, has been studied by the ATLAS experiment, with
95\% CL upper limits of 5--1\% set on \BtbHp for $\mHp = 90$--160\GeV, under the assumption $\BHpcsbar = 1$ and using data at $\sqrt{s}=7\TeV$~\cite{Aad:2013hla}.

If the charged Higgs boson mass exceeds the mass difference between
the top and bottom quark, $\mHp > \mt-\mb$, the charged Higgs boson is predominantly produced by the fusion of 
bottom and top quarks illustrated in Figs.~\ref{fig:feynmanDiagrams} (middle) and (right) for the four-flavour scheme (4FS)
and the five-flavour scheme (5FS), respectively.
In the 4FS, there are no $\cPqb$ quarks in the initial state, causing a different ordering of the 
perturbative terms at any finite order between the 4FS and 5FS~\cite{Flechl:2014wfa,Heinemeyer:2013tqa,Dittmaier:2009,Berger:2003sm}.
The predictions of the 4FS and the 5FS cross sections calculated at next-to-leading order (NLO) are 
combined using the ``Santander matching scheme''~\cite{Santander}.
In the MSSM benchmark scenarios considered, the \Hptaunu decay mode dominates for $\mHp < 220\GeV$~\cite{Heinemeyer:1998yj},
and for large values of both $\mHp$ and \tanbeta, the decay \Hptb becomes dominant 
but the \Hptaunu decay mode still contributes.
For the \Hptaunu decay mode, considering the final state with hadronic $\tau$ lepton and associated $\PW$ boson decays,
the current upper limits of 0.8--0.004\unit{pb} have been set on \heavyLimitTaunuHadr by the ATLAS experiment
for $\mHp = 180$--1000\GeV using data at $\sqrt{s}=8\TeV$~\cite{Aad:2014kga}.

In this paper, a search for the charged Higgs boson is performed in $\Pp\Pp$ collisions at $\sqrt{s} = 8\TeV$.
The data were recorded by the CMS experiment at the LHC and correspond to an integrated luminosity of $19.7 \pm 0.5\fbinv$.
The charged Higgs boson decay modes and final states discussed in this paper are summarized in \reftab{channel_overview}.
Model-independent limits without any assumption on the charged Higgs boson branching fractions are calculated
on \lightLimitTaunuHadr and \heavyLimitTaunuHadr for $\mHp < \mt-\mb$ and $\mHp > \mt-\mb$, respectively,
with the analysis on the \Hptaunu decay mode in the \tauhjets final state.
Additionally, the \Hptaunu and \Hptb decay modes are inclusively studied in the \mutauh, single lepton (\ljets), and
$\ell\ell'$ ($\ell'$ referring to the possible different flavour between the two leptons)  final states for $\mHp > \mt-\mb$.
Combined limits for the \Hptb decay mode are set on \SigmaHpFourandFiveFS 
by assuming either $\BHptaunu = 1$ or $\BHptb = 1$.
The \tauhjets final state is not sensitive to the presence of charged Higgs boson decay modes other than \Hptaunu, because any such decay mode would be estimated inclusively with the background through the measurement from data described in Section~\ref{sec:taunu:hadr:backgrounds:ewkgenuine}.
All the decay modes and final states considered
are used to set exclusion limits in the \mHp--\tanbeta parameter space for different MSSM benchmark scenarios~\cite{Heinemeyer:2013tqa,Carena:2013}. To set these limits, the specific branching fractions predicted by those benchmark scenarios are applied.
This paper includes the first results on the direct charged Higgs boson production for $\mHp > \mt-\mb$ in the \Hptb decay mode.

\begin{table*}[htbp]
\begin{center}
\setlength{\extrarowheight}{1.5pt}
\topcaption{Overview of the charged Higgs boson production processes, decay modes, final states, and mass regions analysed in this paper ($\ell=\Pe,\Pgm$).
All final states contain additional jets from the hadronization of b quarks and missing transverse energy from undetected neutrinos.
The index after each signature denotes the section where it is discussed.
}
\begin{tabular}{l c c }
\hline
Decay mode & Signatures for $\mHp < \mt-\mb$ & Signatures for $\mHp > \mt-\mb$ \\
\hline
 & $\Pp\Pp \to \ttbar\to\cPqb\PH^+\cPaqb\PH^-/\cPqb\PH^+\cPaqb\PW^-$ & \ProdHpFourandFiveFS \\
\Hptaunu & $\Pgt_{\rm h}$+jets$^{(\ref{sec:taunu:hadr:analysis})}$ & 
$\Pgt_{\rm h}$+jets$^{(\ref{sec:taunu:hadr:analysis})}$, 
$\Pgm\Pgt_{\rm h}^{(\ref{sec:taunutb:taumu:analysis})}$, 
$\ell\ell'^{(\ref{sec:taunutb:dilepton:analysis})}$ \\
\Hptb & --- & 
$\Pgm\Pgt_{\rm h}^{(\ref{sec:taunutb:taumu:analysis})}$, 
$\ell\ell'^{(\ref{sec:taunutb:dilepton:analysis})}$, 
$\ell$+jets$^{(\ref{sec:tb:singlelepton:analysis})}$ \\
\hline
\end{tabular}
\label{tab:channel_overview}
\end{center}
\end{table*}

The CMS detector is briefly described in \refsec{cmsdet}, followed by details of
the event reconstruction and simulation in \refdisec{reco}{sim}, respectively.
The event selection together with the background estimation is described in 
Sections~\ref{sec:taunu:hadr:analysis}, \ref{sec:taunutb:taumu:analysis}, \ref{sec:taunutb:dilepton:analysis},
and \ref{sec:tb:singlelepton:analysis}
for the \tauhjets, \mutauh, $\ell\ell'$, and \ljets final states, respectively.
The treatment of statistical and systematic uncertainties is described in \refsec{uncertainties}.
The results are presented in \refsec{results} and summarized in \refsec{conclusion}.

\section{The CMS detector}
\label{sec:cmsdet}

The central feature of the CMS apparatus is a superconducting solenoid of 6\unit{m} internal diameter, providing a magnetic field of 3.8\unit{T}. 
Within the superconducting solenoid volume are a silicon pixel and strip tracker, a lead tungstate crystal electromagnetic calorimeter (ECAL), 
and a brass and scintillator hadron calorimeter (HCAL), each composed of a barrel and two endcap sections. Muons are measured in 
gas-ionization detectors embedded in the steel flux-return yoke outside the solenoid.
Forward calorimeters extend the pseudorapidity coverage provided by the barrel and endcap detectors up to $|\eta| < 5$.
The first level (L1) of the CMS trigger system, composed of custom hardware processors, uses information from the calorimeters
and muon detectors to select the most interesting events in a fixed time interval of less than 4\mus.
The high-level trigger processor farm further decreases the event rate from
around 100\unit{kHz} to around 1\unit{kHz}, before data storage.
A more detailed description of the CMS detector, together with a definition of the coordinate system used and the relevant kinematic 
variables, can be found in Ref.~\cite{Chatrchyan:2008zzk}. 

\section{Event reconstruction}
\label{sec:reco}

In the data collected during 2012, an average of 21 proton-proton interactions occurred per LHC bunch crossing.
To select the primary interaction vertex, the squared sum of the transverse momenta of the charged-particle tracks,
$\sum\pt^2$, associated with each interaction vertex is calculated. The interaction vertex with the largest $\sum\pt^2$ value is taken as
the primary interaction vertex in the event~\cite{Chatrchyan:2014fea}. The other $\Pp\Pp$ collisions are referred to as pileup.

Events are reconstructed with the particle-flow (PF) algorithm~\cite{CMS-PAS-PFT-09-001,CMS-PAS-PFT-10-001}, 
which combines information from all sub-detectors to identify and reconstruct individual electrons, muons, photons, and charged and neutral hadrons.
Electrons are reconstructed from clusters of ECAL energy deposits matched
to hits in the silicon tracker~\cite{Khachatryan:2015hwa}. 
Muons are reconstructed by performing a simultaneous global track fit to hits in the silicon tracker and the muon system~\cite{CMS-PAPERS-MUO-10-004}. 
The energy of photons is directly obtained from the ECAL measurement, corrected for zero-suppression effects. 
The energy of charged hadrons is determined from a combination of their momentum measured in the tracker and the matching 
ECAL and HCAL energy deposits, corrected for zero-suppression effects and for the response function of the calorimeters
to hadronic showers. 
Finally, the energy of neutral hadrons is obtained from the corresponding corrected ECAL and HCAL energy. 
The composite physics objects, such as jets, hadronic tau lepton decays, and missing transverse energy are reconstructed from these PF particles.

Jets are reconstructed from the PF particles clustered by the 
anti-$k_\mathrm{t}$ algorithm~\cite{Cacciari:2008gp, Cacciari:2011ma} with a distance parameter of 0.5. 
The jet momentum is determined as the vectorial sum of all particle momenta in the jet, and is found in the simulation to be 
within 5--10\% of the true momentum over the whole \pt spectrum and detector acceptance. An offset correction is 
applied to take into account the extra energy clustered in jets arising from pileup. Jet energy corrections are derived from simulation, and are confirmed by in situ measurements of the energy balance in dijet and photon+jet events~\cite{CMS-JME-10-011}. Additional selection criteria are applied to each event to remove spurious jet-like 
features originating from isolated noise patterns in certain HCAL regions. 
Jets originating from pileup interactions are removed by a multivariate jet identification algorithm~\cite{CMS-PAS-JME-13-005}.

Jets from the hadronization of b quarks are identified (b tagged) with the ``combined secondary vertex'' algorithm~\cite{CMS-PAS-BTV-13-001,CMS-PAPER-BTV-12-001}.
The algorithm consists of evaluating a likelihood-based discriminator which uses information from reconstructed decay vertices of short-lived mesons 
and transverse impact parameter measurements of charged particles.
In the \tauhjets final state, the algorithm is used to identify b-tagged jets with a mistagging probability, i.e. the probability that a jet from the fragmentation of light quarks ($\cPqu,\cPqd,\cPqs,\cPqc$) or gluons
is misidentified as a b jet, of approximately 0.1\% (``tight'' working point).
In the analyses of the \mutauh and \ljets final states, the b tagging algorithm used has a mistagging probability of 1\% (``medium'' working point),
since the multijet background is smaller than in the \tauhjets final state.
In the analysis with the $\ell\ell'$ final state, the b tagging working point is adjusted to allow a 10\% mistagging probability
to enhance signal acceptance since the multijet background in this analysis is even smaller.
The corresponding probability to identify a b jet is about 50, 70, and 85\%, respectively.
The difference in b tagging efficiency between data and simulation is corrected by applying data-to-simulation scale factors dependent on the jet \pt and the jet pseudorapidity ($\eta$).

The missing transverse momentum vector \ptvecmiss is defined as the projection of the negative vector sum of the momenta of all reconstructed PF particles
in an event onto the plane perpendicular to the 
beams. Its magnitude is referred to as \ETmiss. 
The \MET reconstruction is improved by propagating the jet energy corrections to it.
Further filter algorithms are used to reject events with anomalously large \ETmiss resulting from instrumental effects~\cite{Khachatryan:2014gga}.

The ``hadron-plus-strips'' algorithm~\cite{CMS-PAPER-TAU-11-001} is used to reconstruct hadronically
decaying $\Pgt$ leptons. The algorithm uses the constituents of the
reconstructed jets to identify individual $\Pgt$ decay modes with one
charged and up to two neutral pions, or three charged pions. The
neutral pions are reconstructed by clustering the reconstructed photons in narrow strips
along the azimuthal angle direction taking into account possible broadening of calorimeter
depositions from photon conversions. The $\Pgt_\mathrm{h}$ candidates compatible with electrons or muons
are rejected. 
Jets originating from the hadronization of quarks and gluons are suppressed by requiring that the \tauh candidate
is isolated as described below. 
The $\tauh$ identification efficiency depends on $\pt^{\tauh}$ and $\eta^{\tauh}$, and is on average 50\%
for $\pt^{\tauh} > 20\GeV$ with a probability of approximately 1\% for hadronic jets to be misidentified as a $\tauh$.

Electrons, muons, and hadronically decaying $\Pgt$ leptons
are required to be isolated from other particles
by considering transverse momenta of neutral and charged particles in a
cone $\Delta R=\sqrt{(\Delta \phi) ^{2}+ (\Delta \eta) ^{2}}$, where $\phi$ is the azimuthal angle, around the charged lepton candidate momentum direction.
The isolation variable for electrons,
muons, 
and \tauh 
is defined as:
\begin{eqnarray}
\label{eq:isolation:e}
I^\Pe & = & \sum_{\rm charged}\pt + \max \left(0, \sum_{\rm neut.~hadr.}\pt + \sum_{\gamma}\pt - \rho_{\rm neutral}\, A_{\rm eff.} \right), \\
\label{eq:isolation:mu}
I^\Pgm & = & \sum_{\rm charged}\pt + \max \left(0, \sum_{\rm neut.~hadr.}\pt + \sum_{\gamma}\pt - 0.5\sum_{\rm charged,pileup}\pt \right), \\
\label{eq:isolation:tau}
I^{\tauh} & = & \sum_{\rm charged}\pt + \max \left(0, \sum_{\gamma}\pt - 0.46\sum_{\rm charged,pileup}\pt \right),
\end{eqnarray}
where $\sum_{\rm charged}\pt$ is the scalar sum of the transverse momenta of charged hadrons, electrons, and muons originating from the primary interaction vertex,
and $\sum_{\rm neut.~hadr.}\pt$ and $\sum_{\gamma}\pt$ are the scalar sums over neutral hadron and photon transverse momenta, respectively, in the cone 
$\Delta R$ around the charged lepton candidate momentum direction.
The presence of particles from pileup events is taken into account depending on the charged-lepton type.
For electron candidates, the scalar sum of the \pt of photons and neutral hadrons from pileup events in the isolation cone
is estimated as the product of the neutral-particle transverse momentum density and the effective cone area, 
$\rho_{\rm neutral}\, A_{\rm eff.}$.
The $\rho_{\rm neutral}$ component is evaluated from all photons and neutral hadrons in the event,
and $A_{\rm eff.}$ accounts for the presence of pileup events.
For muons and hadronically decaying $\Pgt$ leptons, the scalar sum of the \pt of photons and neutral hadrons from pileup events
is estimated from the scalar sum of the transverse momenta of charged hadrons from pileup events
in the isolation cone, $\sum_{\rm charged,pileup}\pt$, by multiplying it by the average ratio of neutral- to charged-hadron production 
in inelastic $\Pp\Pp$ collisions. 
Since the contribution from neutral hadrons is ignored when computing the \tauh isolation variable, the pileup correction factor is slightly smaller than that used for correcting the muon isolation variable.

For electrons, an isolation cone size of $\Delta R = 0.3$ or 0.4 is used, depending on the final state.
For muons and hadronically decaying $\Pgt$ leptons, isolation cone sizes of $\Delta R = 0.4$ and 0.5 are used, respectively.
Electrons and muons are considered isolated
if the relative isolation variable $I^\ell_{\rm rel} = I^\ell / p_{\mathrm{T}}^\ell$, where $\ell = \Pe,\Pgm$, is lower than 10--20\%, depending on the final state.
Hadronically decaying $\Pgt$ leptons are considered isolated if $I^{\tauh} < 1\GeV$.

\section{Simulation}
\label{sec:sim}
The signal processes
are generated with \PYTHIA~6.426~\cite{Sjostrand:2006za}.
The $\ttbar$, $\PW$+jets, and $\cPZ$+jets backgrounds are generated using the 
\MADGRAPH~5.1.3.30~\cite{Alwall:1699128} event generator with matrix elements (ME) providing up to four additional
partons, including b quarks. The event generator is interfaced with \PYTHIA to provide the parton showering
and to perform the matching of the soft radiation with the contributions from the ME.
The single top quark production is generated with \POWHEG~1.0~\cite{Nason:2004,Frixione:2007vw,Alioli:2010,Alioli:2009,Re:2011}
and the quantum chromodynamics (QCD) multijet and diboson production processes $\PW\PW$, $\PW\cPZ$, and $\cPZ\cPZ$ are generated
using \PYTHIA. Both the \MADGRAPH and \POWHEG generators are interfaced with \PYTHIA 
for parton shower and hadronization.
The \TAUOLA~27.121.5~\cite{Was:2000st} package is used to generate $\Pgt$ decays for the simulated signal, as well as background samples.

The events are passed through full CMS detector simulation
based on \GEANTfour~\cite{Agostinelli:2002hh,Allison:2006ve},
followed by a detailed trigger simulation and event reconstruction.
Simulated minimum bias events are superimposed upon the hard interactions
to match the pileup distribution observed in data.
The \PYTHIA parameters for the underlying event 
are set according to the Z2* tune, which is derived from the Z1 tune~\cite{Field:2010bc}, which uses the CTEQ5L parton distribution set, whereas Z2* adopts CTEQ6L~\cite{Pumplin:2002vw}.

The number of $\ttbar$ events produced is normalized to the predicted \ttbar production cross section of
$246.7^{+6.2}_{-8.4} \pm 11.4\unit{pb}$ as calculated with the {\sc Top++} v2.0 program
to next-to-next-to-leading order (NNLO) in perturbative QCD, including soft-gluon resummation to next-to-next-to-leading-logarithmic (NNLL) order~\cite{Czakon:2011xx}, 
and assuming $\mt = 173.34\GeV$~\cite{ATLAS:2014wva}.
The first uncertainty originates from the independent variation of the factorization and renormalization scales, $\mu_F$ and $\mu_R$, 
while the second is associated with variations in the parton density functions (PDFs) and strong coupling constant $\alpha_{\mathrm{S}}$, following the PDF4LHC prescription with the 
MSTW2008 68\% CL
NNLO, CT10 NNLO and NNPDF2.3 5-flavour fixed-flavour number (FFN) PDF sets~\cite{Botje:2011sn,Martin:2009bu,Gao:2013xoa,Ball:2012cx}.
The predicted cross section is in good agreement with the measurements by ATLAS and CMS~\cite{Aad:2014kva,Chatrchyan:2013faa}.
The top quark \pt spectrum in data is found to be softer than that predicted using the \MADGRAPH MC generator~\cite{CMS-TOP-11-013}.
To correct for this effect, the \ttbar events are reweighted to make the top quark \pt spectrum in simulation match that
observed in data~\cite{CMS-PAPER-TOP-12-028}. 

The NNLO SM prediction is calculated with {\sc FEWZ} v3.1 for the W+jets and \DY backgrounds~\cite{Melnikov:2006a,Melnikov:2006b}.
The cross section for the t-channel single top quark sample is calculated at next-to-leading order (NLO) in QCD with {\sc Hathor} v2.1~\cite{Aliev:2010zk,Kant:2014oha}
with PDF and $\alpha_{\mathrm{S}}$ uncertainties calculated using the PDF4LHC prescription~\cite{Alekhin:2011sk,Botje:2011sn}.
For the single top quark s-channel and tW-channel cross section, the SM prediction at NNLL in QCD is taken from Refs.~\cite{Kidonakis:2010a,Kidonakis:2010b}.

\section{The \texorpdfstring{\tauhjets}{tau\_h+jets} final state for \texorpdfstring{$\Hptaunu$}{H+ to taunu}}
\label{sec:taunu:hadr:analysis}
In this analysis, a charged Higgs boson is assumed to be produced through the $\ttbar\to\cPqb\PH^+\cPaqb\PH^-$, $\ttbar\to\cPqb\PH^+\cPaqb\PW^-$, and \ProdHpFourandFiveFS processes and
searched for in the \Hptaunu decay mode with a hadronic decay of the $\Pgt$ and 
a hadronic decay of the W boson that originates from the associated $\cPaqt \rightarrow \cPaqb\PW^{-}$ decay.
In these events, the missing transverse momentum is expected to originate from the neutrinos in the decay of the charged Higgs boson, which allows the reconstruction
of the transverse mass, \mT, of the charged Higgs boson: 

\begin{equation}
  \label{eq:mt}
  \mT = \sqrt{2 \PT^{\Pgt_\mathrm{h}} \MET (1-\cos \Delta\phi(\ptvec^{\Pgt_\mathrm{h}},\ptvecmiss))},
\end{equation}

where $\ptvec^{\Pgt_\mathrm{h}}$ denotes the transverse momentum vector of the hadronically decaying $\Pgt$ lepton and $\pt^{\Pgt_\mathrm{h}}$ its magnitude,
and $\Delta\phi$ is the angle between
the $\Pgt_\mathrm{h}$ direction and the \ptvecmiss in the transverse plane. The presence of the two neutrinos from the charged Higgs boson decay smears the expected Jacobian peak somewhat, 
but leaves the kinematic edge at the charged Higgs boson mass intact.
The search is performed as a shape analysis, using the transverse mass to infer the presence of a signal.
The dominant background processes are the SM \ttbar and single top quark
production, and the electroweak (EW) processes: $\PW$+jets, $\cPZ$+jets, and dibosons
($\PW\PW$, $\PW\cPZ$, $\cPZ\cPZ$). The multijet background constitutes a subleading background.

\subsection{Event selection}
\label{sec:taunu:hadr:selection}

Events are selected with a trigger that requires the presence of
a \tauh and large \MET. First the events are required to
have calorimetric $\ETmiss > 40\GeV$ at the first level of the CMS trigger system.
The calorimetric \MET is defined as the \MET calculated from the ECAL and HCAL energy deposits instead of the PF particles.
At the high-level trigger, the
events are required to have calorimetric $\ETmiss > 70\GeV$, and a \tauh 
of $\pt^{\Pgt_\mathrm{h}} > 35\GeV$ and $|\eta^{\tauh}| < 2.5$. The \tauh is required to be loosely isolated, 
to contain at least one track of $\pt > 20\GeV$, and
to have at most two tracks in total, targeting the $\Pgt$ lepton decays into a single charged pion and up to two neutral pions.
The probability for a signal event to be accepted by the trigger amounts to 8--14\% in the \mHp range of 80--160\GeV, and 19--44\% in the \mHp range of 180--600\GeV with all tau decays considered.

The efficiency of the $\Pgt$ part of the trigger is 
evaluated as a function of $\pt^{\Pgt_\mathrm{h}}$ using a ``tag-and-probe'' technique~\cite{CMS-PAPER-TAU-11-001}
from $\DY\to\Pgt_\Pgm \Pgt_\mathrm{h}$ events, where $\Pgt_\Pgm$ refers to a muonic $\Pgt$ lepton decay.
The efficiency of the \MET part of the trigger is evaluated from events with a \ttbar-like final state of \tauhjets selected with a single-$\Pgt$ trigger.
The trigger efficiencies in simulated events are corrected with data-to-simulation scale factors applied
as function of $\pt^{\Pgt_\mathrm{h}}$ for the \tauh part of the trigger
and as function of \MET for the \MET part of the trigger.
The scale factors range between 0.95--1.06 and 0.97--1.02 for the \tauh and \MET parts of the trigger, respectively.

Selected events are required to have at least one \tauh with
$\pt^{\Pgt_\mathrm{h}} > 41\GeV$ within $| \eta | < 2.1$ and to be matched to a trigger-level \tauh object.
These thresholds are chosen to be compatible with the single-muon trigger used for estimate of backgrounds 
with hadronic $\Pgt$ decays from control samples in data, as described in \refsec{taunu:hadr:backgrounds:ewkgenuine}.
Only one charged hadron is allowed to be associated with the \tauh
and its \pt is required to fulfil $\pt > 20\GeV$.
Background events with $\PW\rightarrow\Pgt\Pgngt$ decays are suppressed by requiring
$R_{\Pgt} = p^\text{charged hadron}/p^{\Pgt_\mathrm{h}} > 0.7$.
The $R_{\Pgt}$ observable is sensitive to different polarizations of $\Pgt$ leptons originating from decays of $\PW$ bosons (spin 1) and from decays of $\PH^+$ (spin 0)~\cite{Roy:1999xw}.

A \ttbar-like event topology is selected by requiring at least three jets of $\pt > 30\GeV$ and $| \eta | < 2.4$ in addition to the \tauh 
and by requiring at least one of the selected jets to be identified as originating from the hadronization of a b quark.
To select a fully hadronic final state,
events containing identified and isolated electrons (muons) with $\pt > 15~(10)\GeV$ are rejected.
The electron (muon) candidates are considered to be isolated if the relative isolation $I^\Pe_{\mathrm{rel.}}$ ($I^\Pgm_{\mathrm{rel.}}$), as described in \refsec{reco},
is smaller than 15\% (20\%).

To suppress the multijet background, $\MET > 60\GeV$ is required.
The lower \MET threshold on the PF \MET compared to the calorimetric \MET requirement applied at the high-level trigger
improves the signal acceptance for $\mHp < \mt-\mb$. This approach can be used because 
the PF \MET has better resolution than the calorimetric \MET~\cite{Khachatryan:2014gga}.

In the multijet events selected with the $\tau$+\MET trigger a hadronic jet is misidentified as the \tauh in the event. In addition, the \tauh typically has a recoiling jet
in the opposite direction. The \MET in these events arises from the mismeasurement of the momenta of these jets with the \ptvecmiss direction aligned
with $\vec{p}_{\mathrm{T}}^{\tauh}$.
The best performance for multijet background suppression and signal acceptance is obtained with two-dimensional circular selections
instead of simple selections based on azimuthal angle differences.
The variables used for the azimuthal angle selections are defined as

\begin{equation}
  \begin{array}{l}
  \Rcollmin = {\rm min}\left\{ 
  \sqrt{\left(\Delta\phi(\Pgt_\mathrm{h},\ptvecmiss)\right)^2 + \left(\pi- \Delta\phi({\rm jet}_n,\ptvecmiss)\right)^2 } \right\} , \\ \vspace{0.01em} \\
  \Rbbmin = {\rm min}\left\{ 
  \sqrt{\left(\pi - \Delta\phi(\Pgt_\mathrm{h},\ptvecmiss)\right)^2 + \left(\Delta\phi({\rm jet}_n,\ptvecmiss)\right)^2 } \right\} ,
  \end{array}
\end{equation}

where the index $n$ refers to any of the three highest \pt jets in the event and $\Delta\phi$ denotes the 
azimuthal angle between the reconstructed \ptvecmiss and the \tauh or one of the three highest-\pt jets.
The labels ``coll'' and ``bb'' denote the collinear and back-to-back systems of the \tauh and the \MET, respectively.
The selected events are required to satisfy $\Rcollmin > 0.70$ and $\Rbbmin > 0.70$.

The same event selection is used for all the \mHp values considered.

\subsection{Background measurements}
\label{sec:taunu:hadr:backgrounds}

The background contributions arise from three sources:
\begin{enumerate}
\item Irreducible background from EW processes --- $\PW$+jets, $\cPZ$+jets, and dibosons --- as well as
   SM $\ttbar$ and single top quark production, where the selected \tauh originates from a hadronic decay of a $\Pgt$ lepton (``EW+$\ttbar$ with \tauh'').
\item Reducible background from multijet events with large mismeasured \MET and jets that mimic 
hadronic $\Pgt$ decays.
\item Reducible background from EW+\ttbar events, where an electron, muon, or a jet is misidentified as the \tauh (``EW+$\ttbar$ no \tauh'').
\end{enumerate}
The two largest backgrounds, ``EW+$\ttbar$ with \tauh'' and multijets, are measured from control samples in data, as
explained in Sections~\ref{sec:taunu:hadr:backgrounds:ewkgenuine} and~\ref{sec:taunu:hadr:backgrounds:multijet}.
The contribution from ``EW+$\ttbar$ no \tauh'' is estimated from simulation and is described
in \refsec{taunu:hadr:backgrounds:ewkfake}.

\subsubsection{Measurement of the \texorpdfstring{EW+$\ttbar$}{EW+ttbar} with hadronically decaying \texorpdfstring{$\Pgt$}{tau} leptons background}
\label{sec:taunu:hadr:backgrounds:ewkgenuine}
The \mT distribution for the ``EW+$\ttbar$ with \tauh'' background is modelled via an
embedding technique.
It uses a control data sample of $\Pgm$+jets events selected with a single-$\Pgm$ trigger.
The same jet selection as in the \tauhjets sample is used, and events with electrons or additional muons are rejected.
Then, the selected $\Pgm$ is replaced by a simulated $\Pgt$ lepton decay. The simulated $\Pgt$ lepton momentum is the same as that of the selected $\Pgm$,
and the reconstructed $\Pgt$ decay products are 
merged with the original $\Pgm$+jets event, from which the reconstructed muon is removed.
In these hybrid events, the jets are reclustered and the \MET is recalculated and then the events
are subjected to the same event selection as the \tauhjets sample,
\ie \tauh identification, b tagging, $\MET$ requirement, and the azimuthal angle selections are applied.

To obtain the \mT distribution for the ``EW+$\ttbar$ with \tauh'' background,
the effect of the muon trigger and the muon offline reconstruction need to be unfolded, and the efficiency of the $\tau$+\MET trigger must be taken into account.
First, the weight of each hybrid event is increased by the inverse of the muon trigger and identification efficiencies.
Then, the efficiency of the $\tau$+\MET trigger is applied by weighting the events with the efficiencies of the $\tau$ part of the trigger and
the first trigger level part of the \MET trigger. The rest of the \MET part of the trigger is taken 
into account by applying a requirement on a hybrid calorimetric \MET constructed from
the original event and the simulated $\Pgt$ lepton decay.

After the trigger has been taken into account, further corrections are applied.
In a fraction of the selected $\Pgm$+jets events the $\Pgm$ originates from a decay of a $\Pgt$ lepton, leading to an overestimation of the EW+$\ttbar$ background by a few percent. 
This bias is corrected for by applying to the hybrid events $\pt^{\Pgm}$-dependent correction factors derived from simulated $\ttbar$ events.
A residual difference is seen in the \mT distribution between non-embedded $\Pgt$+jets and embedded $\Pgm$+jets events in simulated \ttbar events.
This difference is corrected by weighting the hybrid events by \mT-dependent correction factors derived from simulated \ttbar events.
The \ttbar events constitute about 85\% of the ``EW+$\ttbar$ with \tauh'' background.

It should be noted that the embedding technique allows the separation of signal from the \Hptaunu decay mode from 
other decay modes, such as \Hptb, where the $\Pgt$ lepton originates from a $\PW$ boson decay.
Namely, in the other charged Higgs boson decays, $\Pgt$ leptons and muons are produced at equal rates 
causing the embedding technique to include the \Hptb signal from data (and other such signals) as part of
the ``EW+\ttbar with \tauh'' background.

\subsubsection{Measurement of the multijet background}
\label{sec:taunu:hadr:backgrounds:multijet}

The multijet background is measured with a ``\tauh misidentification rate'' technique. 
An estimate of the multijet background is obtained by measuring the 
probability of the \tauh candidate to pass the nominal and inverted \tauh isolation criterion.
The misidentification rate is measured in bins of \tauh transverse momentum, in an event sample that is obtained
prior to applying the b tagging, \MET, and \Rbbmin parts of the event selection described in \refsec{taunu:hadr:selection}.
The event sample that passes the nominal \tauh isolation selection contains a nonnegligible contamination from 
EW+\ttbar backgrounds with genuine and misidentified $\Pgt$ leptons.
Therefore, the number of multijet and EW+\ttbar events is determined by a maximum likelihood fit of the \MET distribution.
A fit is performed for each $\pt^{\Pgt_\mathrm{h}}$ bin.
For multijet events, the \MET templates are obtained
from the data sample with inverted \tauh isolation by subtracting a small contribution of simulated EW+\ttbar events.
The \MET templates for the EW+\ttbar events are taken from simulation in the nominal region.
The misidentification rate probabilities $w_j$ are defined as the ratio of the number of multijet events in the isolated sample and the inverted isolation sample.
Their measured values vary between 0.050--0.061 depending on the $\pt^{\Pgt_\mathrm{h}}$ bin with a statistical uncertainty smaller than 3\%.

The measured \tauh misidentification rate probabilities are then applied as weights to multijet events passing 
all nominal event selection criteria, except that the isolation criterion applied on the \tauh is inverted.
The number of multijet events is obtained by subtracting the number of simulated EW+\ttbar events from data.
The estimate for the number of multijet events in a given bin $i$ of the \mT distribution
($N_{i}^{\text{multijet}}$) is obtained by summing these weighted events over the $\pt^{\Pgt_\mathrm{h}}$ bins
according to

\begin{equation} \label{eq:qcd1}
  N_{i}^{\text{multijet}} = \sum_{j} (N^{\text{data, inverted}}_{i,j}-N^{\text{EW+}\ttbar\text{, inverted}}_{i,j}) w_j,
\end{equation}

where $N$ is the number of events and $i$ and $j$ denote \mT and $\pt^{\Pgt_\mathrm{h}}$ bins, respectively.

\subsubsection{The \texorpdfstring{EW+$\ttbar$}{EW+ttbar} with misidentified \texorpdfstring{$\Pgt$}{tau} leptons background}
\label{sec:taunu:hadr:backgrounds:ewkfake}

The ``EW+$\ttbar$ no \tauh'' background originates almost solely from jets that are misidentified
as the \tauh with a small contribution from electrons and muons misidentified as the \tauh.
About 85\% of the ``EW+$\ttbar$ no \tauh'' background events
come from \ttbar and the rest from single top quark production in the $\cPqt\PW$- and $\cPqt$-channels. The
number of selected simulated events in the single top quark samples is small and therefore the \mT distribution for them is estimated
with a procedure where the probability of each event to pass the b tagging
is applied as a per-event weight instead of applying the b tagging selection.
This probability is evaluated for simulated events with the \ttbar-like final state as function of jet \pt and flavour.

\subsection{Event yields}
\label{sec:taunu:hadr:yields}
\begin{figure*}[h!]
\begin{center}
\includegraphics[width=0.40\textwidth]{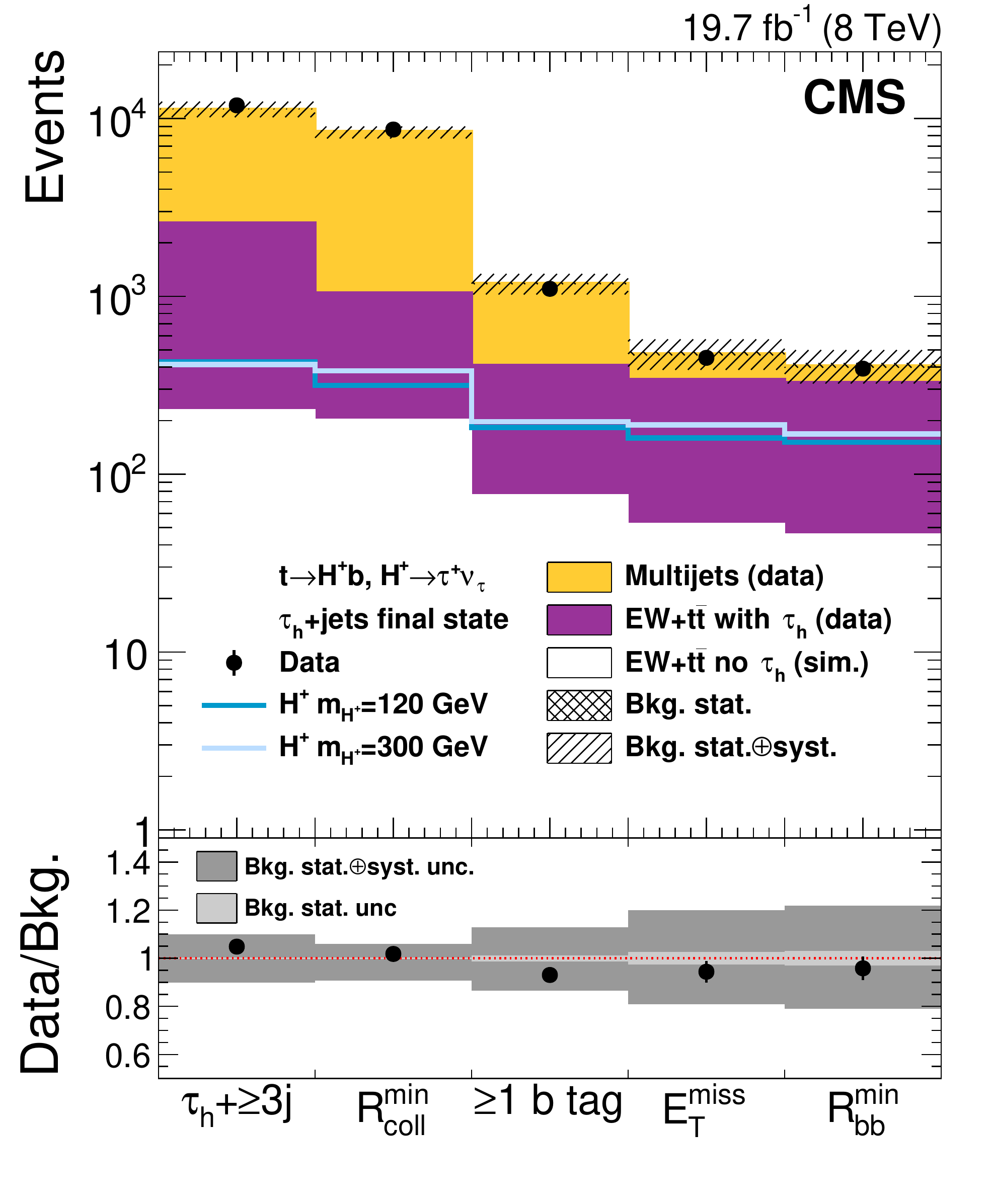}
\caption{The event yield in the \tauhjets final state after each selection step.
For illustrative purposes, the expected signal yields are shown for $\mHp = 120\GeV$ normalized to $\lightLimitTaunuHadr = 0.01$ 
and for $\mHp = 300\GeV$ normalized to $\heavyLimitTaunuHadr = 1$\unit{pb}, which are typical values for the sensitivity of this analysis.
The bottom panel shows the ratio of data over the sum of expected backgrounds and its uncertainties.
The cross-hatched (light grey) area in the upper (lower) part of the figure represents the statistical uncertainty, while
the collinear-hatched (dark grey) area gives the total uncertainty in the background expectation.}
\label{fig:taunu:hadr:selections}
\end{center}
\end{figure*}

Figure~\ref{fig:taunu:hadr:selections} shows the event yields after each selection step starting from
the requirement that a \tauh, no isolated electrons or muons, and at least three jets
are present in the event.
The multijet background and the ``EW+\ttbar with \tauh'' background are shown as
measured from data, while the ``EW+\ttbar no \tauh'' background is shown as estimated from the simulation.
The data agree with the sum of expected backgrounds within the total uncertainties.

\begin{figure*}[h!]
\begin{center}
{\includegraphics[width=0.40\textwidth]{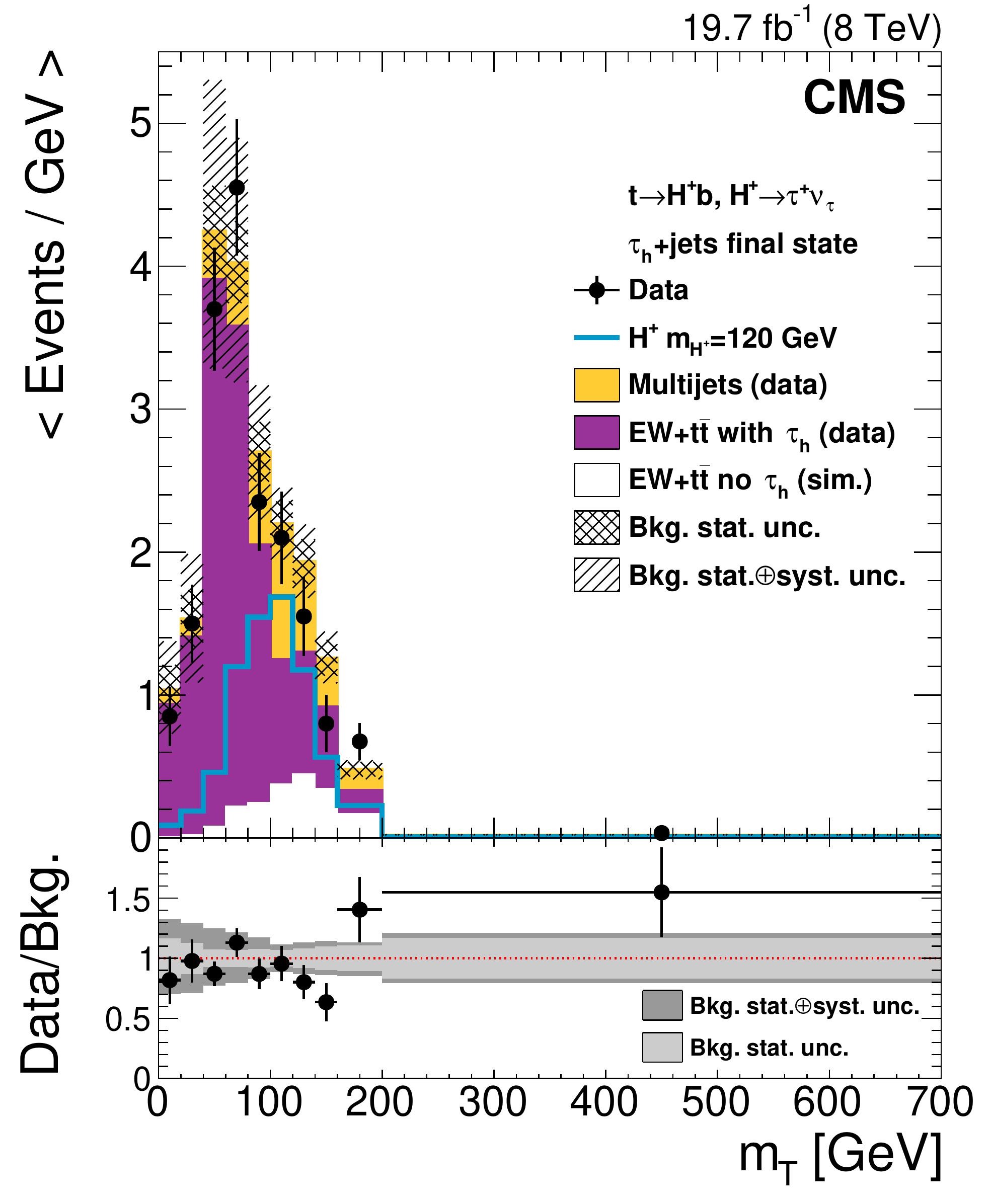}}
\hspace{0.02\textwidth}
{\includegraphics[width=0.40\textwidth]{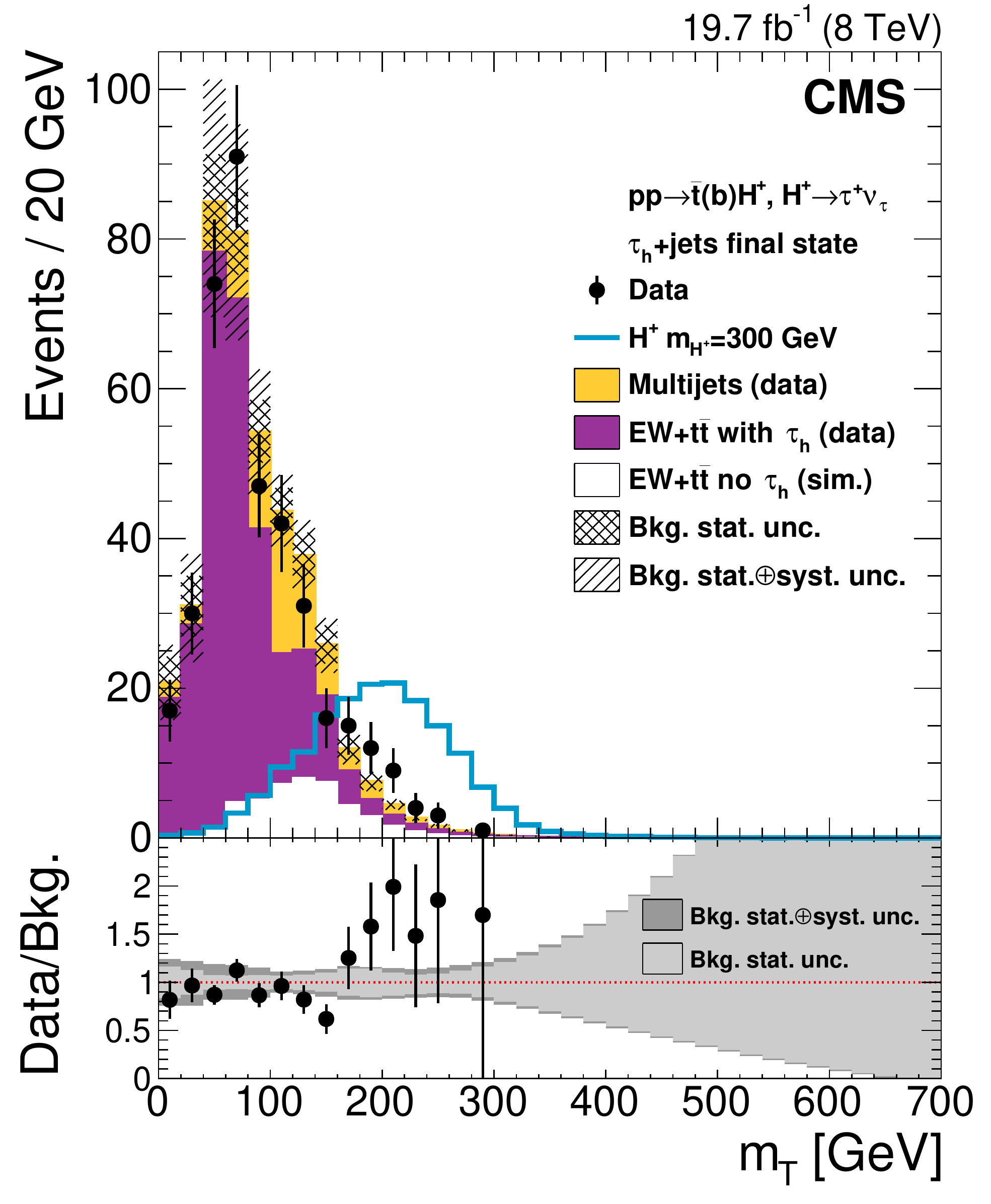}}
\caption{The transverse mass (\mT) distributions in the \tauhjets final state for
the $\PH^+$ mass hypotheses of 80--160\GeV (left) and 180--600\GeV (right). The event selection is the same in both left and right plots,
but in the right plot the background expectation is replaced for $\mT > 160\GeV$ by a fit of the falling part of the \mT distribution.
Since a variable bin width is used in the left plot the event yield in each bin has been divided by the bin width.
For illustrative purposes, the expected signal yields are shown in the left plot for $\mHp = 120\GeV$ normalized to $\lightLimitTaunuHadr = 0.01$
and in the right plot for $\mHp = 300\GeV$ normalized to $\heavyLimitTaunuHadr = 1$\unit{pb}, which are typical values for the sensitivity of this analysis.
The bottom panel shows the ratio of data over the sum of expected backgrounds along with the uncertainties.
The cross-hatched (light grey) area in the upper (lower) part of the figure represents the statistical uncertainty, while
the collinear-hatched (dark grey) area gives the total uncertainty in the background expectation.
}
\label{fig:taunu:hadr:mt}
\end{center}
\end{figure*}

The observed numbers of events after the full event selection are listed in Table~\ref{tab:yield},
along with those expected for the backgrounds and for the charged Higgs boson production.
The systematic uncertainties listed in Table~\ref{tab:yield} are discussed in \refsec{uncertainties}.
The $\mT$ distributions with
all event selection criteria applied are shown in Fig.~\ref{fig:taunu:hadr:mt} for
$\mHp < \mt-\mb$ and $\mHp > \mt-\mb$.
In the $\mHp > \mt-\mb$ region, the limited number of background events in the high-\mT tail
is modelled by fitting an exponential function of the form $p_0 \re^{-p_1(\mT-c)}$, where
$p_0$ and $p_1$ are positive free parameters and where $c=180\GeV$ is the
starting point of the fit.
In the region of $\mT > 160\GeV$ the event yields for the backgrounds are
replaced by those obtained from this fit.
The slight excess of observed events in the $\mT$ spectrum for $\mHp > \mt-\mb$
and limits on the production of the charged Higgs boson extracted from these distributions are discussed in \refsec{results}.

\begin{table*}[htbp]
\begin{center}
\topcaption{Numbers of expected signal and background events with their statistical and systematic uncertainties
listed together with the number of observed events after the full event selection is applied in the \tauhjets final state.
For illustrative purposes, the expected signal yields are shown for $\mHp = 120\GeV$ normalized to $\lightLimitTaunuHadr = 0.01$ 
and for $\mHp = 300\GeV$ normalized to $\heavyLimitTaunuHadr = 1$\unit{pb}, which are typical values for the sensitivity of this analysis.
}
\setlength{\extrarowheight}{1.5pt}
\newcolumntype{x}{D{,}{}{-1}}
  \begin{tabular}{ l x }
  \hline
        Source & N_{\text{events}} (, \pm \text{stat} \pm \text{syst})\\
  \hline
  Signal, $\mHp = 120\GeV$ & 151, \pm   4 ~^{+  17}_{-  18}  \\
  Signal, $\mHp = 300\GeV$ & 168,  \pm  2 \pm  16 \\ 
  \hline
  EW+\ttbar with \tauh (data)    &  283, \pm  12 ~^{+  55}_{-  54}  \\
  Multijet background (data)       &   80, \pm   3 ~^{+  9}_{-  10} \\ 
  EW+\ttbar no \tauh (sim.)	& 47, \pm   2 ~^{+  11}_{-  10} \\ 
  \hline
  Total expected			&  410,  \pm 12 ~^{+  57}_{-  56} \\ 
  Data					& 392, \\
  \hline
  \end{tabular}
\label{tab:yield}
\end{center}
\end{table*}

\section{The \texorpdfstring{\mutauh}{tau+mu} final state for \texorpdfstring{$\Hptaunu$ and $\Hptb$}{H+ to taunu and H+ to tb}}
\label{sec:taunutb:taumu:analysis}

In this analysis, a charged Higgs boson with $\mHp > \mt-\mb$ is
assumed to be produced through \ProdHpFourandFiveFS: this can result in a final state characterized by the presence of two leptons. Here we describe the \mutauh  choice, whereas the $\ell\ell'$ ($\ell=e, \mu$) final state is discussed in Section~\ref{sec:taunutb:dilepton:analysis}.
The \mutauh final state is sensitive to the charged Higgs boson decay modes \Hptaunu and \Hptb.

In the first case, the $\Pgt$ decays hadronically and 
the final state is characterized by 
the leptonic decay of the $\PW$ boson from the $\cPaqt\to\cPaqb\PW^{-}$ decay which results in a muon in the final state.
In the second,
at least one of the W bosons from the top quarks decays to a $\Pgt$ lepton which in turn decays to hadrons, whereas the other decays into a muon.
Selecting the tau decay for one of the W bosons enhances the sensitivity to the \Hptaunu decay mode of the charged Higgs boson.
In this final state, the charged Higgs boson production is characterized by a number of b-tagged jets larger than in the SM backgrounds, 
 and consequently the shape of the b-tagged jet multiplicity distribution is used to infer the presence of a signal.
The dominant SM background processes are from $\ttbar\to\mu\tauh+X$, and other backgrounds where a jet is misidentified as a \tauh 
(mainly lepton+jet \ttbar events and $\PW$+jet production).

\subsection{Event selection}
\label{sec:taunutb:taumu:selection}

The event selection is similar to that used in the measurement of the top quark pair production cross section
in dilepton final states containing a \tauh~\cite{Chatrchyan:2012vs,Khachatryan:2014loa}.
A single-muon trigger with a threshold of $\pt > 24\GeV$ and $|\eta| < 2.1$ is used to select the events.

Events are selected by requiring one isolated muon with
$\pt > 30\GeV$ and $|\eta| < 2.1$, one hadronically decaying $\tau$ with
$\pt > 20\GeV$ and $|\eta| < 2.4$, at least two jets with $\pt > 30\GeV$ and $|\eta| < 2.4$, with at least
one jet identified as originating from the hadronization of a $\cPqb$ quark, and $\ETmiss > 40\GeV$. 
The $\Pgt_\mathrm{h}$ and the muon are required to have opposite electric charges.
The muon candidate is considered to be isolated if the relative isolation, as defined in Section~\ref{sec:reco}, is $I_\text{rel} < 0.12$.
The muon and the $\tau$ are required to be separated from each other and from any selected jet by a distance $\Delta R > 0.4$.
The choice of the radius matches the lepton isolation cone.
Events with an additional electron (muon) with $I_\text{rel} < 0.2$ and $\pt > 15 (10)\GeV$ are rejected.

\subsection{Background estimate}
\label{sec:taunutb:taumu:backgrounds}
There are three main background categories.
The first includes backgrounds that contain a genuine muon and a genuine $\tauh$, and is constituted by \ttbar$\to\mu\Pgt_\mathrm{h}+X$ 
production, associated $\cPqt\PW\to\mu\Pgt_\mathrm{h}+X$ production, $\cPZ\to\tau\tau\to\mu\Pgt_\mathrm{h}$ Drell--Yan production, and $\mathrm{VV}\to\mu\Pgt_\mathrm{h}+X$ processes.
The second category includes backgrounds with a genuine muon and an electron or muon misidentified as a $\Pgt_\mathrm{h}$, 
namely \ttbar$\to\mu\ell +X$, $\cPZ\to\mu\mu$, associated $\cPqt\PW\to\mu\ell +X$ production and $\mathrm{VV}\to\mu\ell +X$ production. 
The third category involves processes with a genuine muon and a jet misidentified as a $\Pgt_\mathrm{h}$, 
which include \ttbar$\to\mu$+jets, $\mathrm{V}$+jets, single top quark, and $\mathrm{VV}\to\mu$+jets events.
Within those categories, all genuine muons come from $\PW/\cPZ$ decays, either direct ($\PW\to\mu\nu$, $\cPZ\to\mu\mu$) 
or via intermediate $\tau$ decays ($\PW\to\tau\nu\to\mu+\MET$, $\cPZ\to\tau\tau\to\mu\tauh+\MET$). 

The backgrounds from the first two categories are estimated using simulation, except for the background due to \DY$\to\Pgt\Pgt$ events with one $\Pgt_\mathrm{h}$ and one $\Pgt$ decaying into a muon, which is estimated by taking for each variable the normalization from simulation and the shape from $\cPZ\to\Pgm\Pgm$ events in data, where each muon has been replaced with reconstructed particles from a simulated $\Pgt$ lepton decay. The procedure is similar to the one described in Section~\ref{sec:taunu:hadr:backgrounds:ewkgenuine}.

The backgrounds containing a jet misidentified as a $\Pgt_\mathrm{h}$  come mostly from $\PW$+jets and from $\ttbar\to\PW^+\PW^-\cPqb\cPaqb\to\Pgm\cPgn\cPq \cPaq' \cPqb\cPaqb$  events, 
and are collectively labeled ``misidentified $\Pgt_\mathrm{h}$'' in the following tables and plots. 
This background is estimated by weighting each event in a $\Pgm+\geq3$~jets control sample by the probability for any jet in the event to mimic a $\Pgt_\mathrm{h}$. 
The contribution from \ttbar$\to\mu\ell+X$ events, where one jet fakes a $\Pgt_\mathrm{h}$, is estimated using simulation and is subtracted from the data driven estimate to avoid double counting.
The probability that a jet is misidentified as a $\Pgt_\mathrm{h}$ is measured from data as a function of jet $\pt$, $\eta$, and jet radius
using $\PW$+jets
and multijet events~\cite{CMS-PAPER-TAU-11-001,Khachatryan:2014loa}.
Here, the estimate of the misidentified $\Pgt_\mathrm{h}$ background is improved with respect to the method used in Ref.~\cite{HIG-11-019} 
by weighting according to the quark and gluon jet compositions (from simulation) the estimates obtained in the W+jet and multijet samples~\cite{Khachatryan:2014loa}.
This data driven estimate is different from the one described in Section~\ref{sec:taunu:hadr:backgrounds:multijet}, where the control region is obtained by inverting isolation requirements on the reconstructed \tauh and only one control region is used. Here, estimating the fake rate in multijet events is not enough: the contamination from $\PW$+jets and  $\ttbar\to\PW^+\PW^-\cPqb\cPaqb\to\Pgm\cPgn\cPq \cPaq' \cPqb\cPaqb$ events must be taken into account as well.
The improvement in the central value of the estimate is verified with a closure test consisting in applying the data driven method to simulated events: the result of the closure test is compatible with the yields obtained from simulation, within the uncertainties. The systematic uncertainty associated to the data driven method is reduced by 30\% with respect to the cited paper.
The misidentified $\Pgt_\mathrm{h}$ background measured from data is consistent with the expectations from simulation.

The fraction of events from SM $\ttbar$ production that is not included in the \ttbar$\to\mu\Pgt_\mathrm{h}+X$ 
or misidentified $\Pgt_\mathrm{h}$ contributions is labeled as ``other \ttbar'' in the following tables and plots.
The \ttbar events are categorized in order to separate the contribution from each decay mode, using the full information on the simulated particles.

The single lepton trigger efficiency and the muon isolation and identification efficiencies are corrected by multiplicative data-to-simulation scale factors that depend on the 
muon $\pt$ and $\eta$.
Those factors are derived using a ``tag-and-probe'' method~\cite{Chatrchyan:2012bra,Khachatryan:2010xn}.
The trigger correction factors vary between 0.96 and 0.99, whereas the corrections to isolation and identification efficiency vary between 0.97 and 0.99. 

\subsection{Event yields}
\label{sec:taunutb:taumu:yields}

The numbers of expected events for the SM backgrounds, the expected number of signal events from the \ProdHpFourandFiveFS
process for $\mHp=250\GeV$ for the decay modes \Hptb and \Hptaunu,
and the number of observed events after all the selection requirements are summarized in Table~\ref{tab:leptonic_ltau_selections_tab1}.
Statistical and systematic uncertainties evaluated as described in Section~\ref{sec:uncertainties} are also shown.
For illustrative purposes, the number of signal events is normalized, assuming a 100\% branching fraction for each decay mode, to a cross section of 1\unit{pb}, which is typical of the cross section sensitivity of this analysis.

\begin{table*}[htbp]
\begin{center}
\topcaption{Numbers of expected events in the \mutauh final state for the SM backgrounds and in the presence of a signal
from \Hptb and \Hptaunu decays for $\mHp=250\GeV$ 
are shown together with the number of observed events after the final event selection.
For illustrative purposes, the number of signal events is normalized, assuming a 100\% branching fraction for each decay mode, to a cross section of 1\unit{pb}, which is typical of the cross section sensitivity of this analysis.
}
\setlength{\extrarowheight}{1.5pt}     
\newcolumntype{x}{D{,}{}{-1}}
\begin{tabular}{l x} 
\hline
Source  & N_{\rm events} (, \pm \text{stat} \pm \text{syst}) \\ 
\hline 
\Hptaunu, $\mHp=250\GeV$	&    176, \pm 10 \pm 13 \\ 
\Hptb, $\mHp=250\GeV$		&   37, \pm 2 \pm 3 \\ 
\hline 
$\ttbar\to\mu\tauh+{\rm X}$         &  2913,  \pm  14 \pm   242 \\ 
Misidentified $\tauh$                   &  1544, \pm14 \pm          175 \\ 
\ttbar~dilepton                &    101,  \pm   10 \pm    27 \\ 
$\cPZ/\gamma^{*}\to\Pe\Pe,\mu\mu$ &    12 ,  \pm   3 \pm     4 \\ 
$\cPZ/\gamma^{*}\to\tau\tau$  &   162, \pm   40 \pm    162 \\ 
Single top quark                 &   150,  \pm  12 \pm    18 \\ 
Dibosons                        &    20,  \pm  3 \pm     2 \\ 
\hline 
Total SM backgrounds	&  4903, \pm  45 \pm   341 \\ 
\hline 
Data &     4839,  \\ 
\hline 
\end{tabular}
\label{tab:leptonic_ltau_selections_tab1}
\end{center}
\end{table*}

\begin{figure*}[htbp]
\begin{center}
{\includegraphics[width=0.5\textwidth]{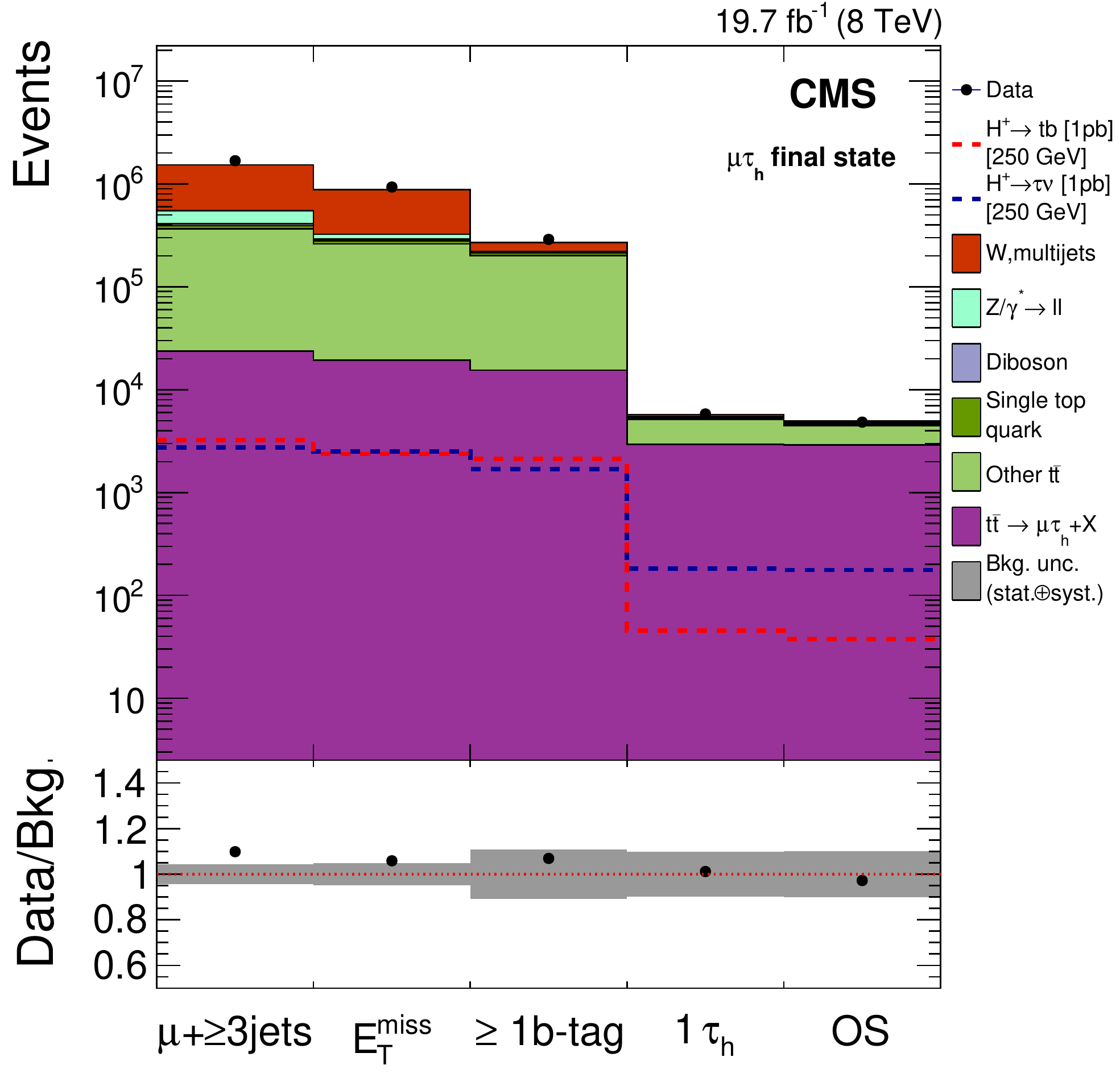}}\hfill
{\includegraphics[width=0.5\textwidth]{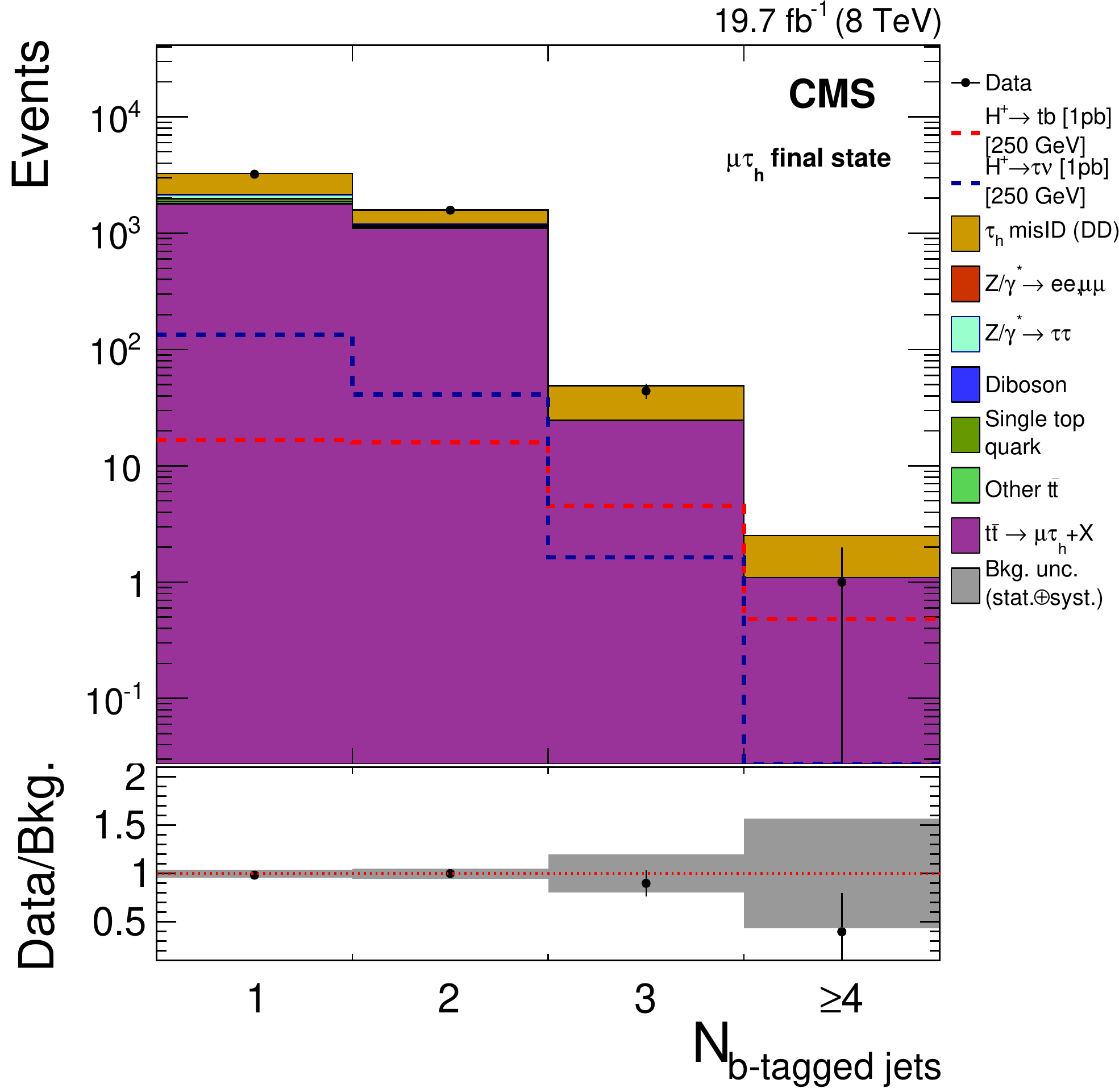}}\\
\caption{
Left: event yields after each selection step, where
OS indicates the requirement to have opposite electric charges for the $\Pgt_\mathrm{h}$ and the $\Pgm$.
The backgrounds are estimated from simulation and normalized to the SM prediction.
Right: the b-tagged jet multiplicity distribution after the full event selection.
As opposed to the left plot, the ``misidentified \tauh'' component is estimated using the data-driven method and labeled ``\tauh misID (DD)``,
while the remaining background contributions are from simulation normalized to the SM predicted values.
For both distributions, the expected event yield in the presence of the \Hptb and \Hptaunu decays is shown as
dashed lines for $\mHp=250\GeV$.
For illustrative purposes, the number of signal events is normalized, assuming a 100\% branching fraction for each decay mode, to a cross section of 1\unit{pb}, which is typical of the cross section sensitivity of this analysis.
$\mathcal{B}(\PH^{+}\to\cPqt\cPaqb)=1$ and $\mathcal{B}(\PH^{+}\to\Pgt^{+}\Pgngt)=1$, respectively.
The bottom panel shows the ratio of data over the sum of the SM backgrounds; the shaded grey area shows the statistical and systematic uncertainties added in quadrature.}
\label{fig:leptonic_ltau_figures}
\end{center}
\end{figure*}

Data and simulated event yields at various steps of the event selection
are shown in Fig.~\ref{fig:leptonic_ltau_figures} (left). Since the background estimate is derived from data only after requiring one \tauh,
the backgrounds here are normalized to the SM prediction obtained from the simulation. 
A good agreement ($\sim1$\% after the full selection) is found between data and the SM background expectations.
The multijet background contribution is negligible at the final selection step.
The expected signal event yields are shown as dashed lines.

The b-tagged jet multiplicity after the full event selection is shown in Fig.~\ref{fig:leptonic_ltau_figures} (right). 
Here the misidentified $\Pgt_\mathrm{h}$ background is derived from data, as discussed in Section~\ref{sec:taunutb:taumu:backgrounds}.
The ratio of the data to the sum of the expected SM background contributions is shown in the bottom panel. 
Limits on the production of the charged Higgs boson are extracted by exploiting this distribution.

\section{The dilepton (\texorpdfstring{$\Pe\Pe/\Pe\Pgm/\Pgm\Pgm$}{ee/emu/mumu}) final states for \texorpdfstring{$\Hptaunu$ and $\Hptb$}{H+ to taunu and H+ to tb}}
\label{sec:taunutb:dilepton:analysis}

In this analysis, a charged Higgs boson with $\mHp > \mt-\mb$ is assumed to be produced through \ProdHpFourandFiveFS and is searched for
in the $\ell\ell'$ final state. Assuming that the top quark produced in association with the charged Higgs boson decays as $\cPaqt\to\ell\nu b$, the dilepton final state is sensitive to charged Higgs boson decay modes \Hptb
(via leptonic decays of the top) or \Hptaunu (via leptonic decays of the tau lepton). 

This leads to a final state similar to the SM \ttbar dilepton final state, with the addition of one or two b jets.
The shape of the b-tagged jet multiplicity distribution is used to infer the presence of a charged Higgs boson signal.
The dominant SM backgrounds are from \ttbar and single top quark production.
An optimization procedure selected the b-tagged jet multiplicity variable as the most discriminating between the signal and the main backgrounds.
\subsection{Event selection}
\label{sec:taunutb:dilepton:selection}

The event selection is similar to that used for the measurement of the SM \ttbar cross section 
and of the ratio 
$\mathcal{B}(\cPqt\to\PW\cPqb)/\mathcal{B}(\cPqt\to\PW \cPq)$ 
in the dilepton channel~\cite{Chatrchyan:2012bra,Khachatryan:2014nda}.
Data were collected with double-lepton triggers ($\Pe\Pe/\Pgm\Pgm/\Pe\Pgm$) 
with \pt thresholds of 17\GeV for the leading lepton and 8\GeV for the other.
After offline reconstruction, events are required to have two isolated, oppositely charged, leptons (one electron and one muon, or two electrons, or two muons) 
with $\pt > 20\GeV$ and $|\eta| < 2.5$~($|\eta| < 2.4$) for electrons (muons), and at least two jets with $\pt > 30\GeV$ and $|\eta| < 2.4$. 
The relative isolation requirement is $I_{\rm rel} < 0.15 (0.20)$ for electrons (muons).
Jets are required to be separated by a distance $\Delta R=0.4$ from the isolated leptons.
A minimum dilepton invariant mass of 12\GeV
is required to reject SM background from low-mass resonances. For the same flavour channels ($\Pe\Pe$, $\Pgm\Pgm$), events with dilepton
invariant mass within 15\GeV from the $\cPZ$ boson mass are vetoed.
In order to account for the presence of neutrinos, $\MET > 40\GeV$ is required.
Finally, at least two b-tagged jets are required.

\subsection{Background estimate}
\label{sec:taunutb:dilepton:backgrounds}

The main background comes from \ttbar events in which both $\PW$ bosons decay leptonically, and surpasses by more than one order of magnitude the sum of the remaining backgrounds.
All backgrounds are estimated from simulation. 
The dilepton trigger efficiency is corrected by a multiplicative data-to-simulation scale factor dependent on the final state, in order
to provide agreement between data and simulation; 
the corresponding scale factors are computed using the ``tag-and-probe'' method, 
and the resulting values are 0.97, 0.95, and 0.92 for the $\Pe\Pe$, $\Pe\Pgm$, and $\Pgm\Pgm$ final states, respectively.
The data-to-simulation scale factors for the lepton identification and isolation efficiencies
are defined using a second ``tag-and-probe'' method with $\cPZ\to\Pe^{+}\Pe^{-}/\Pgm^{+}\Pgm^{-}$ events.
For electrons (muons) with $\pt > 20\GeV$,
they are found to vary between 0.91 (0.97) and 1.0 (0.99).

\subsection{Event yields}
\label{sec:taunutb:dilepton:yields}

\begin{figure}[htp]
\begin{center}
{\includegraphics[width=0.5\textwidth]{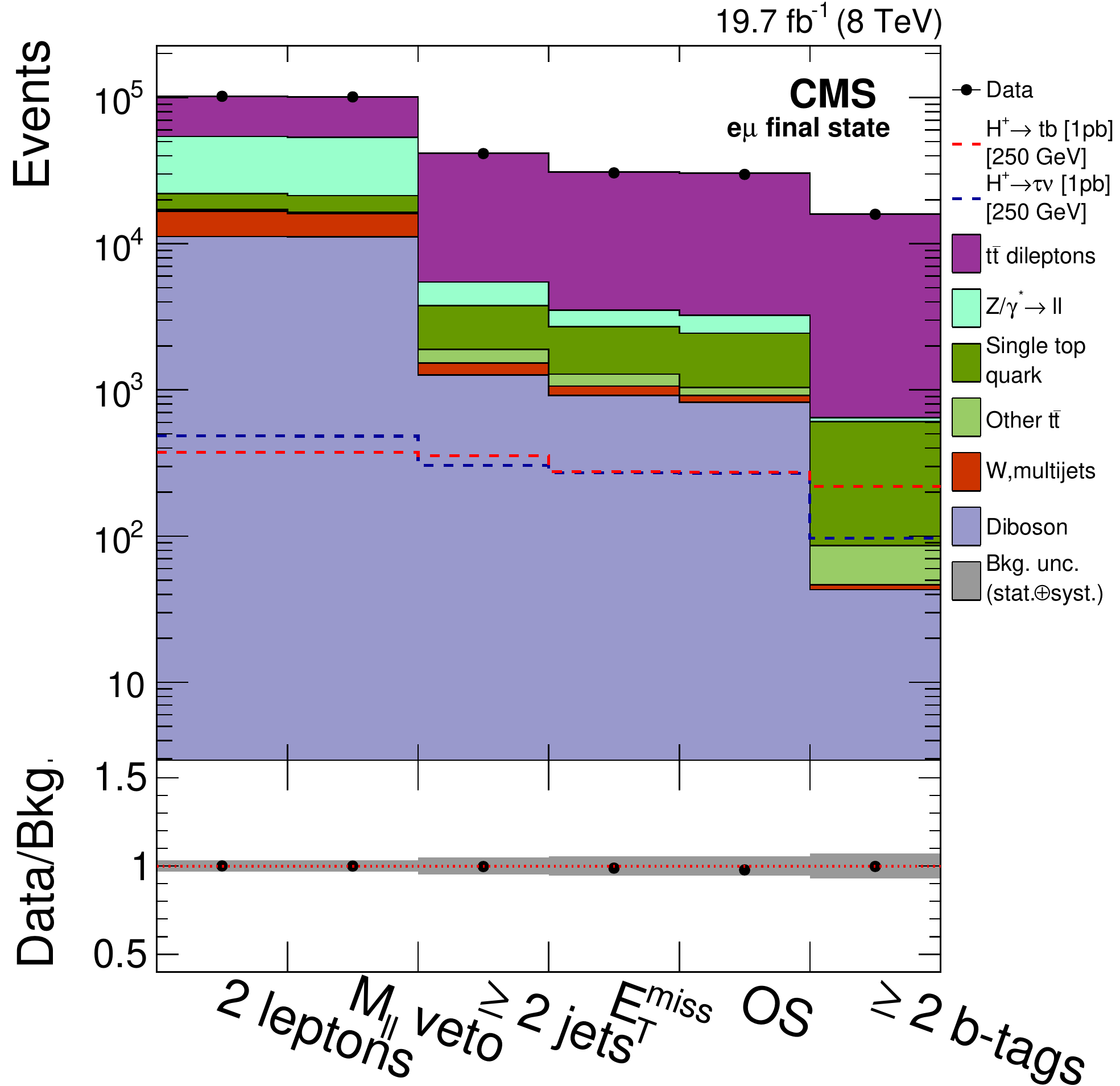}}\hfill
{\includegraphics[width=0.5\textwidth]{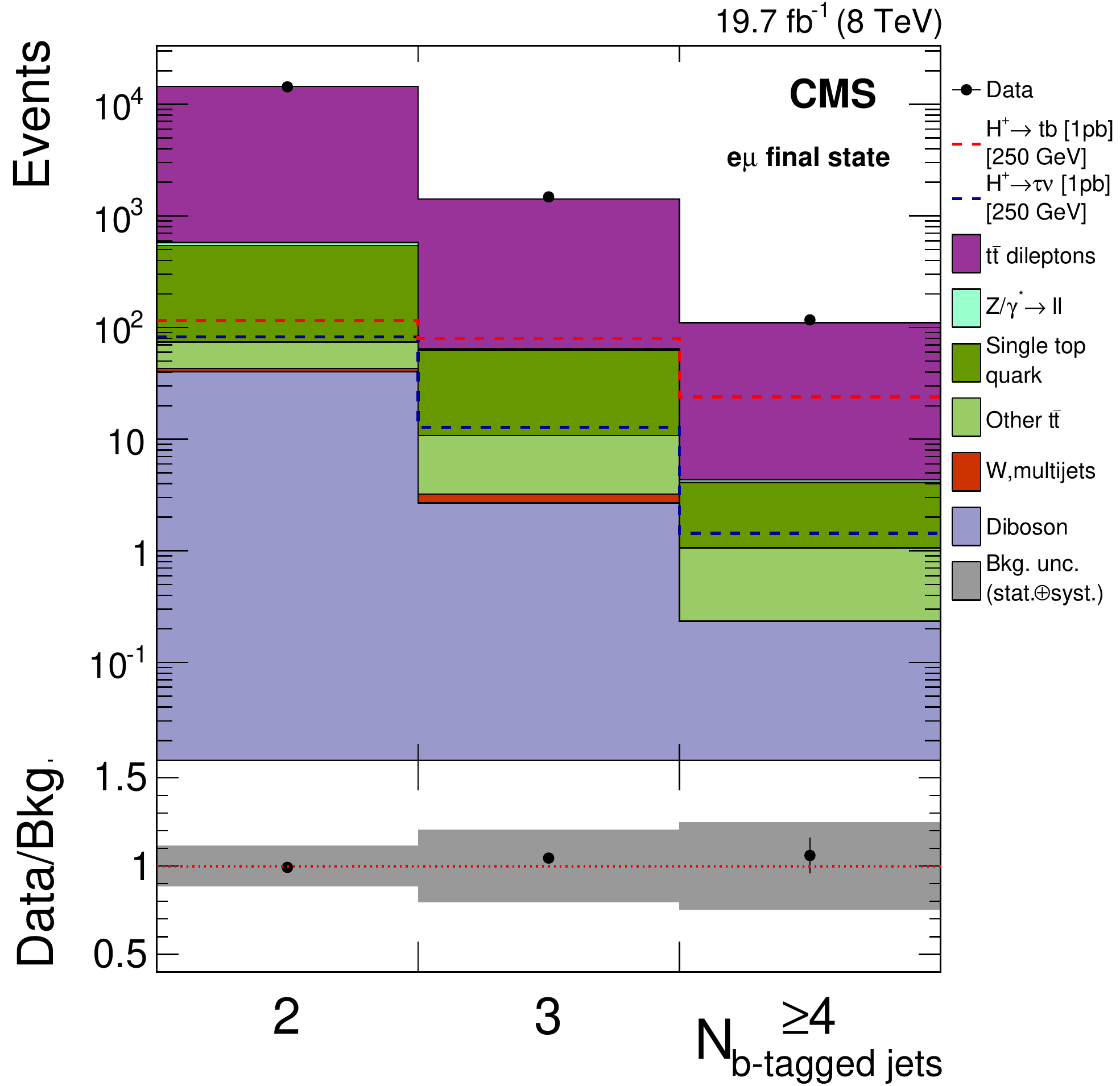}}\\         
\caption{
The event yields at different selection steps (left) and the b-tagged jet multiplicity after the full event selection for the $\Pe\Pgm$ final state (right).
For illustrative purposes, the number of signal events is normalized, assuming a 100\% branching fraction for each decay mode, to a cross section of 1\unit{pb}, which is typical of the cross section sensitivity of this analysis.
The bottom panel shows the ratio of data over the sum of the SM backgrounds; the shaded area shows the statistical and systematic uncertainties added in quadrature.
}
\label{fig:emu_figures}
\end{center}
\end{figure}

The number of data events after each selection requirement are in good agreement with the SM background expectations, 
and are shown in Fig.~\ref{fig:emu_figures} (left), for the $\Pe\Pgm$ final state as a representative example. 

The number of expected events after all selections in the $\ell\ell'$ final state
is summarized in Table~\ref{tab:SummaryEventYieldTauLeptonic} for the SM background processes 
and for a charged Higgs boson with a mass of $\mHp=250\GeV$.
The main background comes from \ttbar production in the dilepton final state, 
including all three lepton flavours.
Backgrounds from \ttbar production in the final states other than ``\ttbar dilepton'' (labelled ``other $\cPqt\cPaqt$") 
and other SM processes result in significantly smaller yields.
Statistical and systematic uncertainties evaluated as described in Section~\ref{sec:uncertainties} are also shown.
The data agree with the sum of expected backgrounds within the total uncertainties.

\begin{table}[htp]
\begin{center}
\topcaption{Number of expected events for the SM backgrounds and for signal events with a charged Higgs boson mass of $\mHp=250\GeV$
in the $\Pe\Pe$, $\Pe\Pgm$, and $\Pgm\Pgm$ dilepton final states after the final event selection. 
For illustrative purposes, the number of signal events is normalized, assuming a 100\% branching fraction for each decay mode, to a cross section of 1\unit{pb}, which is typical of the cross section sensitivity of this analysis.
Event yields are corrected with the trigger and selection efficiencies.
Statistical and systematic uncertainties are shown. 
}
\setlength{\extrarowheight}{1.5pt}
\begin{tabular}{l c c c }
\hline
Source  & \multicolumn{1}{c}{$\Pe\Pe$}  & \multicolumn{1}{c}{$\Pe\mu$}  & \multicolumn{1}{c}{$\mu\mu$} \\
\hline
\Hptaunu, $\mHp=250\GeV$	& $39\pm3\pm3$ & $97\pm4\pm5$ & $40\pm3\pm3$ \\
\Hptb, $\mHp=250\GeV$		& $85\pm3\pm2$ &  $219\pm5\pm5$ & $90\pm3\pm2$ \\
\hline
\ttbar~dilepton  & $5692\pm17\pm520$ & $15296\pm28\pm1364$ & $6332\pm18\pm572$\\
Other~\ttbar & $22\pm4\pm5$ & $40\pm5\pm9$ & $17\pm3\pm5$\\
$\cPZ/\gamma^{*}\to\ell\ell$ & $96\pm7\pm35$ & $36\pm2\pm7$ & $139\pm10\pm42$\\
W+jets, multijets & $6\pm2\pm1$ & $3\pm1\pm1$& \multicolumn{1}{c}{$<1$} \\
Single top quark & $199\pm10\pm21$   &  $522\pm15\pm54$  & $228\pm10\pm26$\\
Dibosons & $15\pm1\pm2$ & $43\pm2\pm6$ & $20\pm1\pm3$\\
\hline
Total SM backgrounds & $6032\pm20\pm521$ & $15941\pm32\pm1365$ & $6736\pm23\pm575$\\
\hline
Data & 6162 & 15902 & 6955\\
\hline
\end{tabular}
\label{tab:SummaryEventYieldTauLeptonic}
\end{center}
\end{table}

The b-tagged jet multiplicity distribution for the $\Pe\Pgm$ final state, shown after the full event selection in 
Fig.~\ref{fig:emu_figures} (right), is used to extract limits on the charged Higgs boson production.

\section{The single-lepton (\texorpdfstring{$\Pe/\Pgm$}{e/mu}+jets) final states for \texorpdfstring{$\Hptb$}{H+ to tb}}
\label{sec:tb:singlelepton:analysis}
In this analysis, a charged Higgs boson with $\mHp > \mt-\mb$ and produced in association with a top quark \ProdHpFourandFiveFS,
is searched for in the decay mode  \Hptb. 
Of the two $\PW$ bosons produced from the top quark decays, one decays leptonically, while the other decays hadronically, leading to the final state signature of one lepton, jets, and \MET. 
These final states are similar to the SM \ttbar semileptonic final states, with the addition of one or two b jets.
While the dilepton analysis (Section ~\ref{sec:taunutb:dilepton:analysis}) uses the shape of the full b~tagged jet multiplicity distribution to check for the presence of a signal, for this analysis, an optimization procedure led to use of  the \HT distribution, defined as the scalar sum of the \pt of all selected jets, subdivided by b~tagged jet multiplicity, to infer the presence of a charged Higgs signal.
Due to the jet composition of the signal, the \HT distribution peaks at higher energies and has a less steeply falling high energy tail than the major backgrounds.
The dominant backgrounds are \ttbar, $\PW$+jets, and single top quark production.

\subsection{Event selection}
\label{sec:tb:ljets:selection}

Data were collected by the single-electron or a single-muon trigger with \pt thresholds of 27 and 24\GeV, respectively.
The offline event selection requires the presence of exactly one isolated
electron (muon) with $\pt > 30~(27)\GeV$ and $|\eta| < 2.5~(2.4)$. The electrons (muons) are
required to be isolated with $I^\ell_{\rm rel} < 0.10$ (0.20), with $I^\ell_{\rm rel}$ defined in Section ~\ref{sec:reco}. 
Events with additional leptons are rejected.
To maintain exclusivity with the other analyses included in this paper, events with one or more hadronic $\tau$ decays with $\pt^{\tauh} > 20\GeV$ and $|\eta_{\tauh}| < 2.4$ are rejected.
In addition, the presence of at least two jets with $\pt > 30\GeV$ and $|\eta| < 2.4$ are required, 
with $\pt > 50\GeV$ for the jet with the highest $\pt$.
At least one of the selected jets is required to be b-tagged.
The \MET must exceed 20\GeV to mimic the presence of a neutrino in the final event signature. 

To account for differences in modelling of the lepton identification and trigger efficiency between simulation and data, 
$\eta$- and $\pt$-dependent scale factors are applied. 
The single-electron trigger correction factor is 0.973 (1.020) for $|\eta|\leq 1.5$ ($1.5 < |\eta|\leq 2.5$) and the single-muon trigger correction factors and  corrections to identification efficiency are similar to those in Section~\ref{sec:taunutb:taumu:backgrounds}.

Events are classified into two categories, a signal region (SR) and a control region (CR). 
The CR is defined by having low reconstructed jet multiplicity,  $2\leq N_\mathrm{jet} \leq 3$, and is used to derive normalizations for dominant backgrounds from data.
The SR is distinguished by its high jet multiplicity, and defined by the  requirement $N_\mathrm{jet} \geq 4$.
These categories are further subdivided according to the b-tagged jet multiplicities, $N_\mathrm{b~tag}$, with the CR split into 3 subcategories ($N_\mathrm{b~tag} =0$, $N_\mathrm{b~tag} =1$, and $N_\mathrm{b~tag} \geq 2$) and the SR split into two ($N_\mathrm{b~tag} =1$ and $N_\mathrm{b~tag} \geq 2$).
Distinguishing between electron and muon channels leads to a total of four SR categories and six CR categories.

\subsection{Background estimate}
\label{sec:tb:ljets:backgrounds}

The following background processes are considered: $\cPqt\cPaqt$, $\PW$+jets, single top quark,
$\cPZ/\gamma^*$+jets, and dibosons ($\PW\PW$, $\PW\cPZ$, and $\cPZ\cPZ$). 

The backgrounds are subdivided into seven independent categories distinguished by their yields and shapes in the signal region.
The six samples: \ttbar, $\PW$+$\cPqc$ (events with one or more $\cPqc$ jet), $\PW$+$\cPqb$ (events with one or more $\cPqb$ jet), $\PW$+light-flavour ($\cPqu, \cPqd, \cPqs, \cPg$) jets, single top quark, and multijets are defined as independent categories.
The small backgrounds with similar \HT distributions from dibosons and $\cPZ/\gamma^*$+jets are merged into the ``$\cPZ/\gamma^*$/VV`` background.
Additional contributions from $\ttbar$+$\PW$ and $\ttbar$+$\cPZ$ are considered negligible.
All \HT distributions are taken from simulation.

For the backgrounds which contribute little to the signal region (single top quark, diboson, Z+jets, and multijet production), the normalizations are taken directly from the simulation. 
For the four remaining processes which provide most of the background in the signal region (\ttbar production, $\PW$+$\cPqc$, $\PW$+$\cPqb$, and $\PW$+light-flavour jets),
the normalization is initially taken from simulation, but is then determined by a simultaneous fit of the background distributions to the data. 
The normalization is allowed to float freely during the limit setting.
Thus, the fit finds the best values for these normalizations, derived using simulated and observed yields from both the control and signal regions.
The values obtained for these normalizations for the electron (muon) channel are 1.01 (1.01) for \ttbar, 2.06 (1.62) for $\PW$+$\cPqc$, 1.90 (1.48) for $\PW$+$\cPqb$, and 1.18 (1.01) for $\PW$+$\mathrm{light}$-flavour jets.
The \ttbar background dominates and constitutes 80\% of events with 1 b-tagged jet and 93\%  of events with 2 or more b-tagged jets, while  $\PW$+$\cPqc$ and $\PW$+$\cPqb$ backgrounds contribute to 8\% and 2\%, respectively. 
Differences in normalizations between electron and muon channels are accounted for in the systematic uncertainties, as noted in Table~\ref{tab:SummarySystematicsLjets}.

A closure test is performed to assess the  validity of the assumption that the normalizations derived from the fit to data are not dependent on the jet multiplicities of the samples. 
A  sample of events with at least four jets, none of which are b-tagged, is used for the closure test. 
The agreement between observed and predicted events, using the post-fit values of the normalizations, across all bins in the high jet multiplicity region is found to be within 10\%. 

\subsection{Event yields}
\label{sec:tb:ljets:yields}

\begin{figure}[htp]
\begin{center}
{\includegraphics[width=0.48\textwidth]{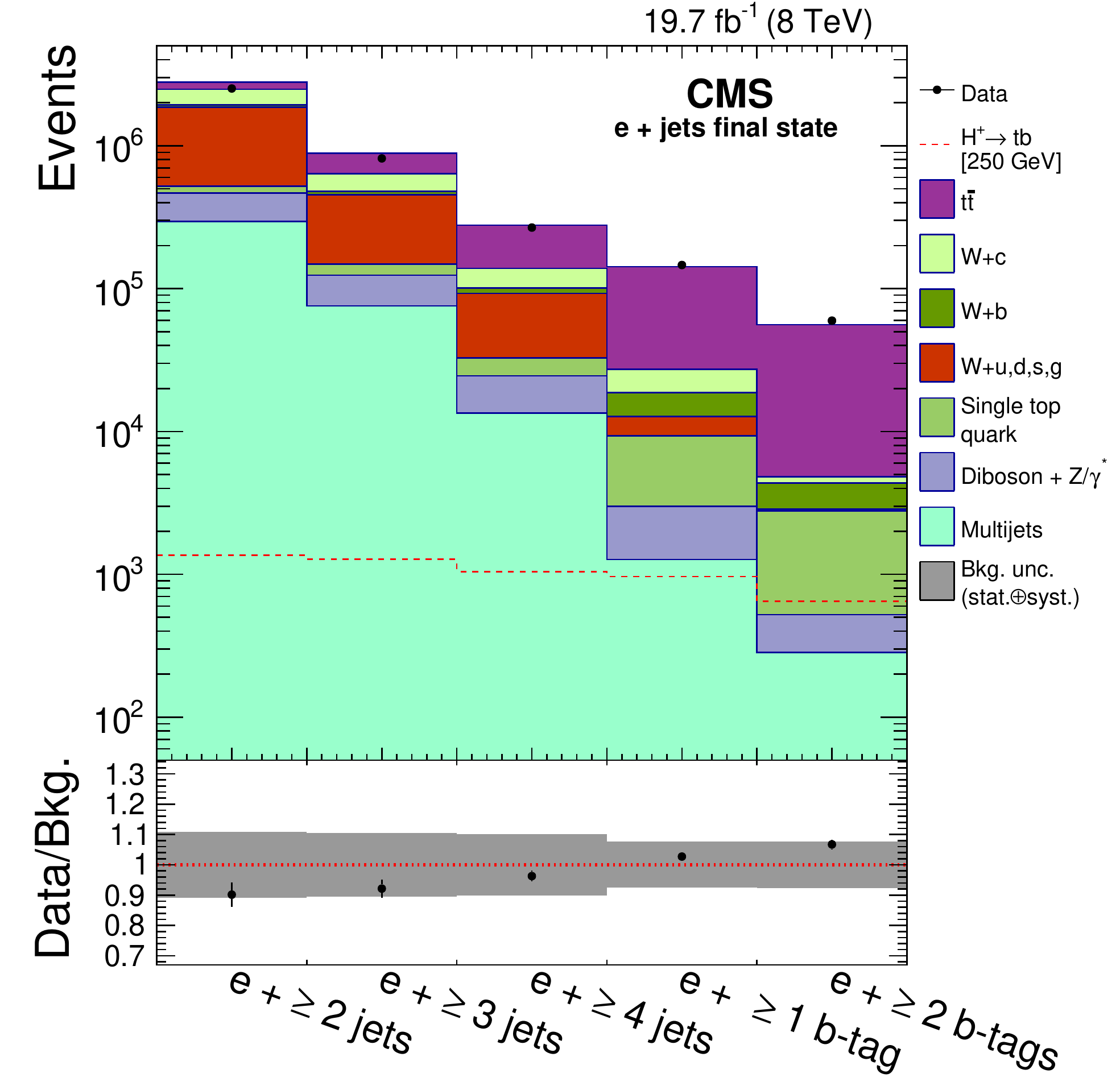}} 
{\includegraphics[width=0.48\textwidth]{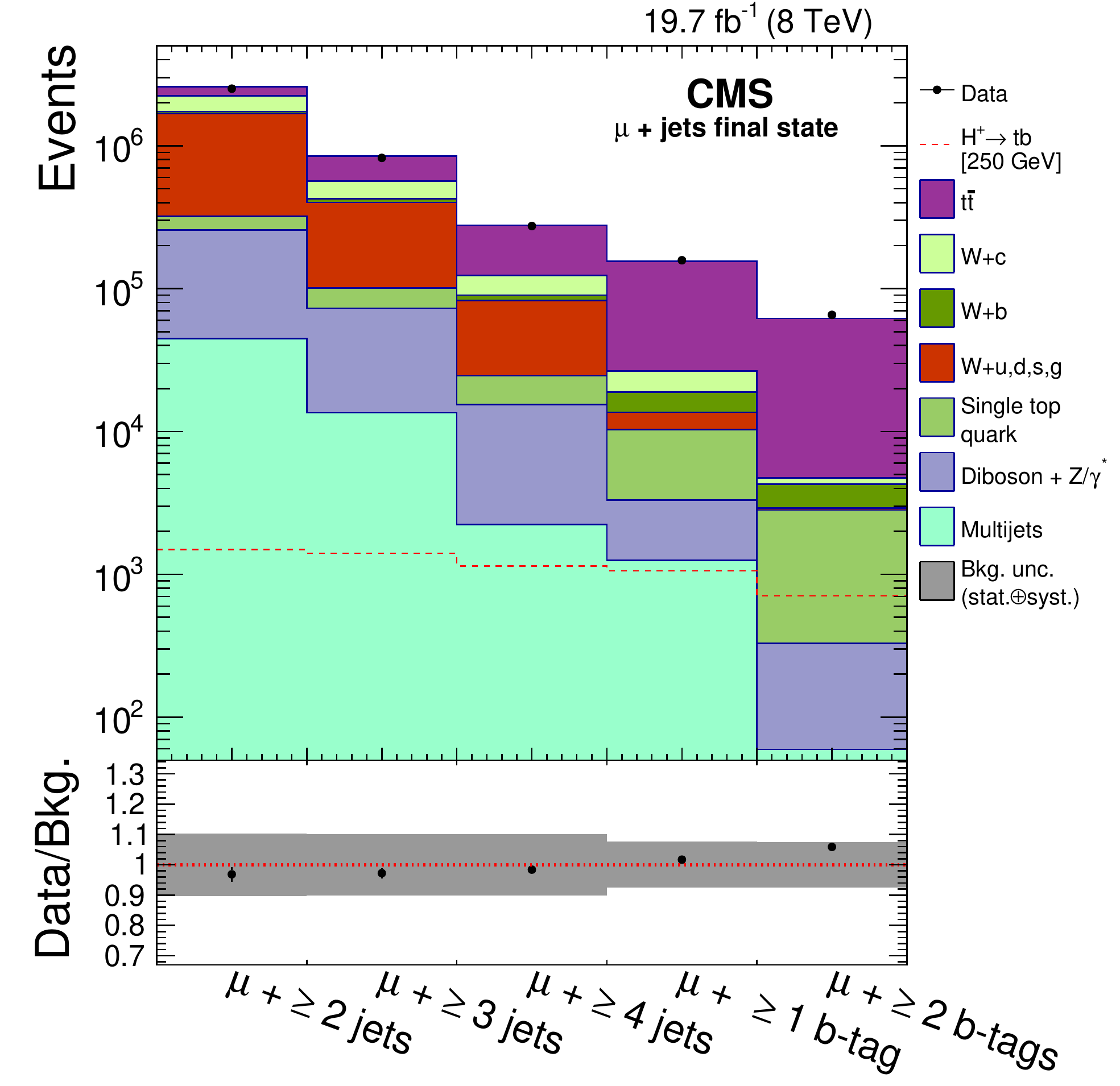}} 
\caption{
Event yields after different selection cuts for both the $\Pe$+jets (left) and $\Pgm$+jets (right) final state.
Expectations for the charged Higgs boson for $\mHp=250\GeV$, for the \Hptb decays, are also shown.
For illustrative purposes, the signal is normalized, assuming $\BtbHp=1$, to a cross section of 1\unit{pb}, which is typical of the cross section sensitivity of this analysis.
The bottom panel shows the ratio of data over the sum of the SM backgrounds with the total uncertainties.
}
\label{fig:cutflow_singlelep}
\end{center}
\end{figure}

The number of data events after different selection cuts are compared to expectations from SM backgrounds and are shown in
Fig.~\ref{fig:cutflow_singlelep}  for both the electron and muon channels. 
Results are in good agreement with SM background expectations.

The number of expected events in the final selection for each subsample can be seen in Table~\ref{tab:SummaryEventYieldLjets}. The number of events for data, SM background processes, and a charged Higgs boson with a mass of $\mHp=250\GeV$ are shown.
The leading contributions to the SM background come from \ttbar events with a semi-leptonic final state, $\PW$ boson production in association with heavy-flavour jets, and single top quark production.
Statistical and systematic uncertainties are evaluated as described in Section~\ref{sec:uncertainties}.

\normalsize
\begin{table}[htp]
	\begin{center}
		\topcaption{Number of expected events for the SM backgrounds and for signal events with a charged Higgs boson mass of $\mHp=250\GeV$ in the \ljets final states after the final event selection.
		Normalizations for $\PW$+light-flavour jets, $\PW+\cPqc$, $\PW+\cPqb$, and \ttbar are derived from data. Normalizations for other backgrounds are based on simulation.
        For illustrative purposes, the signal is normalized, assuming $\BtbHp=1$, to a cross section of 1\unit{pb}, which is typical of the cross section sensitivity of this analysis.
		Statistical and systematic uncertainties are shown. 
		}

	\label{tab:SummaryEventYieldLjets}

	\setlength{\extrarowheight}{1.5pt}
	\footnotesize{
	\begin{tabular}{l c c | c c }

		\hline
		Source  								& $N_\mathrm{b~tag}=1$			& $N_\mathrm{b~tag}\geq 2$ 		& $N_\mathrm{b~tag}=1$ & $N_\mathrm{b~tag}\geq 2$  \\  \hline \hline
        											& \multicolumn{2}{c|}{$\Pe$+jets}   								& \multicolumn{2}{c}{$\Pgm$+jets} \\ \hline
		\Hptb, $\mHp=250\GeV$ 				& 315 $\pm$ 4 $\pm$ 17			& 647 $\pm$ 6 $\pm$ 34			& 348 $\pm$ 5 $\pm$ 19			& 707 $\pm$ 7 $\pm$ 37  \\
		\hline
		\ttbar                            						& 64111 $\pm$ 74 $\pm$ 5174 	& 51059 $\pm$ 66 $\pm$ 4679  	& 71593 $\pm$ 78 $\pm$ 5711  	& 57094 $\pm$ 70 $\pm$ 5160 \\
		$\PW$+$\cPqc$						& 8031 $\pm$ 89 $\pm$ 1047  		& 482 $\pm$ 21 $\pm$ 79		     	& 7156 $\pm$ 77 $\pm$ 11193 	& 460 $\pm$ 18 $\pm$ 92     \\
		$\PW$+$\cPqb$						& 4470 $\pm$ 61 $\pm$ 1206  		& 1486 $\pm$ 35 $\pm$ 404     	& 3926 $\pm$ 53 $\pm$ 1386 	 	& 1364 $\pm$ 32 $\pm$ 484     \\
		$\PW$+u,d,s,g							& 3326 $\pm$ 44 $\pm$ 598      	& 90 $\pm$ 7 $\pm$ 21			& 3231 $\pm$ 39 $\pm$ 581       	& 95 $\pm$ 7 $\pm$ 22              \\
		Single top quark   						& 4059 $\pm$ 42 $\pm$ 463      	& 2253 $\pm$ 30 $\pm$ 274  		& 4496 $\pm$ 44 $\pm$ 524    		& 2493 $\pm$ 32 $\pm$ 295 \\
		$\cPZ/\gamma^*$/VV   					& 1492 $\pm$ 54 $\pm$ 771  		& 237 $\pm$ 21 $\pm$ 130   		& 1792 $\pm$ 60 $\pm$ 942  		& 269 $\pm$ 22 $\pm$ 140 \\
		Multijet background						& 990 $\pm$ 270 $\pm$ 1040	   	& 280 $\pm$ 160 $\pm$ 290	     	& 1220 $\pm$ 480 $\pm$ 1260 	& 59 $\pm$ 34 $\pm$ 60  \\
		\hline
		Total SM backgrounds    					& 86480 $\pm$ 310 $\pm$ 5620 	& 55890 $\pm$ 190 $\pm$ 4720	& 93410 $\pm$ 500 $\pm$ 6240 	& 61836 $\pm$ 95 $\pm$ 5194	 \\
		Data                              					& 86580						& 59637						& 92391						& 65472           \\
		\hline
		
	\end{tabular}
	}
	\end{center}
\end{table}
\normalsize

\begin{figure}[htp]
  \begin{center}

      {\includegraphics[width=0.5\textwidth,angle=0]{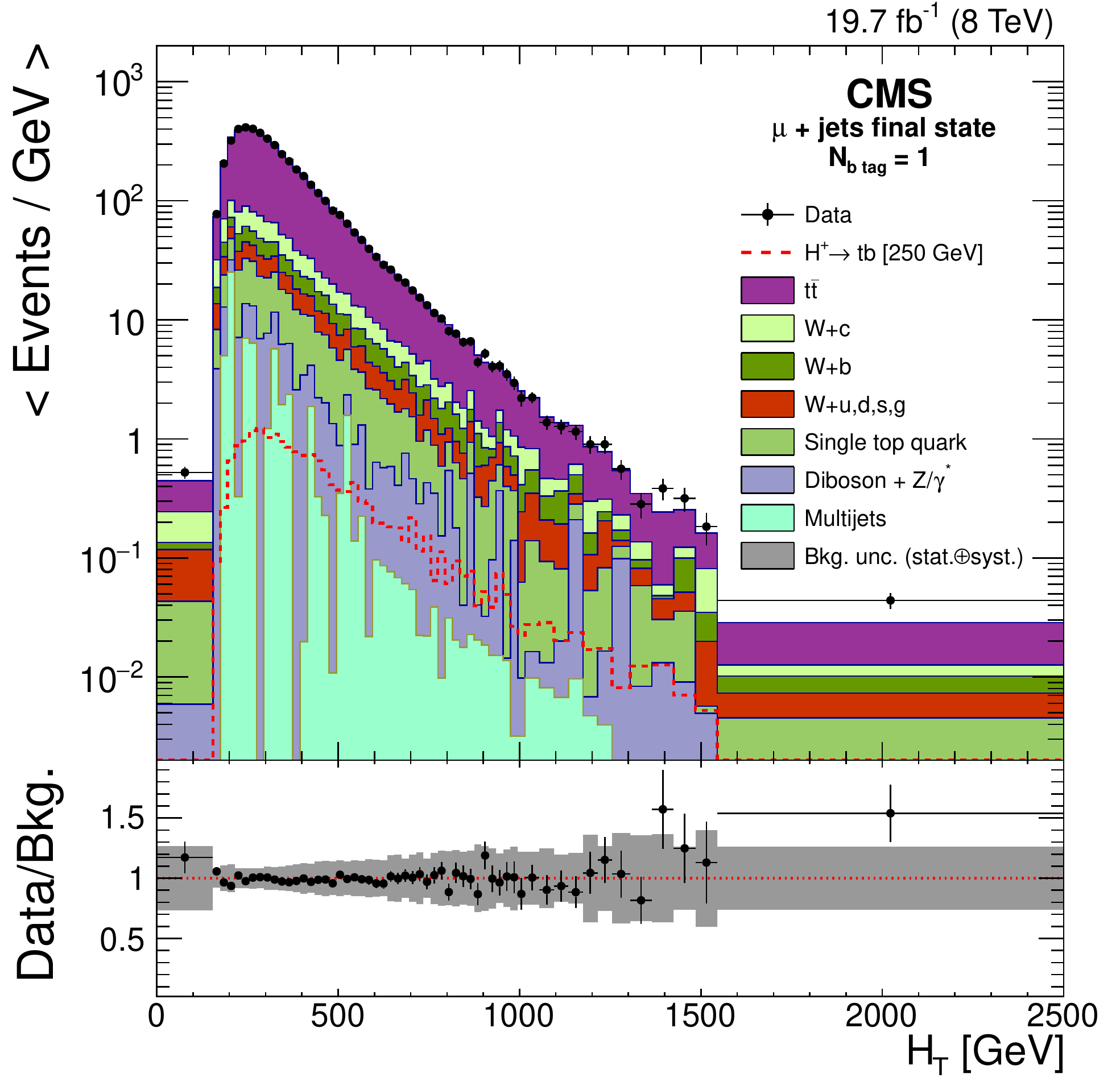}}\hfill
      {\includegraphics[width=0.5\textwidth,angle=0]{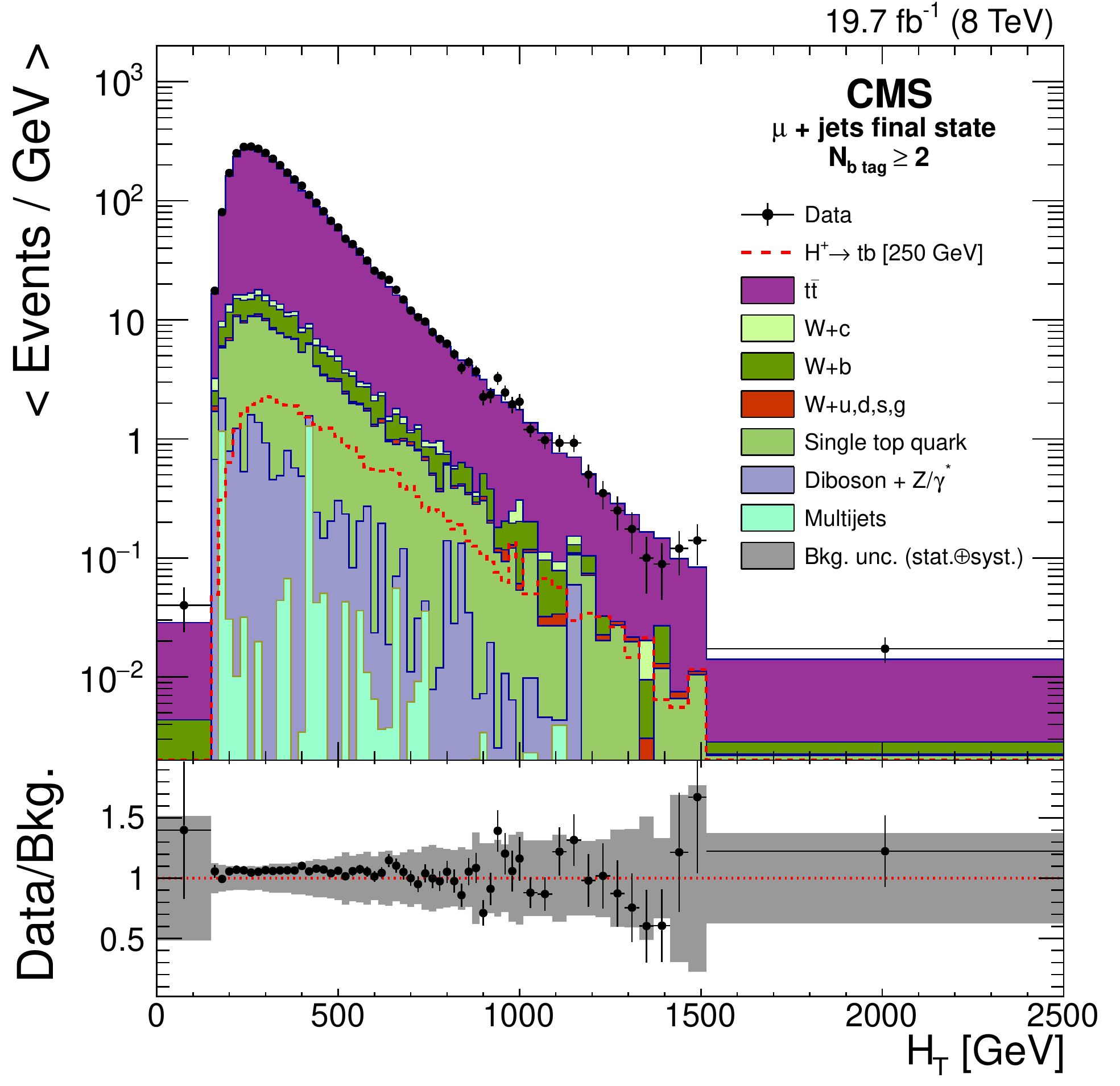}}\\

    \caption{The \HT distributions observed in data and predicted for signal and background in the $\Pgm$+jets channel with $N_\mathrm{b~tag}=1$ (left) and $N_\mathrm{b~tag}\geq 2$ (right). 
Normalizations for \ttbar, $\PW+\cPqc$, $\PW+\cPqb$, and $\PW$+light-flavour jets are derived from data. Normalizations for other backgrounds are based on simulation.
Expectations for the charged Higgs boson for $\mHp=250\GeV$, for the \Hptb decays, are also shown.
For illustrative purposes, the signal is normalized, assuming $\BtbHp=1$, to a cross section of 1\unit{pb}, which is typical of the cross section sensitivity of this analysis.
The bottom panel shows the ratio of data and the sum of the SM backgrounds with the total uncertainties. 
Bin contents are normalized to the bin width.}
\label{fig:ljets_HT_scaled}
  \end{center}
\end{figure}

The \HT distributions for the two signal regions in the muon channel are shown in  
Fig.~\ref{fig:ljets_HT_scaled}.
Limits on the production cross section of the charged Higgs boson are extracted by exploiting these distributions.

\section{Systematic uncertainties}
\label{sec:uncertainties}
The uncertainties common to the analyses are presented in \refsec{common:uncertainties}.
The uncertainties specific to the individual analyses are discussed in \refmultisec{taunu:hadr:uncertainties}{tb:ljets:uncertainties}.

\subsection{Uncertainties common to the analyses}
\label{sec:common:uncertainties}
The sources of systematic uncertainties common to 
the analyses (unless specified otherwise) and affecting simulated samples only are as follows:
\begin{itemize}
\item 
Uncertainties in the lepton trigger, identification, and isolation efficiencies are
calculated from independent samples with a ``tag-and-probe'' method.
The uncertainties in the single electron, single muon, and dilepton triggers amount to 2\%, 2\%, and 3\%, respectively.
For the \tauhjets final state, the treatment is detailed in \refsec{taunu:hadr:uncertainties};
\item 
The uncertainty in the efficiency and identification of electrons is 2\% (1\%) for $\pt > 20~(30)\GeV$. 
For muons, the uncertainty in the efficiency and identification is 1\%;
\item 
The uncertainty in $\tauh$ identification efficiency is estimated to be 6\%~\cite{Khachatryan:2014wca};
\item
The misidentification uncertainty in events with an electron misidentified as the $\tauh$ is 20\% (25\%) for the barrel (endcap);
for events with a muon (jet) misidentified as the $\tauh$ an uncertainty of 30\% (20\%) is estimated~\cite{Khachatryan:2014wca};
\item
The uncertainty in the $\Pgt_\mathrm{h}$ energy scale ($\Pgt_\mathrm{h}$ ES) is estimated by varying the $\Pgt_\mathrm{h}$ momentum by ${\pm}3\%$~\cite{Khachatryan:2014wca};
\item
The uncertainties in the jet energy scale (JES), jet energy resolution (JER), and the contribution to \MET scale from particles not clustered to jets (``unclustered \MET scale'') are estimated independently
according to the prescription described in Ref.~\cite{CMS-JME-10-011}, and found to within 1--6\% for the signal and dominant 
simulated backgrounds in all the analyses. The variations of these quantities are also propagated to the \MET.
The uncertainty in JES is evaluated as a function of jet \pt and jet $\eta$, and takes into account JES variations due to parton flavour;
\item 
The uncertainty arising from b tagging/mistagging efficiencies is estimated according to the description in Ref.~\cite{CMS-PAS-BTV-13-001}. 
Values of 3--20\% are found in the different analyses;
\item 
A 100\% uncertainty is assumed for the reweighting of the top quark \pt spectrum of each top quark in simulated SM \ttbar events, discussed in \refsec{sim}.
The reweighting and uncertainty depends on the top quark decay~\cite{CMS-PAPER-TOP-12-028};
\item 
The uncertainty in pileup event modelling is estimated by varying the total inelastic cross section used to infer the pileup distribution in data by ${\pm} 5$\%;
\item 
Uncertainties in the theoretical cross section normalization described in detail in \refsec{sim};
\item 
For the \mutauh, \ljets, and $\ell\ell'$ final states, the uncertainties 
due to ME and parton shower (PS) matching, and those due to the factorization and renormalization scale choices
are applied only to the dominant simulated \ttbar backgrounds; they are
estimated by varying by a factor of two the threshold between jet production at the ME level and via PS
and by varying by a factor of four the nominal scale given by the momentum transfer of the hard process ($Q^{2}$) in the event;
\item 
For the \mutauh and  $\ell\ell'$ final states, the uncertainty in the b-tagged jet multiplicity distribution 
shapes due to PDF variations is estimated separately for the dominant simulated \ttbar backgrounds by varying independently the components of the PDF parameterization;
\item 
For the \mutauh and $\ell\ell'$ final states, the uncertainty due to the modelling
of the associated heavy-flavour production (\ttbb) is taken into account
by assigning to each bin of the b-tagged jet multiplicity distribution of the \ttbb events
an uncorrelated bin-by-bin uncertainty of 44\%.
This uncertainty is based on the comparison between the observed and predicted ratios of 
$\sigma(\ttbar+\cPqb\cPaqb)/\sigma(\ttbar+\cPq\cPaq)$~\cite{CMS-PAPER-TOP-13-010};
\item 
The uncertainty in the integrated luminosity is estimated to be 2.6\%~\cite{lum-13-001}.
\end{itemize}

\subsection{The \texorpdfstring{\tauhjets}{tau\_h+jets} final state for \texorpdfstring{$\Hptaunu$}{H+ to taunu}}
\label{sec:taunu:hadr:uncertainties}

In the \tauhjets final state, some of the systematic uncertainties related to simulated samples also affect
the background measurements from data.
In the multijet background, a small number of simulated EW+\ttbar events is subtracted
from the data to obtain the number of multijet events.
The uncertainties affecting this small number of simulated events are taken
into account, but their magnitudes are suppressed because they apply to only a fraction of the multijet background and 
a minus sign is assigned for them to denote anticorrelation.
For the ``EW+\ttbar with \tauh'' background, uncertainties 
related to the simulated $\tau$ lepton decays are taken into account.

In addition to the uncertainties already described in \refsec{common:uncertainties},
the following sources of systematic uncertainties are taken into account for the \tauhjets final state:
\begin{itemize}
\item 
The uncertainties in the efficiencies of the $\Pgt$ part and \MET part of the $\Pgt$+\MET trigger
measured from data and simulation are considered separately.
The simulated samples are affected by both sources of uncertainty, while the ``EW+$\ttbar$ with \tauh'' background, obtained 
with the ``embedding'' procedure, is affected only by the uncertainty in the trigger efficiency measured in data.
Furthermore,
for the ``EW+$\ttbar$ with \tauh'' background, the data part of the $\Pgm$ trigger efficiency is also considered, and
a further 12\% uncertainty is applied for approximating the \MET of the high-level trigger by offline calorimeter-based \MET;
\item 
The uncertainty in vetoing events with electrons and/or muons affecting only the simulated samples is estimated from the uncertainty
in the electron and muon reconstruction, identification, and isolation efficiencies as 2\% (1\%) for electrons (muons);
\item 
A 50\% normalization uncertainty for the \mT distribution is assigned for the simulated single top quark samples in the
``EW+$\ttbar$ no \tauh'' background for assigning as event weight the probability to pass b tagging instead of applying the b tagging condition;
\item 
The uncertainties in the ``EW+$\ttbar$ with \tauh'' background measurement method are described in the following.
The uncertainty in the muon identification efficiency in data is found to be small.
The contamination of the $\Pgm$+jets control sample by multijet events is estimated with a $\Pgm$ enriched simulated multijet sample to be at most 2\%, which is taken as a systematic uncertainty.
The fraction of events with $\PW \to \Pgt\Pgngt \to \Pgm\Pgngm\Pgngt$, discussed in \refsec{taunu:hadr:backgrounds:ewkgenuine}, is evaluated from simulated events and found to obey a functional form 
$( 1-a ) \, p_{T}^{-b}$, 
where $a$ and $b$ are positive constants and \pt is the transverse momentum of the selected muon. The systematic uncertainty for correcting the event yield for this effect amounts to 1.2\%.
A 100\% uncertainty is assumed on the event weights accounting for the 
difference between the $\Pgt$+jets and embedded $\Pgm$+jets events from simulated \ttbar events (denoted as ``Non-emb. vs. emb. difference'' in Table~\ref{tab:fullyhadronic_systematics_tab1}) observed in the \mT distribution;
\item 
The uncertainties in the multijet background measurement method are described in the following. 
The statistical uncertainty in the \MET template fit that is performed in each bin of $\pt^{\Pgt_\mathrm{h}}$, as described in \refsec{taunu:hadr:backgrounds}, is estimated to be 3\% in each $p_{\mathrm{T}}^{\tauh}$ bin.
The difference in the \mT distribution shapes between the nominal sample and the sample with inverted \tauh isolation criterion
is taken as a systematic uncertainty. It is evaluated from the ratio of the event yields of the samples with nominal and inverted \tauh isolation criterion as a function of \mT
after requiring the other \tauh selection criteria, the veto against electrons and muons, at least three jets, and the requirement on \Rcollmin.
The statistical uncertainty of the ratio of the event yields is found to account for the difference in the shape and its magnitude is taken as the systematic uncertainty. 
Its value ranges between 5--15\% depending on the bin of the \mT distribution.
\end{itemize}

A summary of the systematic uncertainties is shown in Table~\ref{tab:fullyhadronic_systematics_tab1}.

\begin{table*}[htbp]
\begin{center}
\topcaption{The systematic uncertainties (in \%) on event yields
for the charged Higgs boson signal processes
$\ttbar\to\cPqb\PH^+\cPaqb\PH^-$ ($\PH^+\PH^-$),
$\ttbar\to\cPqb\PH^+\cPaqb\PW^-$ ($\PH^+\PW^-$), and
$\Pp\Pp \to \cPaqt(\cPqb)\PH^{+}$ ($\PH^+$) and
for the background processes.
The uncertainties which depend on the \mT distribution bin are marked with (S) 
and for these the maximum integrated value of the negative or positive variation is displayed. 
Empty cells indicate that an uncertainty does not affect the sample.
The uncertainty values within the rows are considered to be fully correlated and
the values within the columns are considered to be uncorrelated.
A minus sign in front of an uncertainty value means anticorrelation with other values in the same row.}
\small{
\setlength{\extrarowheight}{1.5pt}
\begin{tabular}{l c c c c c c}
\hline
& Signal & Signal & Signal
& Multi- & EW+\ttbar & EW+\ttbar  \\
& $\PH^+\PH^-$ & $\PH^+\PW^-$ & $\PH^+$ 
& jets & with \tauh & no \tauh \\
\hline
$\Pgt$ part of trigger (data)    & 1.5--1.8 & 1.3--1.5 & 1.8--3.0 & $-$0.5      & 1.2      & 1.4 \\
$\Pgt$ part of trigger (simulation) & 0.7--0.8 & 0.6--0.7 & 0.8--1.1 & $-$0.2      &          & 0.8 \\
$\MET$ part of trigger (data)       & 2.6--3.3 & 2.5--2.8 & 2.9--4.2 & $-$1.2      & 2.5      & 2.8 \\
$\MET$ part of trigger (simulation) & 0.1      & 0.1      & 0.1      & $-$0.1      &          & 0.4 \\
Approximation in \MET part of trigger  &          &          &          &          & 12       & \\
Single $\mu$ trigger; data      &          &          &          &          &  $-$0.1     & \\
Veto of events with $\Pe$       & 0.1--0.2 & 0.2--0.3 & 0.2--0.3 & ${}< $$-$0.1 &          & 0.4 \\
Veto of events with $\Pgm$      & 0.1      & 0.1--0.2 & 0.1      & ${}< $$-$0.1 &          & 0.5 \\
\tauh identification (S)        & 6.0      & 6.0    & 5.9--6.0 & $-$0.8    &  6.0     & \\
$\Pe$ misidentification as \tauh (S)        & ${}< $0.1 & ${}< $0.1 & ${}< $0.1 & $-$0.1    &          & 3.3 \\
$\Pgm$ misidentification as \tauh (S)       & ${}< $0.1 & ${}< $0.1 & ${}< $0.1 & ${}< $$-$0.1 &          & 1.1 \\
Jet misidentification as \tauh (S)        & 0.1      & 0.1--0.3 & 0.1      & $-$6.9    &          & 17  \\
\tauh energy scale (S)          & 0.3--2.6 & 2.7--5.2 & 0.3--2.7 & $-$1.8      & 5.8      & 2.0 \\
Jet energy scale                & 2.6--5.2 & 2.0--3.0 & 1.6--2.1 & $-$1.4      &          & 3.2 \\
Jet energy resolution           & 1.1--1.8 & 0.5--1.3 & 0.7--1.5 & $-$0.2      &          & 3.2 \\
Unclustered \MET energy scale   & 0.1--0.4 & 0.1--0.9 & 0.1--0.4 & $-$0.5      &          & 1.5 \\
b-jet tagging (S)               & 5.9--20  & 4.7--5.3 & 4.6--5.4 & $-$3.5      &          & 5.0 \\
Top quark \pt modelling (S)     &          &          &          & $^{+5.6}_{-6.8}$ &    & $^{+11}_{-6.6}$ \\
Pileup modelling                & 0.1--0.9 & 0.1--0.8 & 0.1--0.6 & $-$0.1      &          & 2.9 \\
$\Pgm$ identification; data     &          &          &          &          & ${}< $$-$0.1 & \\
Multijet contamination          &          &          &          &          & 2.0      &     \\
$\PW \to \Pgt\Pgngt \to \Pgm\Pgngm\Pgngt$ fraction & &   &          &          & 1.2      &     \\
Non-emb. vs. emb. difference (S)    &          &          &          &          & $^{+14}_{-12}$ & \\
Multijet \mT distribution shape (S) &      &          &          & 4.6      &          &     \\
Multijet template fit           &          &          &          & 3.0      &          &     \\
Probabilistic \mT in single top quark &          &          &          &          &          & 6.8 \\
\ttbar cross section, scale     & $^{+2.5}_{-3.4}$ & $^{+2.5}_{-3.4}$ & & $^{+1.0}_{-0.7}$ & & $^{+2.2}_{-2.9}$ \\
\ttbar cross section, PDF+$\alpha_\mathrm{S}$ & 4.6 & 4.6 &      & $-$1.6     &          & 4.0 \\
Single top quark cross section        &          &          &          &          &          & 1.0 \\
$\PW$+jets, $\cPZ/\gamma^*$, VV cross section    &          &          &          &          &          & 0.1 \\
Integrated luminosity                      & 2.6      & 2.6      & 2.6      & $-$0.8      &          & 2.6 \\
\hline
\end{tabular}  
}
\label{tab:fullyhadronic_systematics_tab1}
\end{center}
\end{table*}

In the region where the background yields are taken from the exponential fit on \mT,
the statistical uncertainties in the background distributions are given by the uncertainties on the fit parameters while
the relative values of the systematic uncertainties are kept the same like in the unfitted \mT distribution.

The dominant systematic uncertainties for signal arise from \tauh identification, \tauh energy scale, b tagging,
and the theoretical \ttbar cross section uncertainty for $\mHp < \mt-\mb$.
For the backgrounds, the dominant uncertainties are those in \tauh identification, jet$\to\tauh$ misidentification, treatment of the \MET part of the trigger,
and the difference between the transverse mass shapes of the $\Pgt$+jets and embedded $\Pgm$+jets events.
In the region $\mHp > 300\GeV$ the sensitivity of the analysis
is driven solely by the signal acceptance and the uncertainties in the signal.

\subsection{The \texorpdfstring{\mutauh}{tau+mu} final state for \texorpdfstring{$\Hptaunu$ and $\Hptb$}{H+ to taunu and H+ to tb}}
\label{sec:taunutb:taumu:uncertainties}

The dominant sources of systematic uncertainties 
are the $\tauh$ identification and misidentification, the top quark $p_{T}$ modelling, and the prediction of the \ttbar cross section.
In addition to the uncertainties described in \refsec{common:uncertainties}, 
an uncertainty associated with the misidentified \tauh background estimated from data is evaluated 
as half of the maximum variation between the ``W+jet'' and ``multijet'' estimates discussed in Section~\ref{sec:taunutb:taumu:backgrounds}. 
The statistical uncertainty associated with the number of events in the control region to which the final estimate is applied amounts to 1\% 
and is taken into account in the limit computation.

The systematic uncertainties for the signal and background samples are summarized in Table~\ref{tab:SummarySystematicsTauHadHPS}. 
The diboson and Drell--Yan background yields are small compared to the uncertainty on the \ttbar background, and consequently are not used in the limit computation. 
Results are not sensitive to the inclusion of those backgrounds.

\begin{table}[htp]
\begin{center}
\topcaption{
The systematic uncertainties (in \%) for the \mutauh final state for backgrounds, and for
signal events from \texorpdfstring{$\Hptb$}{H+ to tb} decays for $\mHp=250\GeV$. 
These systematic uncertainties are given as the input to the exclusion limit calculation.
The uncertainties that depend on the b-tagged jets multiplicity distribution bin are marked with (S)
and for these the maximum integrated value of the negative or positive variation is displayed.
Empty cells indicate that an uncertainty does not affect the sample.
The uncertainty values within the rows are considered to be fully correlated and
the values within the columns are considered to be uncorrelated.
The uncertainties in the cross sections are to be considered uncorrelated for different samples and fully correlated for different final states of the same sample (e.g. the different \ttbar decays).
}
\footnotesize{
\setlength{\extrarowheight}{1.5pt}
\hspace*{-1cm}
\newcolumntype{.}{D{.}{.}{-1}}
\begin{tabular}{l . . . c c }
  \hline
& \multicolumn{1}{c}{Signal} & \multicolumn{1}{c}{$\ttbar\to\Pgm\tauh+{\rm X}$} & 
\multicolumn{1}{c}{$\ttbar\ \text{dilepton}$} & \multicolumn{1}{c}{$\tauh\ \text{mis-id}$} & single top quark  \\

  \hline

  Single $\Pgm$ trigger 		& 2.0 & 2.0 & 2.0 &   \\
  $\Pe$ identification		& 2.0 & 2.0 & 2.0 &  & 2.0  \\
  $\Pgm$ identification		& 1.0 & 1.0 & 1.0 &  & 1.0  \\
  \tauh identification	&  6.0  &  6.0    &       &     &  6.0  \\
  $\Pe$ misidentification as \tauh	&     &       & 3.0    &     &       \\
  $\Pgm$ misidentification as \tauh	&     &       & 3.0    &     &       \\
  Jet  misidentification as \tauh	&     &       & \multicolumn{1}{c}{$20$}    &      &      \\
  \tauh energy scale (S)             & 0.6 & 2.4 & 4.4 &  & 4.1 \\
  Jet energy scale (S)               & 2.5 & 1.9 & 2.6 &  & 3.9 \\
  Jet energy resolution (S)          & 0.8 & 0.1 & 1.6 &  & 0.2 \\
  Unclustered \MET energy scale (S)  & 0.8 & 0.1 & 1.8 &  & 0.2 \\
  b tagging (S)			&  1.8  &  1.8      &  2.7     &     &  3.2     \\
  udsg$\to$b mistagging (S)    &         {<}0.1  &   {<} 0.1     &   {<} 0.1    &     &  0.1    \\
  Top quark \pt modelling (S)		&     & 5.4 & 5.2 &     &        \\
  Pileup modelling 	&  4.0  &  2.0    &  8.0    &     &  2.0    \\
  Misidentified $\tauh$ background &     &       &       &  \multicolumn{1}{c}{$11$} &       \\
  Cross sections	        &     &  \multicolumn{1}{c}{$^{+2.5}_{-3.4}\pm4.6$} &   
  \multicolumn{1}{c}{$^{+2.5}_{-3.4}\pm4.6$}  &     &  8.0     \\
  Matching scale (S) 	&     & \multicolumn{1}{c}{$12$}     & 5.1     &     &         \\
  Fact./renorm. scale (S)       	&     & 3.4     & 7.5     &     &        \\
  PDF effect on shape	&     & \multicolumn{1}{c}{shape only}     & \multicolumn{1}{c}{shape only}     &     &       \\
  Heavy flavours (S) 	&     & {<}0.1 & {<}0.1   &     &        \\

  Integrated luminosity			&  2.6  &  2.6    &    2.6   &    &  2.6    \\
\hline
\end{tabular}
}
\label{tab:SummarySystematicsTauHadHPS}
\end{center}
\end{table}

\subsection{Dilepton (\texorpdfstring{$\Pe\Pe/\Pe\Pgm/\Pgm\Pgm$}{ee/emu/mumu}) final states for \texorpdfstring{$\Hptaunu$ and $\Hptb$}{H+ to taunu and H+ to tb}}
\label{sec:taunutb:dilepton:uncertainties}

The main sources of systematic uncertainties are the unclustered \MET scale, the b tagging efficiency, and the prediction of the \ttbar cross section.

The systematic uncertainties for signal and background events are summarized in Table~\ref{tab:SummarySystematicsDilepton}.
The diboson, \DY, ``other \ttbar'', and W+jets backgrounds yields are small compared to the uncertainty on the \ttbar background, 
and consequently are not used in the limit computation. Results are not sensitive to the inclusion of those backgrounds.

\begin{table}[htp]
\begin{center}
\topcaption{
The systematic uncertainties (in \%) for backgrounds, and for signal events from \texorpdfstring{$\Hptb$}{H+ to tb} decays for the dilepton channels
for a charged Higgs boson mass $\mHp=250\GeV$. The $\Pe\Pgm$ final state is shown as a representative example.
These systematic uncertainties are given as the input to the exclusion limit calculation.
The uncertainties that depend on the b-tagged jets multiplicity distribution bin are marked with (S)
and for these the maximum integrated value of the negative or positive variation is displayed.
Empty cells indicate that an uncertainty does not affect the sample.
The uncertainty values within the rows are considered to be fully correlated and
the values within the columns are considered to be uncorrelated.
The uncertainties in the cross sections are to be considered uncorrelated for different samples and fully correlated for different final states of the same sample (e.g. the different \ttbar decay channels).
}
\small{
\hspace*{-1cm}
\setlength{\extrarowheight}{1.5pt} 
\newcolumntype{.}{D{.}{.}{-1}}
\begin{tabular}{l . . . . }
\hline
  & \mathrm{Signal} & \multicolumn{1}{c}{\ttbar~dilepton}  & \multicolumn{1}{c}{$\cPZ/\gamma^{*}\to\ell\ell$} &  \multicolumn{1}{c}{single top quark}\\
  \hline
    $\Pe\Pgm$ trigger         & 3.0 & 3.0  & 3.0  & 3.0  \\
    $\Pe$ identification            & 2.0 & 2.0  & 2.0  & 2.0  \\
    $\Pgm$ identification              & 1.0 & 1.0  & 1.0  & 1.0  \\
    Jet energy scale (S)               & 1.4 & 1.1 & 1.7   & 1.4  \\
    Jet energy resolution (S)          & 0.3 & 0.3 & 0.4   & 0.4  \\
    Unclustered \MET energy scale (S)  & 1.3 & 2.1 & 11.7  & 2.6  \\
    b tagging (S)                  & 2.4 & 3.7  & \multicolumn{1}{c}{$10$}   & 4.3   \\
    udsg$\to$b mistagging (S) & 2.3 & 3.6  & \multicolumn{1}{c}{$10$}   & 4.4   \\
    Top quark \pt modelling (S)           &   & 3.8          & &  \\
    Pileup modelling                        & 0.6  &  0.4   & 1.2   & 1.2   \\
    Cross sections                     &    &  \multicolumn{1}{c}{$^{+2.5}_{-3.4}\pm4.6$ } &  4.0   & 8.0  \\
    Matching scale (S)                &    &  7.7       & &  \\
    Fact./renorm. scale (S)                   &    &  8.4       & &  \\
    PDF shape                  &    &  \multicolumn{1}{c}{shape only}  & &  \\
    Heavy flavours (S)                &    &  {<} 0.1    & &  \\

    Integrated luminosity            &   2.6  &  2.6 & 2.6  & 2.6  \\
\hline
\end{tabular}
}
\label{tab:SummarySystematicsDilepton}
\end{center}
\end{table}

\subsection{Single-lepton (\texorpdfstring{$\Pe/\Pgm$}{e/mu}+jets) final states for \texorpdfstring{$\Hptb$}{H+ to tb}}
\label{sec:tb:ljets:uncertainties}

In addition to the uncertainties described earlier in this section,
the following systematic uncertainties specific to the \ljets final states, affecting the simulated samples only, are as follows:
\begin{itemize}
\item
The normalizations for \ttbar, $\PW$+$\cPqc$, $\PW$+$\cPqb$, and $\PW$+light-flavour backgrounds are left unconstrained.
Statistical and systematic uncertainties are applied to yields in the control regions described in \refsec{reco}. These uncertainties are based on deviations of the
fitted normalization factor when varying multijet and $\cPZ/\gamma^*$+jets contributions by a factor of two, signal contamination by a factor of five, 
and by requiring either two or three jets in the control region. The total uncertainty in the normalization factors ranges between 5--35\%.
\item
A 50\% uncertainty~\cite{Aad:2011cx,Campbell:2011bn,Campbell:2012dh}
is applied to the $\cPZ/\gamma^*$+jets and diboson backgrounds due to their small contribution to the signal region;
\item
A 100\% systematic uncertainty is applied to the QCD cross section normalization. 
This accounts for the maximal variation in the QCD normalization when left unconstrained in the background-only fit to data while constraining normalizations for other backgrounds to their systematic uncertainties.
\end{itemize}

The systematic uncertainties for signal and background events are summarized in \reftab{SummarySystematicsLjets}. 

\normalsize
\begin{table}[h!tb]
	\begin{center}
\topcaption{
The systematic uncertainties (in \%) for backgrounds, and for signal events from \texorpdfstring{$\Hptb$}{H+ to tb} decays for
the $\ell$+jets channels for a charged Higgs boson mass $\mHp=250\GeV$.
The uncertainties that depend on the shape of the \HT distribution bin are marked with (S)
and for these the maximum integrated value of the negative or positive variation is displayed.
Empty cells indicate that an uncertainty does not affect the sample.
The uncertainty values within the rows are considered to be fully correlated, with the exception of cross section and data-driven normalization, which are considered to be uncorrelated.
The uncertainty values within the columns are considered to be uncorrelated.
Uncertainties labelled with a "*" are only present in the CR with an implicit unconstrained parameter correlated across all bins (Sec. ~\ref{sec:tb:ljets:backgrounds}).
The values for these are assigned prior to the setting of limits. 
\label{tab:SummarySystematicsLjets}
}
\scriptsize{
\begin{tabular}{lcccccccc}
     \hline
													& \Hptb		& \ttbar		& $\PW$+$\cPqc$	& $\PW$+$\cPqb$	& $\PW$+u,d,s,g	& single top quark 	&  \DY/VV			&  Multijets \\
   	\hline
    Single-$\Pe$ trigger        								& 2.0   		& 2.0			& 2.0    			& 2.0    			& 2.0				& 2.0     			& 2.0     			& 2.0    \\
    Single-$\Pgm$ trigger         								& 2.0   		& 2.0			& 2.0    			& 2.0    			& 2.0				& 2.0     			& 2.0     			& 2.0    \\
    $\Pe$ identification      									& 1.0   		& 1.0			& 1.0    			& 1.0    			& 1.0				& 1.0     			& 1.0     			& 1.0    \\
    $\Pgm$ identification       								& 1.0   		& 1.0			& 1.0    			& 1.0    			& 1.0				& 1.0     			& 1.0     			& 1.0    \\
    Jet energy scale (S)            								& 4.0   		& 6.4			& 15     			& 11				& 14				& 9.2     			& 27     			& 49    \\
    Jet energy resolution (S)             							& 0.1  		& 0.3			& 1.7    			& 2.3				& 1.4				& 0.8      			& 2.3     			& 6.9    \\
    b tagging (S)          										& 3.9  		& 1.3			& 14     			& 6.2				& 11				& 0.7      			& 5.4     			& 16    \\
    Top quark \pt modelling (S)          							&       		& 3.5			&        			&				&				&				&        			&       \\
    Pileup modelling (S)       								& 1.2  		& 0.7			& 2.3     			& 0.5				& 0.4				& 0.7      			& 3.7     			& 7.0    \\
    Normalization from data, \Pe+jets							&			& 5.5*		& 4.9*			& 25*			& 9.6*			&				&				& \\
    Normalization from data, $\mu$+jets						&			& 5.2*		& 10*			& 34*			& 10*			&				&				& \\
    Cross section     										&     			&			&      				&				&				& 8.0     			& 50       			& 100       \\
    Fact./renorm. scales (S)       								&      			& 7.3			&       			&				&				&				&        			&       \\    
    $Q^2$ scale (S)    										&      			& 7.6			&       			&				&				&				&        			&       \\
    Integrated Luminosity		 							& 2.6   		&			& 		    		&				&				& 2.6      			& 2.6      			& 2.6     \\
    \hline
   \end{tabular}
  }
 \end{center}
\end{table}

\normalsize

\section{Results}
\label{sec:results}
A statistical analysis of the \mT (Fig.~\ref{fig:taunu:hadr:mt}), b-tagged jet multiplicity (Fig.~\ref{fig:leptonic_ltau_figures} (right) and Fig.~\ref{fig:emu_figures} (right)), 
and \HT (Fig.~\ref{fig:ljets_HT_scaled}) distributions
has been performed using a binned maximum likelihood fit.
The data agree with the SM prediction and consequently
95\% CL upper limits on charged Higgs boson production are derived using the modified
frequentist \CLs criterion~\cite{Junk:1999kv,CLs} with a test
statistic based on the profile likelihood ratio with asymptotic approximation~\cite{Cowan:2010js,cms:Note-2011-005}.

The systematic uncertainties described in \refsec{uncertainties} are incorporated via nuisance
parameters following the frequentist paradigm.
Correlations between the different sources of systematic uncertainty are taken into account.
Uncertainties affecting the shape of the \mT, b-tagged jet multiplicity, or \HT distributions are represented by nuisance
parameters whose variation results in a continuous perturbation of the distribution~\cite{Conway-PhyStat}. 

\subsection{Model-independent limits on charged Higgs boson production (\texorpdfstring{\Hptaunu}{H+ -> taunu})}
\label{sec:results:fully_model_independent}
In the analysis of the \Hptaunu decay mode with the \tauhjets final state
no assumption on the charged Higgs boson branching fractions is needed
because subtracting the background from ``EW+\ttbar with \tauh'' will remove any potential \Hptb and other such signals from data due to the embedding technique described in \refsec{taunu:hadr:backgrounds:ewkgenuine}.
For $\mHp = 80$--$160\GeV$, the charged Higgs boson is produced most copiously through \ttbar production
which can produce one ($\ttbar\to\cPqb\PH^+\cPaqb\PW^-$) or two charged Higgs bosons ($\ttbar\to\cPqb\PH^+\cPaqb\PH^-$) if $\BtbHp > 0$. 
Furthermore, the presence of the charged Higgs boson suppresses the $\ttbar\to\cPqb\PW^+\cPaqb\PW^-$ yield compared to the SM prediction. 
Consequently, the number of events in a given bin of the $\mT$ distribution depends on the signal strength parameter $\mu$ according to:

\begin{equation}
  \label{eq:results:nlight}
  N(\mu) = \mu^2 \, s(\PH^+\PH^-) + 2\mu(1-\mu) \, s(\PH^+\PW^-) + (1-\mu)^2 \, b(\PW^+\PW^-) + b,
\end{equation}

where $\mu = \lightLimitTaunuHadr$, 
$s(\PH^+\PH^-)$ and $s(\PH^+\PW^-)$ are the number of expected signal events for the
$\ttbar\to\cPqb\PH^+\cPaqb\PH^-$ and $\ttbar\to\cPqb\PH^+\cPaqb\PW^-$
processes, respectively; 
$b(\PW^+\PW^-)$ is the expected number of events from the portion of $\ttbar\to\cPqb\PW^+\cPaqb\PW^-$ background that
is estimated with simulation, and $b$ is the expected number of other background events.
The number of signal and $\ttbar\to\cPqb\PW^+\cPaqb\PW^-$ background events is normalized
to the SM predicted cross section
and by setting $\lightLimitTaunuHadr = 1$ for a top quark decaying to a charged Higgs boson.

For $\mHp = 180$--$600\GeV$, 
the number of events in a given bin of the $\mT$ distribution depends on the signal strength parameter 
according to:

\begin{equation}
  \label{eq:results:nheavy}
  N(\mu) = \mu\, \varepsilon_\mathrm{s}\mathcal{L} + b,
\end{equation}

where $\mu = \heavyLimitTaunuHadr$, 
$\varepsilon_\mathrm{s}$ is the event selection efficiency for signal events, 
$\mathcal{L}$ is the integrated luminosity,
and $b$ is the expected number of background events.

\begin{figure*}[htbp]
\begin{center}
{\includegraphics[width=0.40\textwidth]{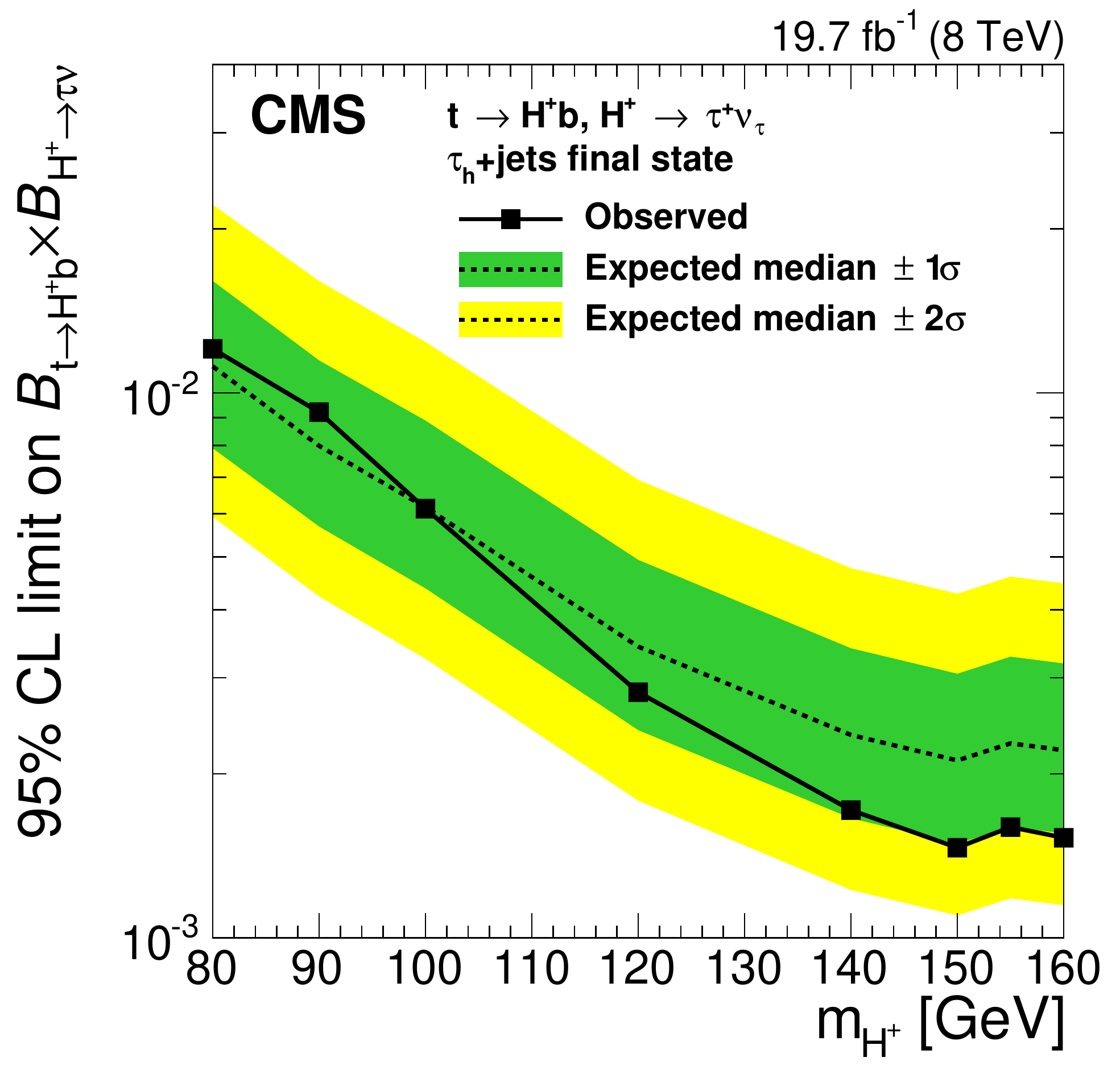}}
\hspace{0.02\textwidth}
{\includegraphics[width=0.40\textwidth]{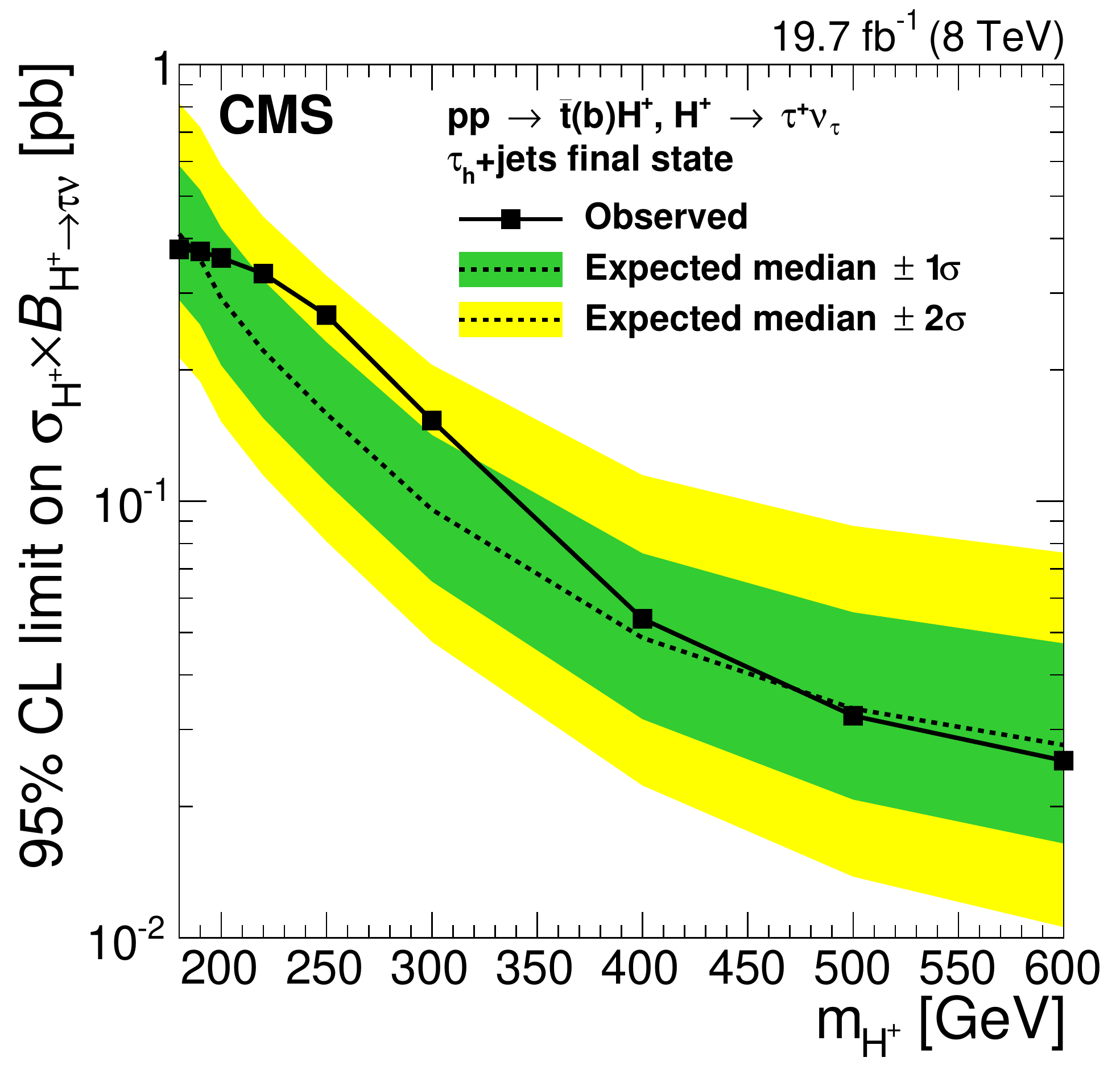}} \\
\caption{Expected and observed 95\% CL model-independent upper limits on \lightLimitTaunuHadr with $\mHp = 80$--$160\GeV$ (left),
and on \heavyLimitTaunuHadr with $\mHp = 180$--$600\GeV$ (right)
for the \Hptaunu search in the \tauhjets final state.
The regions above the solid lines are excluded.}
\label{fig:taunu:hadr:brlimit}
\end{center}
\end{figure*}

The upper limits on \lightLimitTaunuHadr  and on \heavyLimitTaunuHadr
are shown in \reffig{taunu:hadr:brlimit} for the \Hptaunu decay mode with the \tauhjets final state
for the ranges $\mHp = 80$--$160\GeV$ and $\mHp = 180$--$600\GeV$, respectively.
The numerical values of the limits are given in Table~\ref{tab:taunu:hadr:brlimit}. 
At $\mHp = 250\GeV$ an excess of data is observed with a local p-value of 0.046 corresponding to significance of $1.7 \sigma$.

\begin{table*}[htbp]
\begin{center}
\topcaption{Expected and observed 95\% CL model-independent upper limits on 
\lightLimitTaunuHadr for $\mHp = 80$--$160\GeV$ (top), and on \heavyLimitTaunuHadr for $\mHp = 180$--$600\GeV$ (bottom),
for the \Hptaunu search in the \tauhjets final state.}
\begin{tabular}{ c c c c c c c } 
\hline
\mHp & \multicolumn{5}{ c }{Expected limit} & Observed \\
\protect [\GeV] & $-2\sigma$  & $-1\sigma$ & median & $+1\sigma$ & $+2\sigma$  & limit \\ 
\hline 
\\
\multicolumn{7}{c}{95\% CL upper limit on \lightLimitTaunuHadr}\\ 
\hline
 80 & 0.0059 & 0.0079 & 0.0112 & 0.0160 & 0.0221 & 0.0120 \\ 
 90 & 0.0042 & 0.0057 & 0.0080 & 0.0115 & 0.0160 & 0.0092 \\ 
100 & 0.0033 & 0.0044 & 0.0062 & 0.0089 & 0.0124 & 0.0061 \\ 
120 & 0.0018 & 0.0024 & 0.0034 & 0.0049 & 0.0069 & 0.0028 \\ 
140 & 0.0012 & 0.0017 & 0.0024 & 0.0034 & 0.0048 & 0.0017 \\ 
150 & 0.0011 & 0.0015 & 0.0021 & 0.0031 & 0.0043 & 0.0015 \\ 
155 & 0.0012 & 0.0016 & 0.0023 & 0.0033 & 0.0046 & 0.0016 \\ 
160 & 0.0011 & 0.0016 & 0.0022 & 0.0032 & 0.0045 & 0.0015 \\ 
\hline
\\
\multicolumn{7}{ c }{95\% CL upper limit on \heavyLimitTaunuHadr [pb]}\\ 
\hline 
180 & 0.213 & 0.289 & 0.409 & 0.587 & 0.816 & 0.377 \\ 
190 & 0.188 & 0.254 & 0.358 & 0.516 & 0.719 & 0.373 \\ 
200 & 0.152 & 0.205 & 0.291 & 0.423 & 0.587 & 0.361 \\ 
220 & 0.114 & 0.155 & 0.221 & 0.321 & 0.448 & 0.332 \\ 
250 & 0.081 & 0.110 & 0.159 & 0.231 & 0.328 & 0.267 \\ 
300 & 0.048 & 0.065 & 0.096 & 0.142 & 0.205 & 0.153 \\ 
400 & 0.022 & 0.032 & 0.049 & 0.076 & 0.115 & 0.054 \\ 
500 & 0.014 & 0.021 & 0.033 & 0.056 & 0.088 & 0.032 \\ 
600 & 0.011 & 0.016 & 0.028 & 0.047 & 0.076 & 0.025 \\
\hline
\end{tabular} 
\label{tab:taunu:hadr:brlimit}
\end{center}
\end{table*}

\subsection{Limits on charged Higgs boson production with branching fraction assumed}
\label{sec:results:partially_model_independent}
In the presence of a charged Higgs boson
and for $\mHp = 180$--$600\GeV$, 
the analyses of the \mutauh, \ljets, and $\ell\ell'$ final states 
have sensitivity to both \Hptaunu and \Hptb decays.
Consequently, a model-independent limit can neither be provided for 
$\SigmaHpFourandFiveFS\,\BHptaunu$ 
nor for 
$\SigmaHpFourandFiveFS\,\BHptb$.
Nevertheless, one can test models by fixing $\BHptaunu$ and $\BHptb$.
In this section, results are reported for a model with $\BHptb = 1$, to which the \tauhjets analysis is blind because of the estimates of the backgrounds 
from data like described in \refsec{taunu:hadr:backgrounds:ewkgenuine}.
For $\BHptaunu = 1$, the sensitivity of the \mutauh and $\ell\ell'$ final states analyses is found to be substantially weaker than that obtained in the \tauhjets analysis.

\begin{figure*}[htbp]
\begin{center}
{\includegraphics[width=0.40\textwidth]{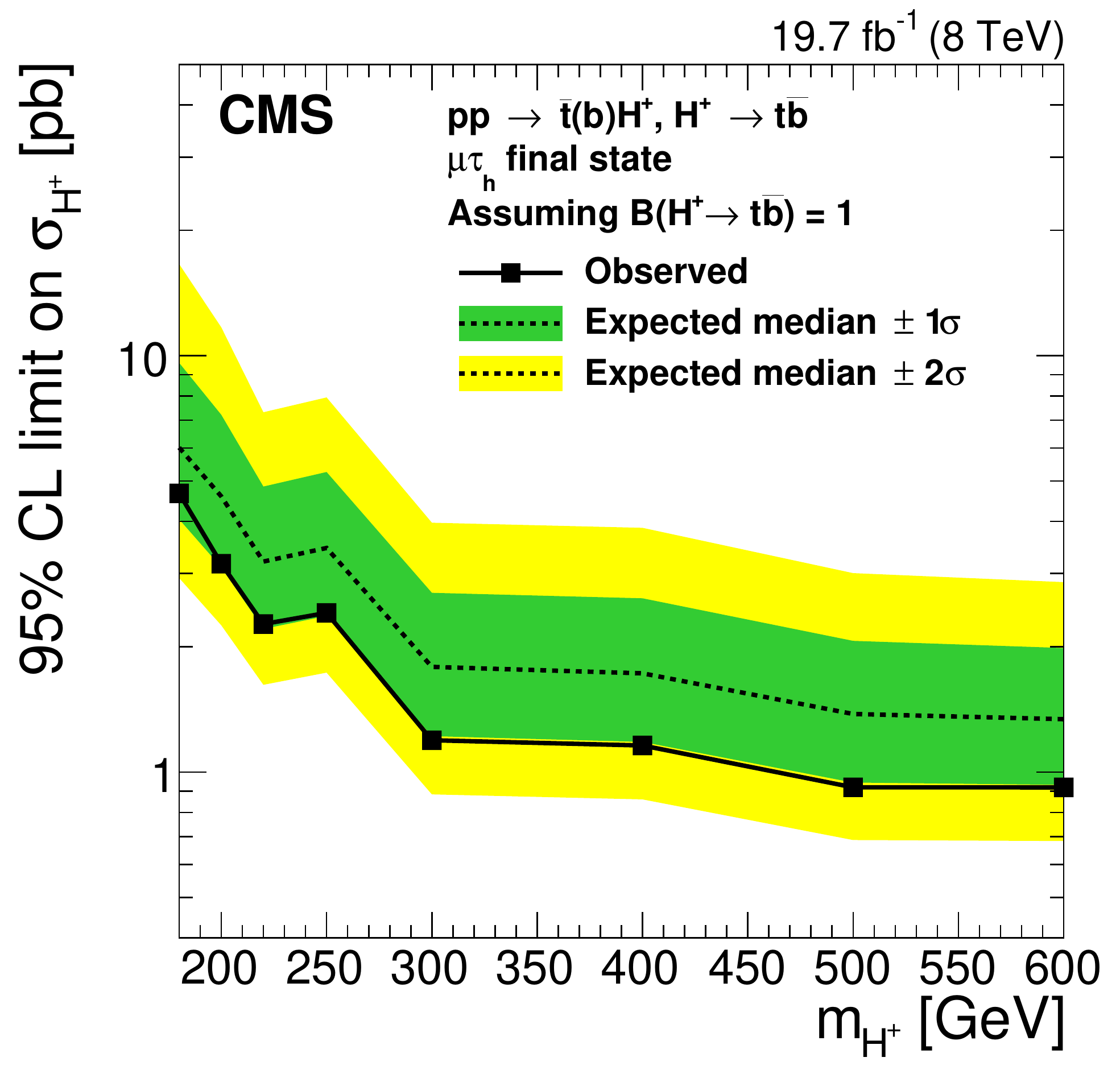}}
\hspace{0.02\textwidth}
{\includegraphics[width=0.40\textwidth]{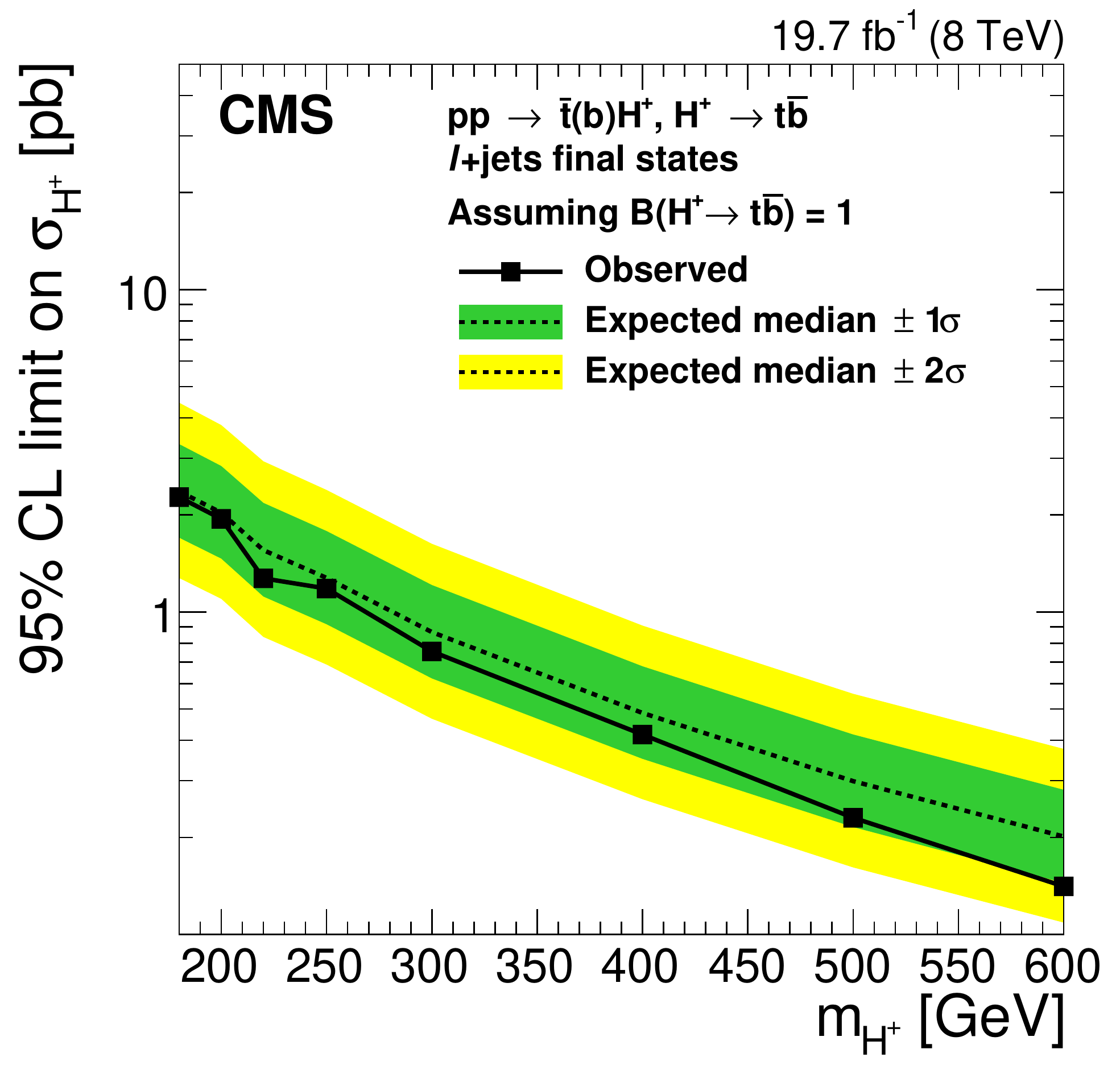}} \\    
{\includegraphics[width=0.40\textwidth]{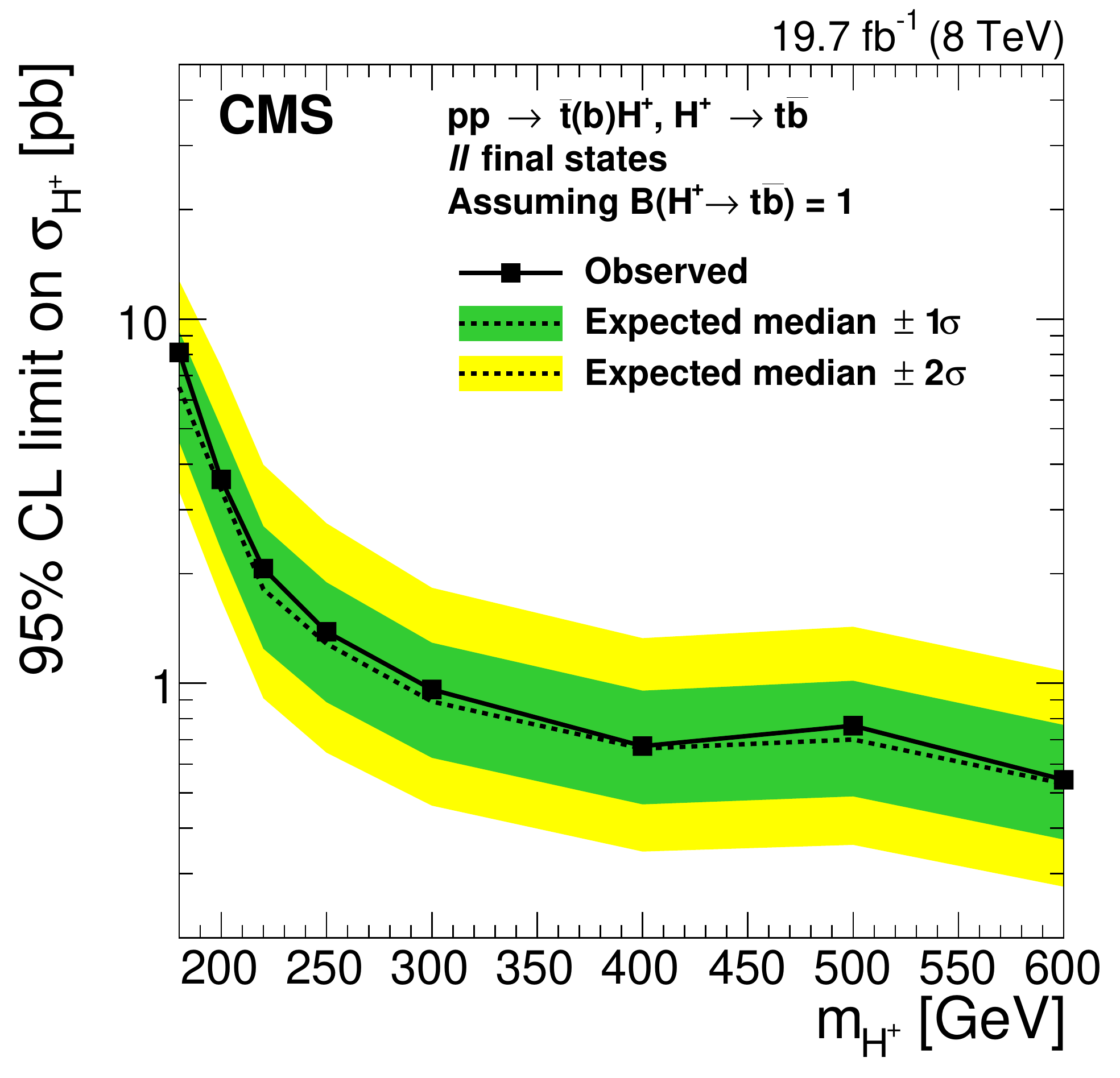}} \\
\caption{Expected and observed 95\% CL upper limits on \SigmaHpFourandFiveFS for the \mutauh (upper left), \ljets (upper right), and $\ell\ell'$ final states (bottom)
assuming $\BHptb = 1$.
The regions above the solid lines are excluded.}
\label{fig:sigmabrlimit_taumudileptons}
\end{center}
\end{figure*}

Equation~(\ref{eq:results:nheavy}) is used to derive the limits by counting the number of events in bins of the b-tagged jet multiplicity 
distribution for the \mutauh and $\ell\ell'$ final states, and in bins of the \HT distribution for the \ljets final state.
The upper limits on \SigmaHpFourandFiveFS assuming $\BHptb = 1$ are shown in \reffig{sigmabrlimit_taumudileptons} for the \mutauh (top left), \ljets (top right), and $\ell\ell'$ (bottom) final states.

\begin{figure*}[htbp]
\begin{center}
\includegraphics[width=0.40\textwidth]{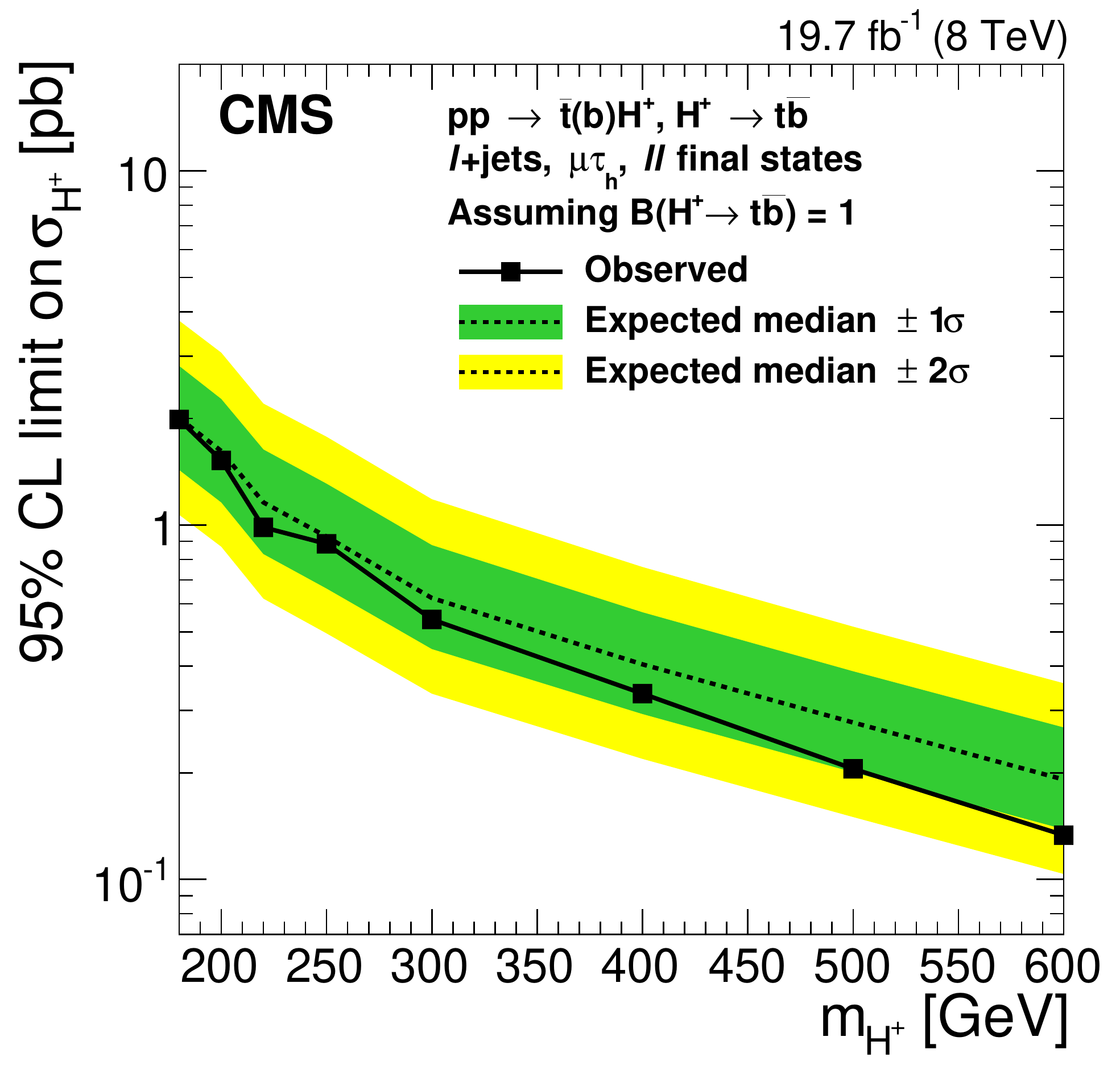} \\
\caption{Expected and observed 95\% CL upper limits on \SigmaHpFourandFiveFS for
the combination of the \mutauh, \ljets, and $\ell\ell'$ final states assuming $\BHptb = 1$.
The region above the solid line is excluded.}
\label{fig:sigmabrlimit}
\end{center}
\end{figure*}

The upper limit on \SigmaHpFourandFiveFS 
for the combination of the \mutauh, \ljets, and $\ell\ell'$ final states
is shown in \reffig{sigmabrlimit}. The numerical values are reported in \reftab{combined:brlimit}.
In the combination, the sensitivity is driven by the \ljets final state.

\begin{table*}[htbp]
\begin{center}
\topcaption{Expected and observed 95\% CL upper limits on 
\heavyLimitTb assuming $\BHptb = 1$ for the combination of the \mutauh, \ljets, and $\ell\ell'$ final states.}
\begin{tabular}{ c c c c c c c } 
\hline
\mHp & \multicolumn{5}{ c }{Expected limit [pb]} & Observed limit [pb] \\
\cline{2-6} 
\protect [\GeV]   & $-2\sigma$  & $-1\sigma$ & median & $+1\sigma$ & $+2\sigma$  & limit \\ 
\hline 
\multicolumn{7}{ c }{95\% CL upper limit on \SigmaHpFourandFiveFS with $\BHptb = 1$}\\
\hline 
180 & 1.07 & 1.43 & 2.01 & 2.81 & 3.78 & 1.99 \\
200 & 0.87 & 1.16 & 1.62 & 2.27 & 3.07 & 1.52 \\
220 & 0.62 & 0.83 & 1.16 & 1.64 & 2.20 & 0.99 \\
250 & 0.49 & 0.66 & 0.93 & 1.31 & 1.78 & 0.89 \\
300 & 0.33 & 0.45 & 0.62 & 0.88 & 1.18 & 0.54 \\
400 & 0.22 & 0.29 & 0.40 & 0.57 & 0.76 & 0.33 \\
500 & 0.15 & 0.20 & 0.28 & 0.39 & 0.52 & 0.21 \\
600 & 0.10 & 0.14 & 0.19 & 0.27 & 0.36 & 0.13 \\
\hline
\end{tabular} 
\label{tab:combined:brlimit}
\end{center}
\end{table*}

\subsection{Combined limits on \texorpdfstring{\tanbeta}{tan beta} in MSSM benchmark scenarios}
\label{sec:results:tanbeta}

Using all decay modes and final states, exclusion regions
have been set in the \mHp--\tanbeta plane
according to the LHC Higgs cross section working group prescription
for different MSSM benchmark scenarios~\cite{Heinemeyer:2013tqa,Carena:2013}:
``updated \mhmax'', ``$m_\unit{h}^\unit{mod+}$'', ``$m_\unit{h}^\unit{mod-}$'', ``light stop'', ``light stau'', ``tau-phobic'', and ``low-$M_{\PH}$'' scenarios.
These MSSM benchmark scenarios are compatible with the properties of the recently discovered neutral scalar boson and with the
current bounds on supersymmetric particle masses, and they are specified using low-energy MSSM parameters, \ie no particular soft SUSY-breaking scenario is assumed. The updated \mhmax scenario and $m_\unit{h}^\unit{mod}$ scenarios 
allow the discovered scalar boson to be interpreted as the
light CP-even Higgs boson in large parts of the \mHp--\tanbeta plane. 
The light stop scenario leads to a suppressed rate for the Higgs boson production by gluon fusion,
and the light stau scenario enhances the decay rate of the light CP-even Higgs boson to photons. A tau-phobic scenario has suppressed couplings to down-type fermions. 
In the low-$M_{\PH}$ scenario, the discovered scalar boson is assumed to be the heavy CP-even Higgs boson and \mA is fixed to be 110\GeV causing \mHp to be 132\GeV.

\begin{figure*}[h!]
\begin{center}
{\includegraphics[width=0.40\textwidth]{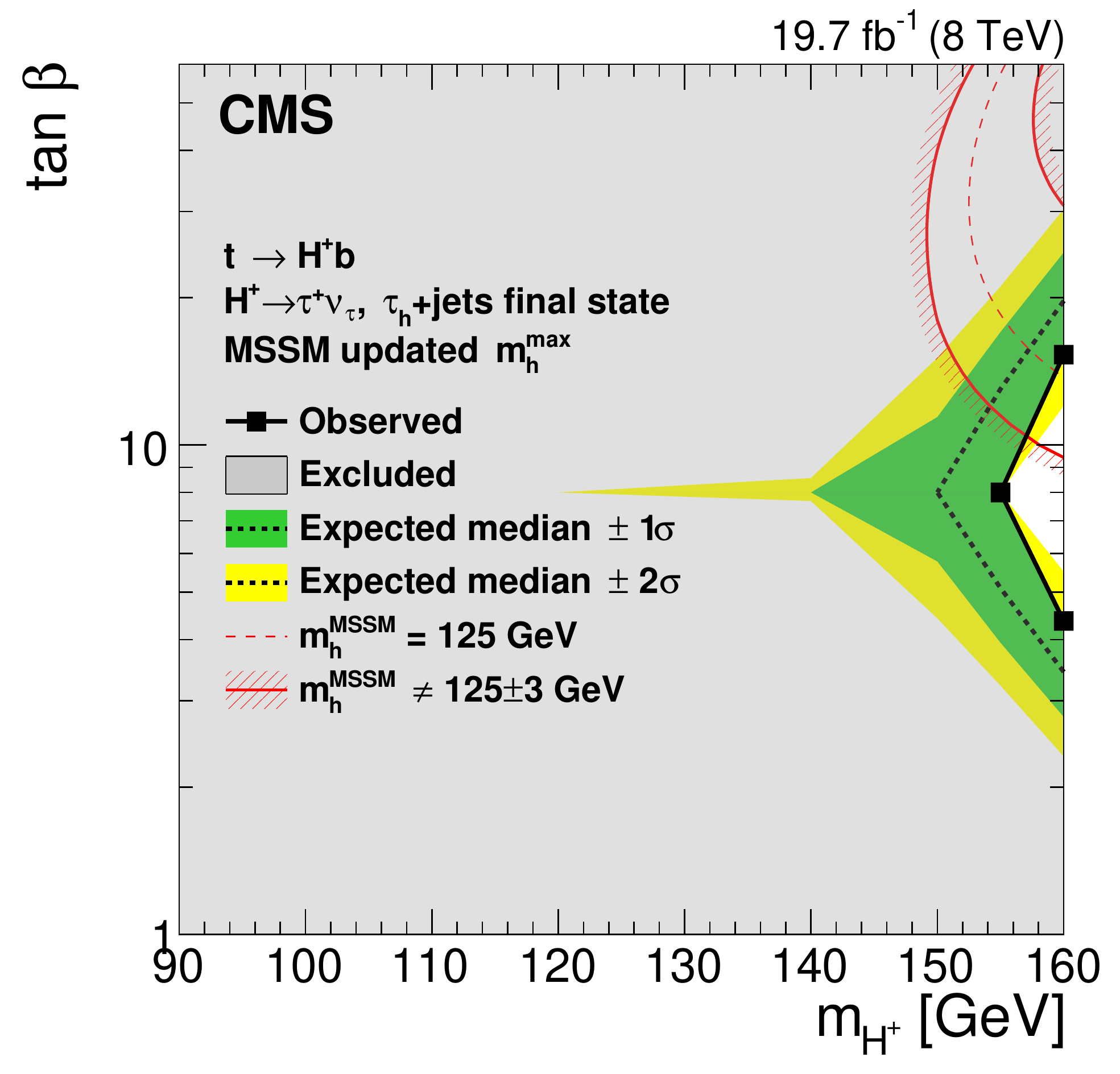}}
\hspace{0.02\textwidth}
{\includegraphics[width=0.40\textwidth]{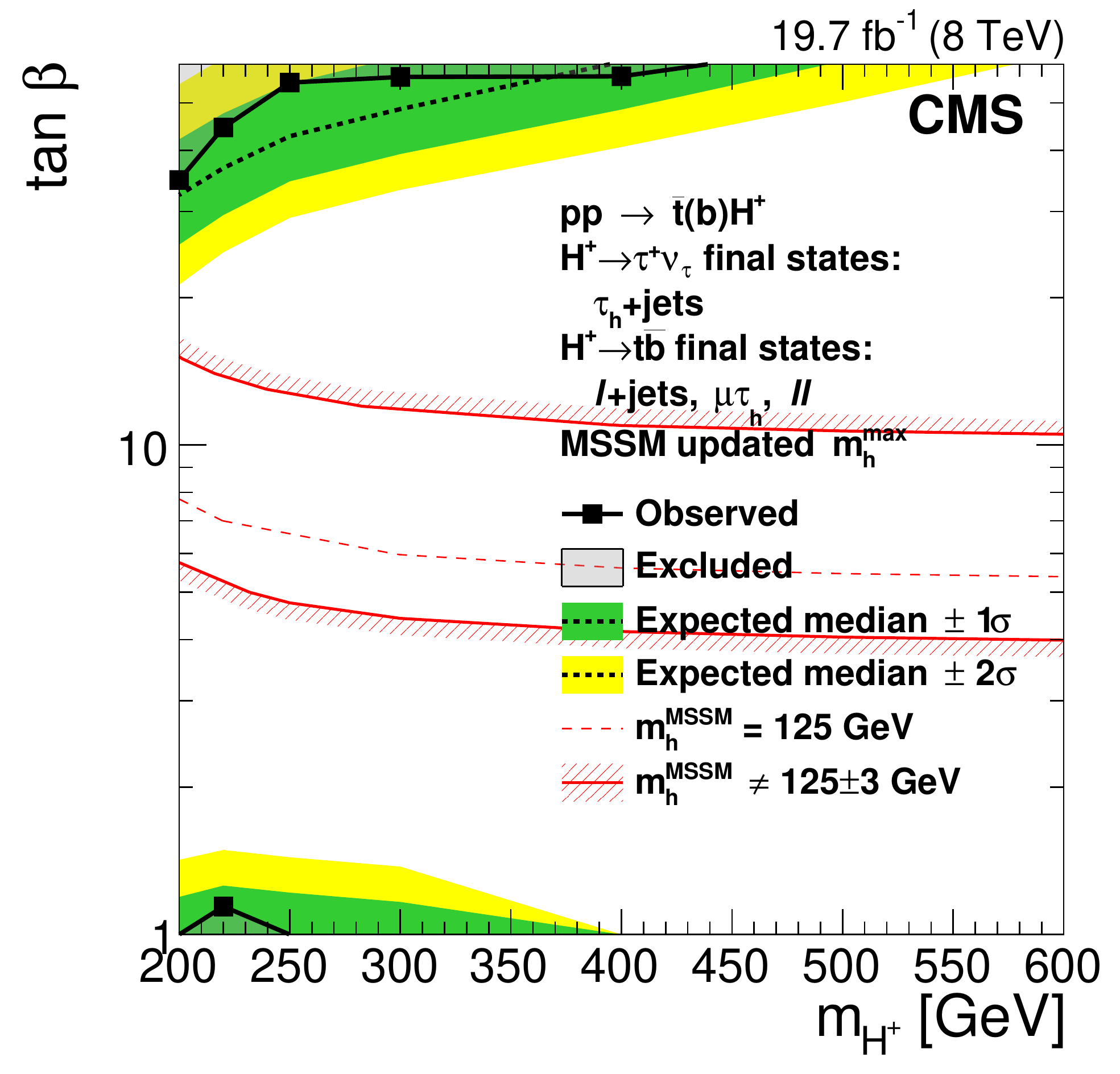}} \\
{\includegraphics[width=0.40\textwidth]{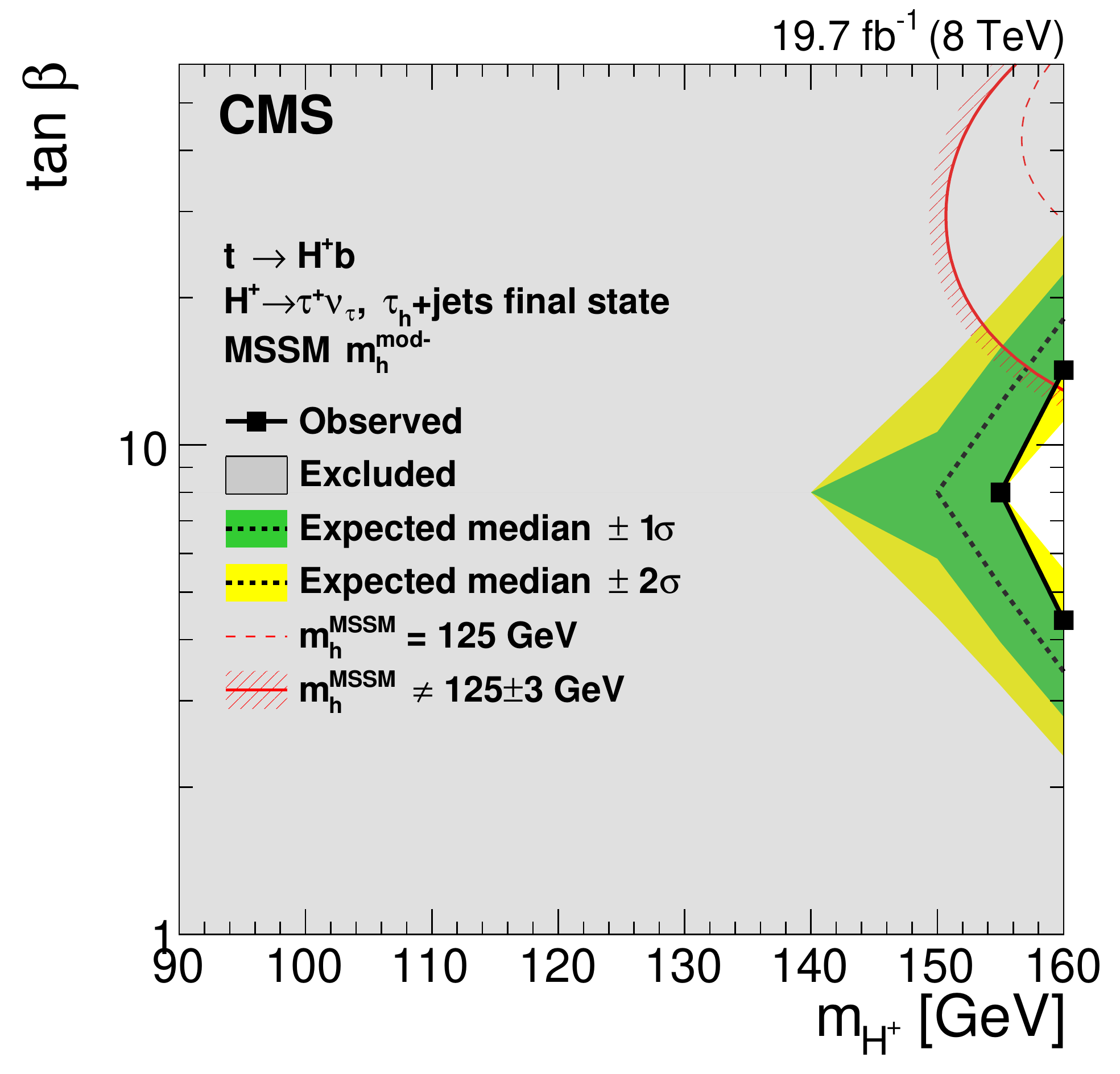}}
\hspace{0.02\textwidth}
{\includegraphics[width=0.40\textwidth]{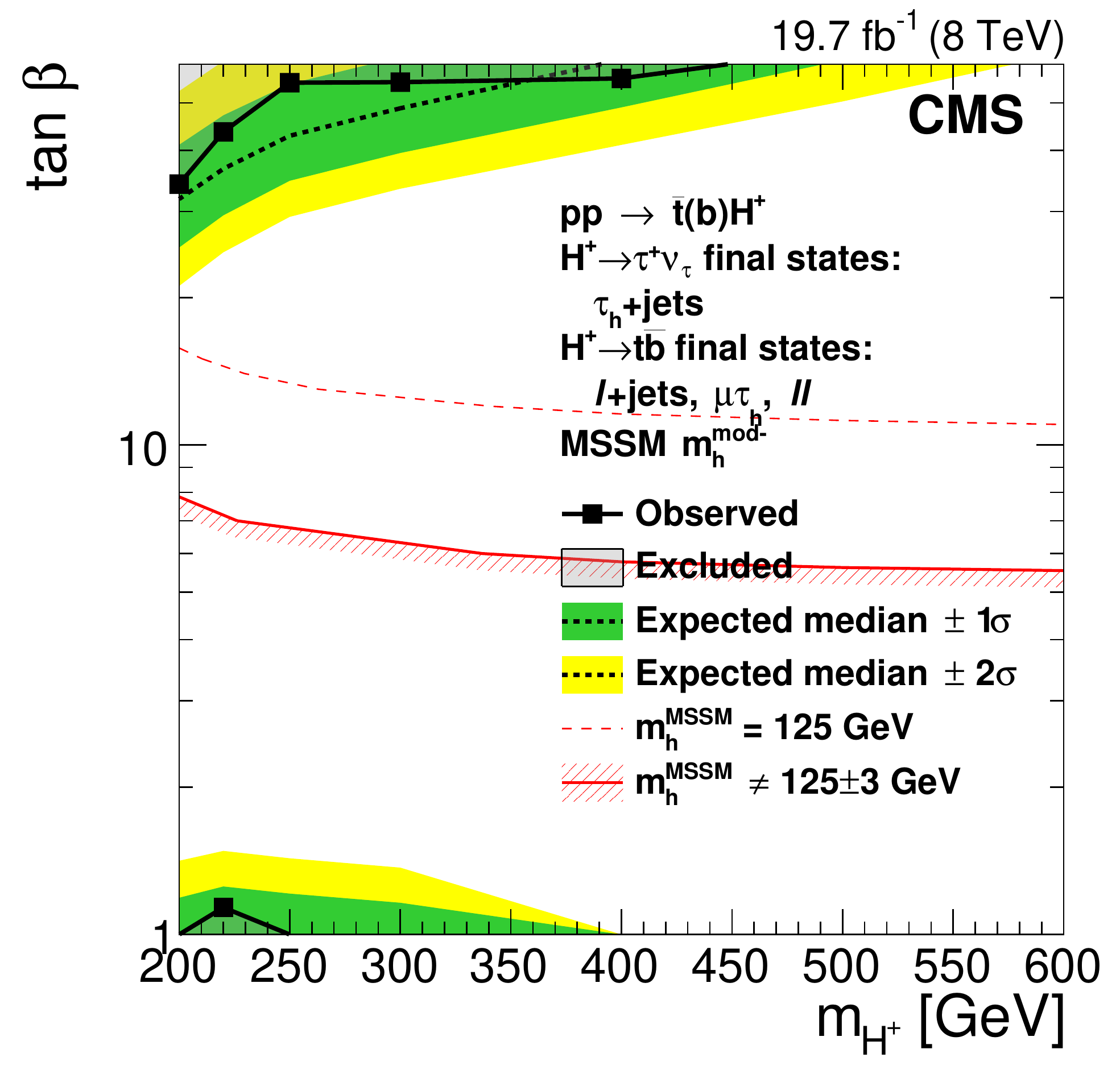}}
\caption{Exclusion region in the MSSM \mHp--\tanbeta parameter space 
         for $\mHp = 80$--$160\GeV$ (left column) and for $\mHp = 180$--$600\GeV$ (right column)
         in the
         updated MSSM \mhmax scenario (top row) and 
         \mhmodm
         scenarios~\cite{Carena:2013,Heinemeyer:2013tqa} (bottom row).
	 In the upper row plots the limit is derived from the \Hptaunu search with the \tauhjets final state, and in the lower row plots
	 the limit is derived from a combination of all the charged Higgs boson decay modes and final states considered.
         The $\pm 1 \sigma$ and $\pm 2 \sigma$ bands around the expected limit are also shown.
         The light-grey region is excluded.
         The red lines depict the allowed parameter space for the assumption that the
         discovered scalar boson is the lightest CP-even MSSM Higgs boson with a mass $\mh=125\pm3\GeV$, 
         where the uncertainty is the theoretical uncertainty in the Higgs boson mass calculation.}
\label{fig:tanbetalimit:others}
\end{center}
\end{figure*}

\begin{figure*}[h!]
\begin{center}
\includegraphics[width=0.40\textwidth]{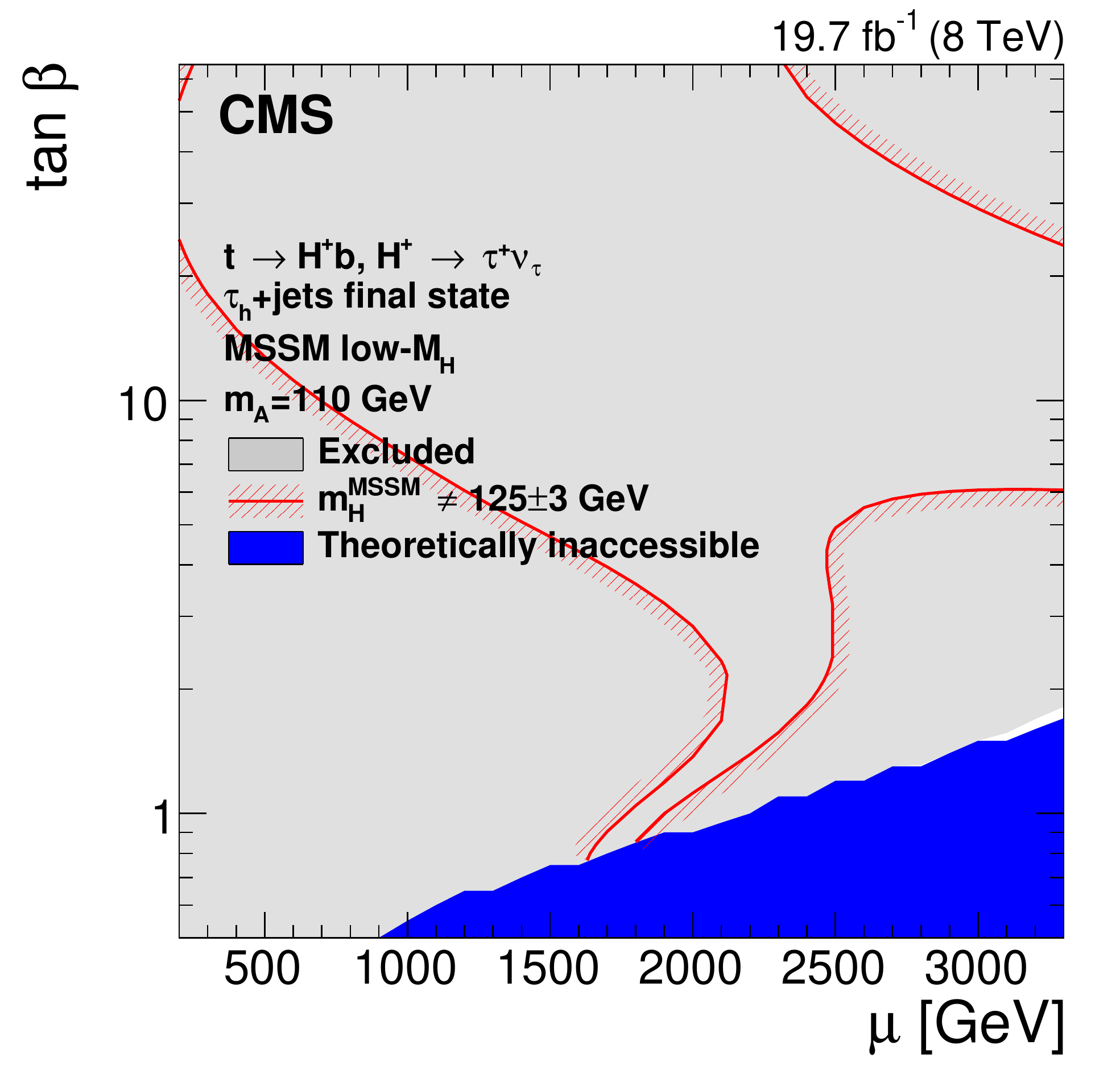}
\caption{Exclusion region in the MSSM Higgsino mass parameter (\Pgm) vs. \tanbeta parameter space 
         in the low-$M_\unit{H}$ scenario~\cite{Carena:2013,Heinemeyer:2013tqa} with $\mA = 110\GeV$
         for the \Hptaunu search with the \tauhjets final state.
         The light-grey region is excluded and the blue region is theoretically inaccessible.
         The area inside the red lines is the allowed parameter space for the assumption that the
         discovered scalar boson is the heavy CP-even MSSM Higgs boson with a mass $m_\PH=125\pm3\GeV$,
         where the uncertainty is the theoretical uncertainty in the Higgs boson mass calculation.}
\label{fig:tanbetalimit:lowmh}
\end{center}
\end{figure*}

Figure~\ref{fig:tanbetalimit:others} shows the limits on the updated \mhmax and $m_\unit{h}^\unit{mod-}$ scenarios.
For $\mHp = 90$--$160\GeV$, the analysis of the \Hptaunu decay mode with the \tauhjets final state described in
\refsec{taunu:hadr:analysis} is taken as input. 
The mass range starts here from $\mHp = 90\GeV$,
as the lower values of a charged Higgs boson mass are not accessible in the considered MSSM scenarios.
For $\mHp = 200$--$600\GeV$, a combination of all decay modes and final states 
is used to set the limits. In this combination, the signal yields from the \Hptaunu and \Hptb decay modes are defined
by the branching fractions predicted by the model. If the limit on the charged Higgs boson production for a given $\mHp$--$\tanbeta$ point
is smaller than the cross section predicted by the model~\cite{Flechl:2014wfa,Heinemeyer:2013tqa,Dittmaier:2009,Berger:2003sm}, the point is excluded. 
The mass range is chosen to start from $\mHp=200\GeV$ to avoid the interference region where a charged Higgs boson
is produced both from off-shell top quark decays and through direct production.
In all these scenarios except for the low-$M_{\PH}$ and light stop scenarios, a lower bound of about $155\GeV$ on the charged Higgs boson mass has been set assuming $m_{\rm h}=125\pm3\GeV$. The light stop scenario is excluded for $\mHp < 160\GeV$ assuming $m_{\rm h}=125\pm3\GeV$.
For $\mHp > \mt - \mb$, the \Hptb decay mode searches yield a lower limit on \tanbeta while the upper limit on \tanbeta is dominated by the results from the analysis of the \Hptaunu decay mode with the \tauhjets final state.
The low-$M_{\PH}$ scenario is completely excluded (\reffig{tanbetalimit:lowmh}) assuming the heavy CP-even MSSM Higgs boson mass is $m_\PH=125\pm3\GeV$.

In Figs.~\ref{fig:tanbetalimit:others}--\ref{fig:tanbetalimit:lowmh}, theoretical systematic uncertainties affecting the expected signal event yields are added to the limit computation, modelled as nuisance parameters, in addition to the
uncertainties discussed in \refsec{uncertainties}.
The uncertainty in the branching fractions of the charged Higgs boson is estimated from the decay width uncertainties as
in Ref.~\cite{Denner:2011} by scaling each partial width separately while fixing all others to their central values. This results in individual
theoretical uncertainties for each branching fraction. The width uncertainties comprise 
the uncertainty from missing higher order corrections to beyond LO EW diagrams (5\%),
missing higher order corrections to 
NLO QCD (2\%),
and $\Delta_{\rm b}$-correction uncertainties (3\%)~\cite{Dittmaier:2012vm}.
The  $\Delta_b$-correction arises from
the presence of squarks and gluino contributions in the charged Higgs boson Yukawa coupling to top and bottom quarks~\cite{Dittmaier:2011ti,Hofer:2009xb}.

For $\mHp = 90$--160\unit{\GeV}, the theoretical uncertainties in the signal yield include the uncertainties in the branching fractions 
for $\cPqt\to\PH^{+}\cPqb$ and $\PH^{+}\to\Pgt^{+}\Pgngt$
totalling 0.1--5.0\% depending on \mHp and \tanbeta.
Additionally, an uncertainty of 3\% is added to the simulated \ttbar background to take into account higher order corrections to the \ttbar cross section.
For $\mHp = 200$--$600\GeV$, the charged Higgs boson production cross section uncertainty and the uncertainty in the branching ratios are considered.
The cross section uncertainty varies between 22--32\% depending on \mHp, \tanbeta, and the MSSM benchmark scenario.
The uncertainty in \BHptaunu  varies between 0.4--5.0\% for $\tanbeta = 10$--60 depending
on \mHp and the MSSM benchmark scenario. The \BHptb uncertainty varies between 0.1--5.0\% for $\tanbeta = 1$--10 depending on \mHp and the MSSM benchmark scenario.
The theoretical branching fraction uncertainties for a given \mHp--\tanbeta point are summed linearly according to the LHC Higgs cross section working group prescription~\cite{Denner:2011,Dittmaier:2012vm}, but the cross section and branching fraction uncertainties are treated as independent nuisances. 
The expected limit improves by no more than 2\% if the theoretical uncertainties are treated in the statistical model as independent sources.

\section{Summary}
\label{sec:conclusion}
A search is performed for a charged Higgs boson with the CMS detector using a data sample 
corresponding to an integrated luminosity of $19.7 \pm 0.5\fbinv$ in proton-proton collisions at $\sqrt{s} = 8\TeV$.
The charged Higgs boson production in \ttbar decays and in \ProdHpFourandFiveFS 
is studied assuming \Hptaunu and \Hptb decay modes, using the \tauhjets, \mutauh, \ljets, and $\ell\ell'$ final states.
Data are found to agree with the SM expectations.

Model-independent limits without an assumption on the charged Higgs boson branching fractions are derived for 
the \Hptaunu decay mode in the \tauhjets final state.
Upper limits at 95\% CL of $\lightLimitTaunuHadr =1.2$--0.15\%
and $\heavyLimitTaunuHadr =0.38$--0.025\unit{pb} are set
for charged Higgs boson mass ranges $\mHp = 80$--$160\GeV$ and $\mHp = 180$--$600\GeV$, respectively.

Assuming $\BHptb = 1$, a 95\% CL upper limit of $\SigmaHpFourandFiveFS=2.0$--0.13\unit{pb} is set for a combination of the
\mutauh, \ljets, and $\ell\ell'$ final states 
for $\mHp = 180$--$600\GeV$.
This is the first experimental result on the \Hptb decay mode.
Here, cross section $\sigma(\Pp\Pp \to \cPqt(\cPqb)\PH^{\pm})$ stands for the sum $\sigma(\Pp\Pp \to \overline{\cPqt}(\cPqb)\PH^{+}) + \sigma(\Pp\Pp \to \cPqt(\overline{\cPqb})\PH^{-})$.

The results are interpreted in different MSSM benchmark scenarios
and used to set exclusion limits in the \mHp--\tanbeta parameter spaces. 
In the various models, a lower bound on the charged Higgs boson mass of about $155\GeV$ is set assuming $m_{\rm h}=125\pm3\GeV$.
The light-stop scenario is excluded for $\mHp < 160\GeV$ assuming $m_{\rm h}=125\pm3\GeV$, 
and the low-$M_\PH$ scenario defined in Refs.~\cite{Carena:2013,Heinemeyer:2013tqa} is completely excluded assuming $m_{\rm H}=125\pm3 \GeV$.

\section*{Acknowledgements \label{sec:acknowledgements}} 

\hyphenation{Bundes-ministerium Forschungs-gemeinschaft Forschungs-zentren} We congratulate our colleagues in the CERN accelerator departments for the excellent performance of the LHC and thank the technical and administrative staffs at CERN and at other CMS institutes for their contributions to the success of the CMS effort. In addition, we gratefully acknowledge the computing centres and personnel of the Worldwide LHC Computing Grid for delivering so effectively the computing infrastructure essential to our analyses. Finally, we acknowledge the enduring support for the construction and operation of the LHC and the CMS detector provided by the following funding agencies: the Austrian Federal Ministry of Science, Research and Economy and the Austrian Science Fund; the Belgian Fonds de la Recherche Scientifique, and Fonds voor Wetenschappelijk Onderzoek; the Brazilian Funding Agencies (CNPq, CAPES, FAPERJ, and FAPESP); the Bulgarian Ministry of Education and Science; CERN; the Chinese Academy of Sciences, Ministry of Science and Technology, and National Natural Science Foundation of China; the Colombian Funding Agency (COLCIENCIAS); the Croatian Ministry of Science, Education and Sport, and the Croatian Science Foundation; the Research Promotion Foundation, Cyprus; the Ministry of Education and Research, Estonian Research Council via IUT23-4 and IUT23-6 and European Regional Development Fund, Estonia; the Academy of Finland, Finnish Ministry of Education and Culture, and Helsinki Institute of Physics; the Institut National de Physique Nucl\'eaire et de Physique des Particules~/~CNRS, and Commissariat \`a l'\'Energie Atomique et aux \'Energies Alternatives~/~CEA, France; the Bundesministerium f\"ur Bildung und Forschung, Deutsche Forschungsgemeinschaft, and Helmholtz-Gemeinschaft Deutscher Forschungszentren, Germany; the General Secretariat for Research and Technology, Greece; the National Scientific Research Foundation, and National Innovation Office, Hungary; the Department of Atomic Energy and the Department of Science and Technology, India; the Institute for Studies in Theoretical Physics and Mathematics, Iran; the Science Foundation, Ireland; the Istituto Nazionale di Fisica Nucleare, Italy; the Ministry of Science, ICT and Future Planning, and National Research Foundation (NRF), Republic of Korea; the Lithuanian Aca\hyphenation{Bundes-ministerium Forschungs-gemeinschaft Forschungs-zentren} We congratulate our colleagues in the CERN accelerator departments for the excellent performance of the LHC and thank the technical and administrative staffs at CERN and at other CMS institutes for their contributions to the success of the CMS effort. In addition, we gratefully acknowledge the computing centres and personnel of the Worldwide LHC Computing Grid for delivering so effectively the computing infrastructure essential to our analyses. Finally, we acknowledge the enduring support for the construction and operation of the LHC and the CMS detector provided by the following funding agencies: the Austrian Federal Ministry of Science, Research and Economy and the Austrian Science Fund; the Belgian Fonds de la Recherche Scientifique, and Fonds voor Wetenschappelijk Onderzoek; the Brazilian Funding Agencies (CNPq, CAPES, FAPERJ, and FAPESP); the Bulgarian Ministry of Education and Science; CERN; the Chinese Academy of Sciences, Ministry of Science and Technology, and National Natural Science Foundation of China; the Colombian Funding Agency (COLCIENCIAS); the Croatian Ministry of Science, Education and Sport, and the Croatian Science Foundation; the Research Promotion Foundation, Cyprus; the Ministry of Education and Research, Estonian Research Council via IUT23-4 and IUT23-6 and European Regional Development Fund, Estonia; the Academy of Finland, Finnish Ministry of Education and Culture, and Helsinki Institute of Physics; the Institut National de Physique Nucl\'eaire et de Physique des Particules~/~CNRS, and Commissariat \`a l'\'Energie Atomique et aux \'Energies Alternatives~/~CEA, France; the Bundesministerium f\"ur Bildung und Forschung, Deutsche Forschungsgemeinschaft, and Helmholtz-Gemeinschaft Deutscher Forschungszentren, Germany; the General Secretariat for Research and Technology, Greece; the National Scientific Research Foundation, and National Innovation Office, Hungary; the Department of Atomic Energy and the Department of Science and Technology, India; the Institute for Studies in Theoretical Physics and Mathematics, Iran; the Science Foundation, Ireland; the Istituto Nazionale di Fisica Nucleare, Italy; the Ministry of Science, ICT and Future Planning, and National Research Foundation (NRF), Republic of Korea; the Lithuanian Academy of Sciences; the Ministry of Education, and University of Malaya (Malaysia); the Mexican Funding Agencies (CINVESTAV, CONACYT, SEP, and UASLP-FAI); the Ministry of Business, Innovation and Employment, New Zealand; the Pakistan Atomic Energy Commission; the Ministry of Science and Higher Education and the National Science Centre, Poland; the Funda\c{c}\~ao para a Ci\^encia e a Tecnologia, Portugal; JINR, Dubna; the Ministry of Education and Science of the Russian Federation, the Federal Agency of Atomic Energy of the Russian Federation, Russian Academy of Sciences, and the Russian Foundation for Basic Research; the Ministry of Education, Science and Technological Development of Serbia; the Secretar\'{\i}a de Estado de Investigaci\'on, Desarrollo e Innovaci\'on and Programa Consolider-Ingenio 2010, Spain; the Swiss Funding Agencies (ETH Board, ETH Zurich, PSI, SNF, UniZH, Canton Zurich, and SER); the Ministry of Science and Technology, Taipei; the Thailand Center of Excellence in Physics, the Institute for the Promotion of Teaching Science and Technology of Thailand, Special Task Force for Activating Research and the National Science and Technology Development Agency of Thailand; the Scientific and Technical Research Council of Turkey, and Turkish Atomic Energy Authority; the National Academy of Sciences of Ukraine, and State Fund for Fundamental Researches, Ukraine; the Science and Technology Facilities Council, UK; the US Department of Energy, and the US National Science Foundation.

Individuals have received support from the Marie-Curie programme and the European Research Council and EPLANET (European Union); the Leventis Foundation; the A. P. Sloan Foundation; the Alexander von Humboldt Foundation; the Belgian Federal Science Policy Office; the Fonds pour la Formation \`a la Recherche dans l'Industrie et dans l'Agriculture (FRIA-Belgium); the Agentschap voor Innovatie door Wetenschap en Technologie (IWT-Belgium); the Ministry of Education, Youth and Sports (MEYS) of the Czech Republic; the Council of Science and Industrial Research, India; the HOMING PLUS programme of the Foundation for Polish Science, cofinanced from European Union, Regional Development Fund; the OPUS programme of the National Science Center (Poland); the Compagnia di San Paolo (Torino); the Consorzio per la Fisica (Trieste); MIUR project 20108T4XTM (Italy); the Thalis and Aristeia programmes cofinanced by EU-ESF and the Greek NSRF; the National Priorities Research Program by Qatar National Research Fund; the Rachadapisek Sompot Fund for Postdoctoral Fellowship, Chulalongkorn University (Thailand); and the Welch Foundation, contract C-1845.demy of Sciences; the Ministry of Education, and University of Malaya (Malaysia); the Mexican Funding Agencies (CINVESTAV, CONACYT, SEP, and UASLP-FAI); the Ministry of Business, Innovation and Employment, New Zealand; the Pakistan Atomic Energy Commission; the Ministry of Science and Higher Education and the National Science Centre, Poland; the Funda\c{c}\~ao para a Ci\^encia e a Tecnologia, Portugal; JINR, Dubna; the Ministry of Education and Science of the Russian Federation, the Federal Agency of Atomic Energy of the Russian Federation, Russian Academy of Sciences, and the Russian Foundation for Basic Research; the Ministry of Education, Science and Technological Development of Serbia; the Secretar\'{\i}a de Estado de Investigaci\'on, Desarrollo e Innovaci\'on and Programa Consolider-Ingenio 2010, Spain; the Swiss Funding Agencies (ETH Board, ETH Zurich, PSI, SNF, UniZH, Canton Zurich, and SER); the Ministry of Science and Technology, Taipei; the Thailand Center of Excellence in Physics, the Institute for the Promotion of Teaching Science and Technology of Thailand, Special Task Force for Activating Research and the National Science and Technology Development Agency of Thailand; the Scientific and Technical Research Council of Turkey, and Turkish Atomic Energy Authority; the National Academy of Sciences of Ukraine, and State Fund for Fundamental Researches, Ukraine; the Science and Technology Facilities Council, UK; the US Department of Energy, and the US National Science Foundation.
Individuals have received support from the Marie-Curie programme and the European Research Council and EPLANET (European Union); the Leventis Foundation; the A. P. Sloan Foundation; the Alexander von Humboldt Foundation; the Belgian Federal Science Policy Office; the Fonds pour la Formation \`a la Recherche dans l'Industrie et dans l'Agriculture (FRIA-Belgium); the Agentschap voor Innovatie door Wetenschap en Technologie (IWT-Belgium); the Ministry of Education, Youth and Sports (MEYS) of the Czech Republic; the Council of Science and Industrial Research, India; the HOMING PLUS programme of the Foundation for Polish Science, cofinanced from European Union, Regional Development Fund; the OPUS programme of the National Science Center (Poland); the Compagnia di San Paolo (Torino); the Consorzio per la Fisica (Trieste); MIUR project 20108T4XTM (Italy); the Thalis and Aristeia programmes cofinanced by EU-ESF and the Greek NSRF; the National Priorities Research Program by Qatar National Research Fund; the Rachadapisek Sompot Fund for Postdoctoral Fellowship, Chulalongkorn University (Thailand); and the Welch Foundation, contract C-1845.

\bibliography{auto_generated}

\cleardoublepage \appendix\section{The CMS Collaboration \label{app:collab}}\begin{sloppypar}\hyphenpenalty=5000\widowpenalty=500\clubpenalty=5000\textbf{Yerevan Physics Institute,  Yerevan,  Armenia}\\*[0pt]
V.~Khachatryan, A.M.~Sirunyan, A.~Tumasyan
\vskip\cmsinstskip
\textbf{Institut f\"{u}r Hochenergiephysik der OeAW,  Wien,  Austria}\\*[0pt]
W.~Adam, E.~Asilar, T.~Bergauer, J.~Brandstetter, E.~Brondolin, M.~Dragicevic, J.~Er\"{o}, M.~Flechl, M.~Friedl, R.~Fr\"{u}hwirth\cmsAuthorMark{1}, V.M.~Ghete, C.~Hartl, N.~H\"{o}rmann, J.~Hrubec, M.~Jeitler\cmsAuthorMark{1}, V.~Kn\"{u}nz, A.~K\"{o}nig, M.~Krammer\cmsAuthorMark{1}, I.~Kr\"{a}tschmer, D.~Liko, T.~Matsushita, I.~Mikulec, D.~Rabady\cmsAuthorMark{2}, B.~Rahbaran, H.~Rohringer, J.~Schieck\cmsAuthorMark{1}, R.~Sch\"{o}fbeck, J.~Strauss, W.~Treberer-Treberspurg, W.~Waltenberger, C.-E.~Wulz\cmsAuthorMark{1}
% ...
\vskip\cmsinstskip
% ...
1:~~Also at Vienna University of Technology, Vienna, Austria\\
2:~~Also at CERN, European Organization for Nuclear Research, Geneva, Switzerland\\
% ...

\end{sloppypar}
\end{document}